\newcommand{\pcmq}{\mbox{cm$^{-2}$}}
\newcommand{\psec}{\mbox{s$^{-1}$}}

\newcommand{\ptev}{\mbox{TeV$^{-1}$}}

\newcommand{\ftev}{\mbox{ph~\pcmq~\psec~\ptev}}

\newcommand{\eunit}{\mbox{erg~\pcmq~\psec}}

\newcommand{\fermi}{\textit{Fermi}\xspace}
\newcommand{\hess}{H.E.S.S.\xspace}
\def\degree{\ensuremath{^\circ}}

\documentclass{aa}  

\usepackage{graphicx}
\usepackage{txfonts}
\usepackage[draft]{hyperref}
\usepackage{amsmath}
\usepackage{natbib}
\usepackage[switch]{lineno}
\usepackage{xspace}
\usepackage{xcolor} 
\DeclareMathOperator{\sgn}{sgn}
\begin{document} 
\bibpunct{(}{)}{;}{a}{}{,} 

\title{Analysis of the H.E.S.S.~public data release with ctools}

\author{
J. Kn\"odlseder\inst{1} \and
L. Tibaldo\inst{1} \and
D. Tiziani\inst{2} \and
A. Specovius\inst{2} \and
J. Cardenzana\inst{1} \and
M. Mayer\inst{3} \and
N. Kelley-Hoskins\inst{4} \and
L. Di Venere\inst{5} \and
S. Bonnefoy\inst{4} \and
A. Ziegler\inst{2} \and
S. Eschbach\inst{2} \and
P. Martin\inst{1} \and
T. Louge\inst{1} \and
F. Brun\inst{6} \and
M. Haupt\inst{4} \and
R. B\"uhler\inst{4}
}
\institute{
Institut de Recherche en Astrophysique et Plan\'etologie, Universit\'e de Toulouse, CNRS, CNES, UPS,
9 avenue Colonel Roche, 31028 Toulouse, Cedex 4, France
\and
Universit\"at Erlangen-N\"urnberg, Physikalisches Institut, Erwin-Rommel-Str. 1, 91058 Erlangen, Germany
\and
Department of Physics, Humboldt University Berlin, Newtonstr. 15, 12489 Berlin, Germany
\and
Deutsches Elektronen-Synchrotron, Platanenallee 6, 15738 Zeuthen, Germany
\and
Istituto Nazionale di Fisica Nucleare, Sezione di Bari, via Amendola 173, I-70126 Bari, Italy,
\and
CEA/IRFU/DPhP, CEA Saclay, Universit\'e Paris-Saclay, B\^at. 141, 91191 Gif-sur-Yvette, France
}

\date{Received 3 June, 2019; Accepted 9 October, 2019}

\abstract
{
The ctools open-source software package was developed for the scientific analysis of astronomical
data from Imaging Air Cherenkov Telescopes (IACTs), such as H.E.S.S., VERITAS, MAGIC, and the
future Cherenkov Telescope Array (CTA).
To date, the software has been mainly tested using simulated CTA data; however, upon the public
release of a small set of H.E.S.S.~observations of the Crab nebula, MSH 15--52, RX~J1713.7--3946,
and PKS 2155--304 validation using real data is now possible.
We analysed the data of the H.E.S.S.~public data release using ctools version 1.6 and compared
our results to those published by the H.E.S.S.~Collaboration for the respective sources.
We developed a parametric background model that satisfactorily describes the expected
background rate as a function of reconstructed energy and direction for each observation.
We used that model, and tested all analysis methods that are supported by ctools, including novel
unbinned and joint or stacked binned analyses of the measured event energies and reconstructed
directions, and classical On-Off analysis methods that are comparable to those used by the
H.E.S.S.~Collaboration.
For all analysis methods, we found a good agreement between the ctools results and the
H.E.S.S.~Collaboration publications considering that they are not always directly comparable 
due to differences in the datatsets and event processing software.
We also performed a joint analysis of H.E.S.S. and \fermi-LAT data of the Crab nebula,
illustrating the multi-wavelength capacity of ctools.
The joint Crab nebula spectrum is compatible with published literature values within
the systematic uncertainties.
We conclude that the ctools software is mature for the analysis of data from existing IACTs, as well 
as from the upcoming CTA. 
}


\maketitle

\section{Introduction}
\label{sec:intro}

Gamma-ray photons are powerful probes for the most extreme and violent phenomena in the
Universe.
Advancements made in the past two decades to the observation of very-high-energy (VHE) gamma
rays ($\ga100$ GeV) that use ground-based Imaging Air Cherenkov Telescopes (IACTs) have revealed
an unexpected ubiquity of sources that attest to the acceleration of particles to relativistic energies
\citep[e.g.][]{holder2012}.
Studying these VHE gamma rays provides unique insights into the acceleration physics, the nature
of the accelerated particles, relativistic particle propagation, the impact of the particles on the
source environment, and the distribution of particle accelerators in the Universe.
VHE gamma rays also probe the intergalactic medium and allow for the assessment of its content in
infrared radiation and magnetic fields. 
In addition, VHE gamma rays have the potential to probe the particle nature of dark matter and
physics beyond the standard model
\citep{CTAScienceBook}.

Making observations of VHE gamma rays using IACTs is complicated by the presence of a large
background due to cosmic-ray induced air showers that are difficult or sometimes impossible to
distinguish from the air showers generated by gamma rays.
In classical IACT data analyses, the residual background is taken into account by estimating its contribution
from gamma-ray source free regions in the same field of view, or, more rarely, from independent `Off'
observations.
Several techniques for background estimation exist and depending on the analysis aims, different
methods are advocated \citep[e.g.][]{berge2007}.
Limiting factors are that the methods generally rely on assumed symmetries of the background rates
in the field of view that are not necessarily verified, and that they generally require large enough
source-free regions for the background estimation in the same field of view.
Also, the handling of complex source morphologies and confused or overlapping sources is not
always straightforward \citep[e.g.][]{hess2018a}.

We propose an alternative data analysis approach that relies on a parametrised joint modelling of
the spatial and spectral event distributions, comprising model components for gamma-ray sources and
background.
The model is adjusted to the data by using a fitting algorithm, and estimates for all free model
parameters are obtained by maximising the likelihood function, given the model and the data.
Source confusion, as well as complex source morphologies, can be consistently treated with
this method, and uncertainties in the residual background can be treated as nuisance
parameters, reducing the systematic uncertainties in the analysis.
The challenge, however, is to devise a parametrised model that describes the residual
background with sufficient accuracy to allow for a reliable determination of gamma-ray
source characteristics.

We implemented these analysis methods in ctools, an open-source software package
developed in the context of the Cherenkov Telescope Array (CTA) project \citep{ctools2016, knoedlseder2016}.
The ctools are inspired by science analysis software available for existing high-energy 
astronomy instruments, and they follow the modular ftools model developed by the High 
Energy Astrophysics Science Archive Research Center.\footnote{
  \url{https://heasarc.gsfc.nasa.gov}}
The latest release of this software can be downloaded from
\url{http://cta.irap.omp.eu/ctools/}
which also provides user documentation and analysis tutorials.
In this paper we use the ctools release version 1.6.
The software has been extensively used on simulated CTA data
\citep{huetten2016, petropoulou2017, acero2017, defranco2017, burtovoi2017, balazs2017,
          patricelli2018, huetten2018, romano2018, landoni2019, tavecchio2019, yang2019,
          CTAScienceBook}
but so far had not yet been validated on existing IACT data.
Moreover, no systematic comparison between the classical and multi-component likelihood fitting
methods was performed on an existing IACT data set.
Upon the recent release of a small dataset obtained by the H.E.S.S.~telescopes into the public domain
\citep{hess2018b}, we can now address these issues.
This validation is an important milestone in the qualification of ctools for the upcoming CTA.
It also paves the way towards broader usage of ctools for the analysis of current IACT data.

\section{Data analysis}
\label{sec:analysis}

\subsection{Public data release}

The H.E.S.S.~public data release includes 48.6 hours of observing time, and comprises
observations of the Crab nebula, MSH 15--52, RX~J1713.7--3946, PKS 2155--304, and empty-field
regions performed with the H.E.S.S.~I array \citep{hess2018b}.
The data are typically divided into 28-minute-long observations during which the telescopes track a
position in the sky.
In total, 105 such observations are included in the data release.
Each observation comprises an event list and corresponding instrument response information,
it also uses the `Heidelberg calibration' and a Hillas event reconstruction, and applies standard cuts
and a spectral data quality selection (see \citealt{aharonian2006a} and \citealt{hess2018b} for details).
The data are provided in FITS format.\footnote{
  \url{http://gamma-astro-data-formats.readthedocs.org}}
Table \ref{tab:hess_release} summarises the key parameters of the datasets that are analysed in this
paper.

\begin{table}[!h]
\caption{Key parameters of datasets analysed in this paper.
$N_{\rm obs}$ is the number of observations,
$T_{\rm obs}$ is the exposure time in hours,
$T_{\rm live}$ is the live time in hours, and
$E_{\rm thres}$ is the minimum safe energy threshold.
\label{tab:hess_release}}
\centering
\begin{tabular}{l r r r r}
\hline\hline
Source name &
\multicolumn{1}{c}{$N_{\rm obs}$} &
\multicolumn{1}{c}{$T_{\rm obs}$} &
\multicolumn{1}{c}{$T_{\rm live}$} &
\multicolumn{1}{c}{$E_{\rm thres}$} \\
& & \multicolumn{1}{c}{h} & \multicolumn{1}{c}{h} & \multicolumn{1}{c}{GeV} \\
\hline
Crab nebula & 4 & 1.87 & 1.75 & 661 \\
MSH 15--52 & 20 & 9.13 & 8.33 & 380 \\
RX~J1713.7--3946 & 15 & 7.01 & 6.29 & 201 \\
PKS 2155--304 (flare) & 15 & 7.04 & 6.58 & 251 \\
PKS 2155--304 (steady) & 6 & 2.81 & 2.61 & 251 \\
Empty fields & 45 & 20.69 & 19.23 & 251 \\
\hline
\end{tabular}
\end{table}

\subsection{Data preparation}

We prepared and analysed the data using a set of Python scripts that we released on GitHub.\footnote{
  \url{https://github.com/jknodlseder/ctools-hess-dr1}}
After downloading\footnote{
  \url{https://www.mpi-hd.mpg.de/hfm/HESS/pages/dl3-dr1/hess_dl3_dr1.tar.gz}}
the H.E.S.S.~data we generated for each target an observation definition file that collects
all observations and the associated instrument response information in a single file.
These observation definition files are available for download.\footnote{
  \url{http://cta.irap.omp.eu/ctools/_downloads/hess_dl3_dr1_obs.tar.gz}}
For each target we then run the {\tt ctselect} tool on each of the files, and select all events within
$2\degree$ of the pointing direction with reconstructed energies above the safe energy threshold,
which is the energy at which the energy bias equals 10\%, at an offset of $1\degree$ in the field of
view \citep{hess2018b}.
The values of the safe energy thresholds are stored in the effective area component of the instrument
response function of each observation, and limiting the analysis to energies above this threshold
ensured that we used only events for which the instrument response information is accurate.

\subsection{Empty-field observations and background model}
\label{sec:bkgmodel}

We used the 45 empty-field observations that are included in the H.E.S.S.~public data release
to develop a parametric model that is suitable to describe the spatial and spectral distribution of
the H.E.S.S.~background events in the H.E.S.S.~public data release.
We used the factorisation
\begin{equation}
B(\vec{p}', E') = B_\mathrm{spatial}(\vec{p}'|E') \times B_\mathrm{spectral}(E')
\end{equation}
to model the background event distribution, where $\vec{p}'$ is the reconstructed event direction
in the field of view and $E'$ is the reconstructed event energy.
The first term represents the energy-dependent spatial component of the background model,
while the second term represents the field-of-view-averaged background spectrum.

\begin{figure}[!t]
\centering
\includegraphics[width=8.8cm]{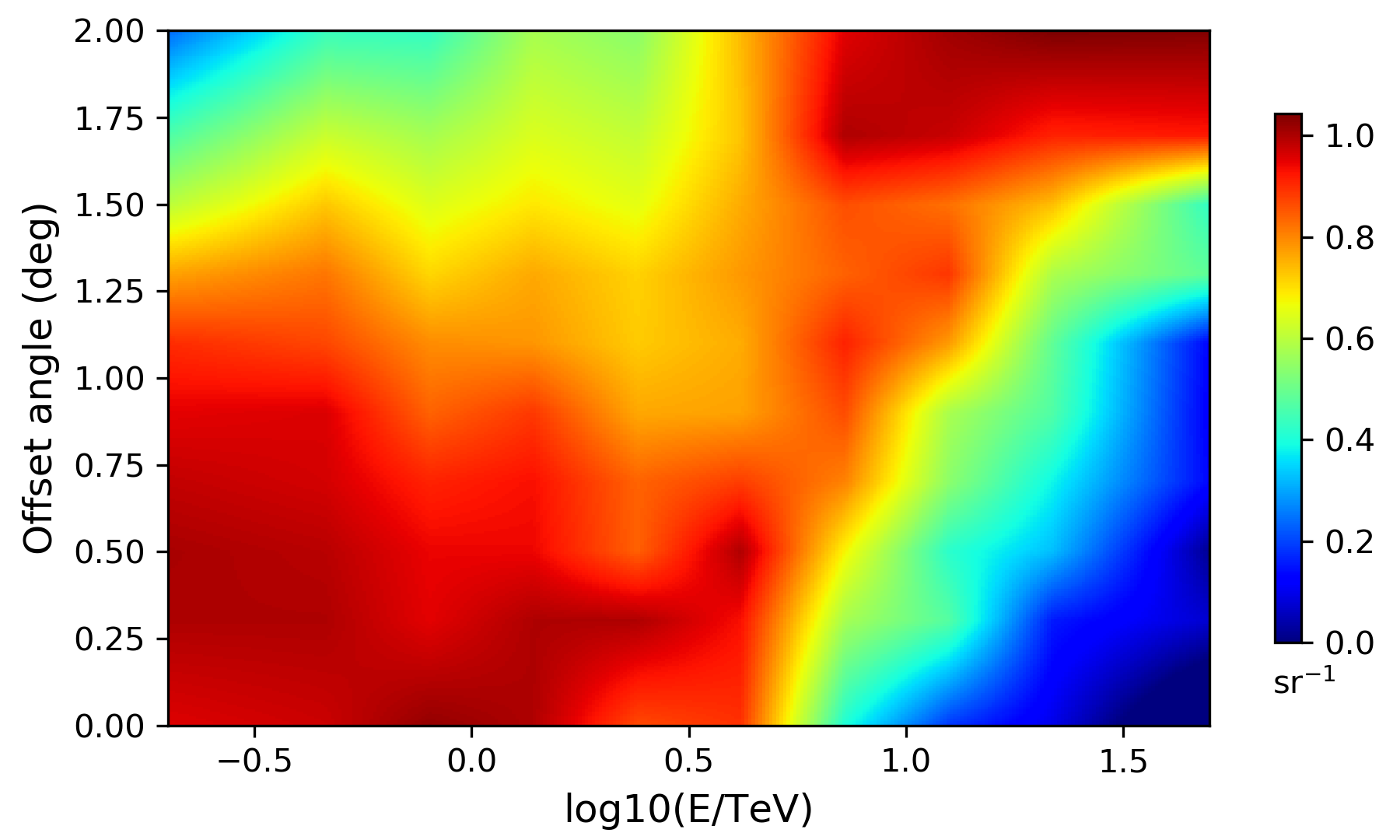}
\caption{
Background lookup table derived from all 45 empty-field observations.
The table was interpolated bi-linearly to visualise the function
$B_\mathrm{lookup}(\theta,E') $
that was used for background modelling.
\label{fig:bkg_lookup}
}
\end{figure}

Since the spatial distribution of background events in the H.E.S.S. data depends in a rather
complex way on event energy \citep{berge2007}, we derive the spatial component from a two-dimensional
lookup table $B_\mathrm{lookup}(\theta,E')$ that gives the background rate per unit solid angle as a function
of offset angle $\theta$ from the pointing direction and event energy $E'$.
We generated this lookup table by filling the events from the 45 empty-field observations into a
histogram spanned by ten $\theta$ bins of equal size and ten logarithmically-spaced energy bins,
and dividing the content of each bin by the solid angle of the corresponding $\theta$ interval.
The background spatial distribution is known to depend on the zenith angle of the observations
\citep{berge2007}.
However, with the limited statistics of the 45 empty-field observations of the public dataset 
we do not find significant variations in the background rate as a function of $\theta$ for different
zenith angles within the relevant energy range of the analysis.
Therefore, we neglect effects related to the zenith angle in the generation of the lookup table.
This aspect should be further investigated by using larger datasets.
The lookup table covers a $\theta$ range of $0\degree$ to $2\degree$ and an energy range
of $0.2$ TeV to $50$ TeV.
For each energy bin of the lookup table, all histogram values were divided by the maximum histogram
value of the energy bin, so that the maximum background rate value for a given energy bin is 1.
Figure \ref{fig:bkg_lookup} shows the resulting lookup table that we used throughout this
paper.
To evaluate $B_\mathrm{lookup}(\theta,E')$ at a specific pair of $\theta$ and $E'$ values, the
lookup table is bi-linearly interpolated in offset angle and the logarithm of energy.

\begin{figure}[!t]
\centering
\includegraphics[width=8.8cm]{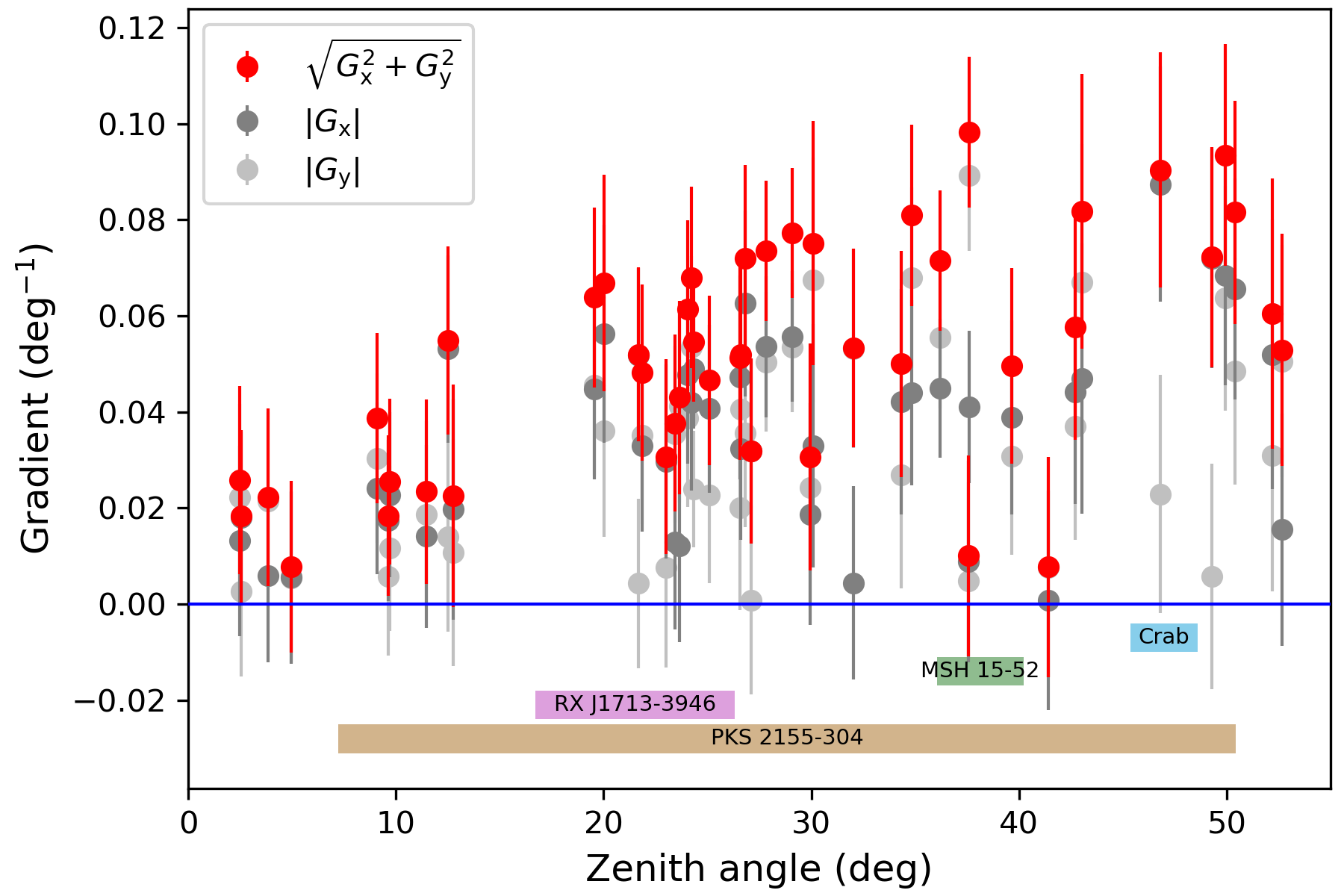}
\caption{
Fitted spatial background gradients as a function of zenith angle for the empty-field observations.
The red points show the total gradient
$(G_\mathrm{x}^2 + G_\mathrm{y}^2)^{1/2}$,
the grey points show the absolute values of the gradients
$G_\mathrm{x}$ and $G_\mathrm{y}$.
The horizontal bars indicate the intervals of zenith angle covered by the source observations.
\label{fig:off_nabla}
}
\end{figure}

The lookup table provides background rates that are azimuthally symmetric with respect to the pointing
direction.
Some observations, however, have non-negligible gradients in the background-event distribution
over the field of view.
Hence we multiply the lookup table by the function
\begin{equation}
B_\mathrm{gradient}(\vec{p}') = 1 + G_\mathrm{x} x + G_\mathrm{y} y ,
\label{eq:bkg_gradient}
\end{equation}
where $G_\mathrm{x}$ and $G_\mathrm{y}$ are
the gradients in the $x$ and $y$ direction, and
$x = \theta \cos \phi$ and
$y = \theta \sin \phi$
are the nominal field-of-view coordinates, with $\phi$ being the azimuth angle around the
pointing direction $x=y=0$.\footnote{
  The field-of-view coordinate system has the pointing direction as origin and is spanned
  by the detector coordinates $x$ and $y$. The offset angle is computed using
  $\theta=(x^2+y^2)^{1/2}$, the azimuth angle using $\phi=\arctan{y/x}$.}
Figure \ref{fig:off_nabla} shows the fitted $G_\mathrm{x}$ and $G_\mathrm{y}$
parameters, as well as the total gradient $(G_\mathrm{x}^2 + G_\mathrm{y}^2)^{1/2}$,
as a function of zenith angle for the 45 empty-field observations.
It illustrates that this correction is indeed required by the data.
Specifically, spatial gradients become more important for larger observation zenith angles, which
is expected to some extent since for larger zenith angles the atmospheric depth variation
over the field of view becomes larger.
For information we indicate also as horizontal bars the zenith angle ranges corresponding to
observations of the four sources that are included in the H.E.S.S.~public data release.
No obvious trends are seen for the spatial gradients as a function of azimuth angle.
To summarise, the energy-dependent spatial distribution of the background is described by
\begin{equation}
B_\mathrm{spatial}(\vec{p}'|E') = B_\mathrm{lookup}(\theta,E') \times B_\mathrm{gradient}(\vec{p}') ,
\end{equation}
with two parameters $G_\mathrm{x}$ and $G_\mathrm{y}$ adjusted for each observation.

\begin{figure*}[!t]
\centering
\includegraphics[width=\textwidth]{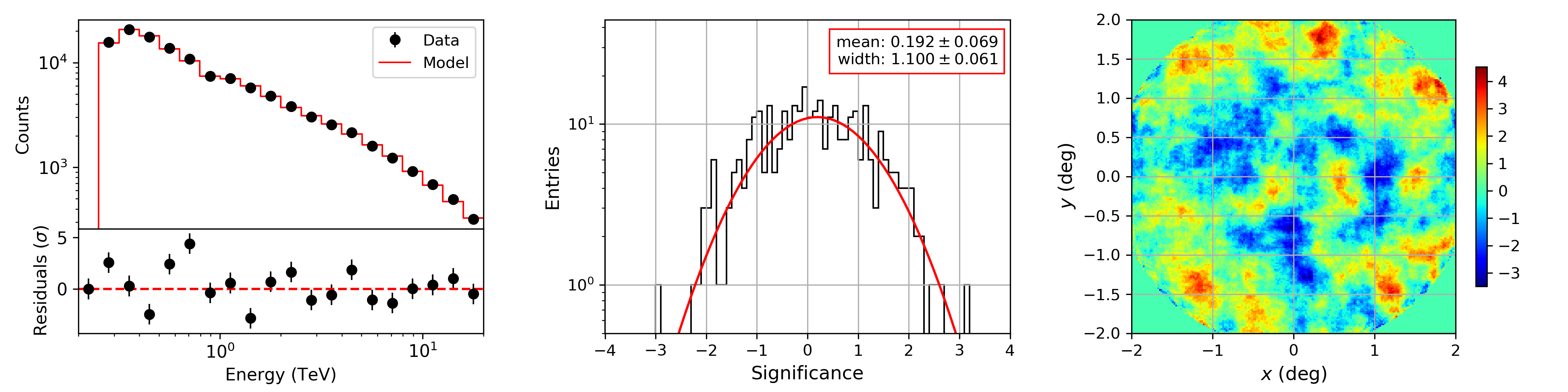}
\caption{
Stacked residuals of all 45 empty-field observations after fitting the {\tt csbkgmodel} background model
to each individual observation.
The left panel shows the residual count spectrum after summing over the entire field of view.
The centre panel shows the histogram of significances, determined after summing over energy and by
sampling the events into bins of $0.2\degree \times 0.2\degree$. 
The right panel shows a residual map in the field-of-view coordinate system.
The residual map is summed over all energies and was computed for a correlation radius of $0.2\degree$.
Residuals are shown in units of significance expressed in Gaussian $\sigma$ and are computed
using Equations \ref{eq:sigma} and \ref{eq:sigma0}.
\label{fig:off}
}
\end{figure*}

The spectral distribution of the background events varies substantially between observations,
and we therefore adopted a piece-wise broken power law
\begin{equation}
B_\mathrm{spectral}(E') = B_i \left( \frac{E'}{E'_i} \right)^{-\Gamma_i} \,\, \mathrm{for} \,\, E'_i \le E' < E'_{i+1} ,
\end{equation}
with $\Gamma_i = - \ln{(B_i/B_{i+1})} / \ln{(E'_i/E'_{i+1})}$ to accommodate this diversity.
The $B_i$ are the background rates at the node energies $E'_i$ which are determined by model
fitting.
We found that eight logarithmically-spaced energy nodes between the minimum and maximum
energy of the analysis interval provide a satisfactory description of the energy dependence
of the background rates of the H.E.S.S.~observations
(e.g.~left panel of Fig.~\ref{fig:off}).
$B_\mathrm{spectral}(E')$ is hence described by the eight parameters $B_0$ to $B_7$,
and in total, our background model $B(\vec{p}', E')$ has ten free parameters for each observation.

\begin{figure*}[!t]
\centering
\includegraphics[width=\textwidth]{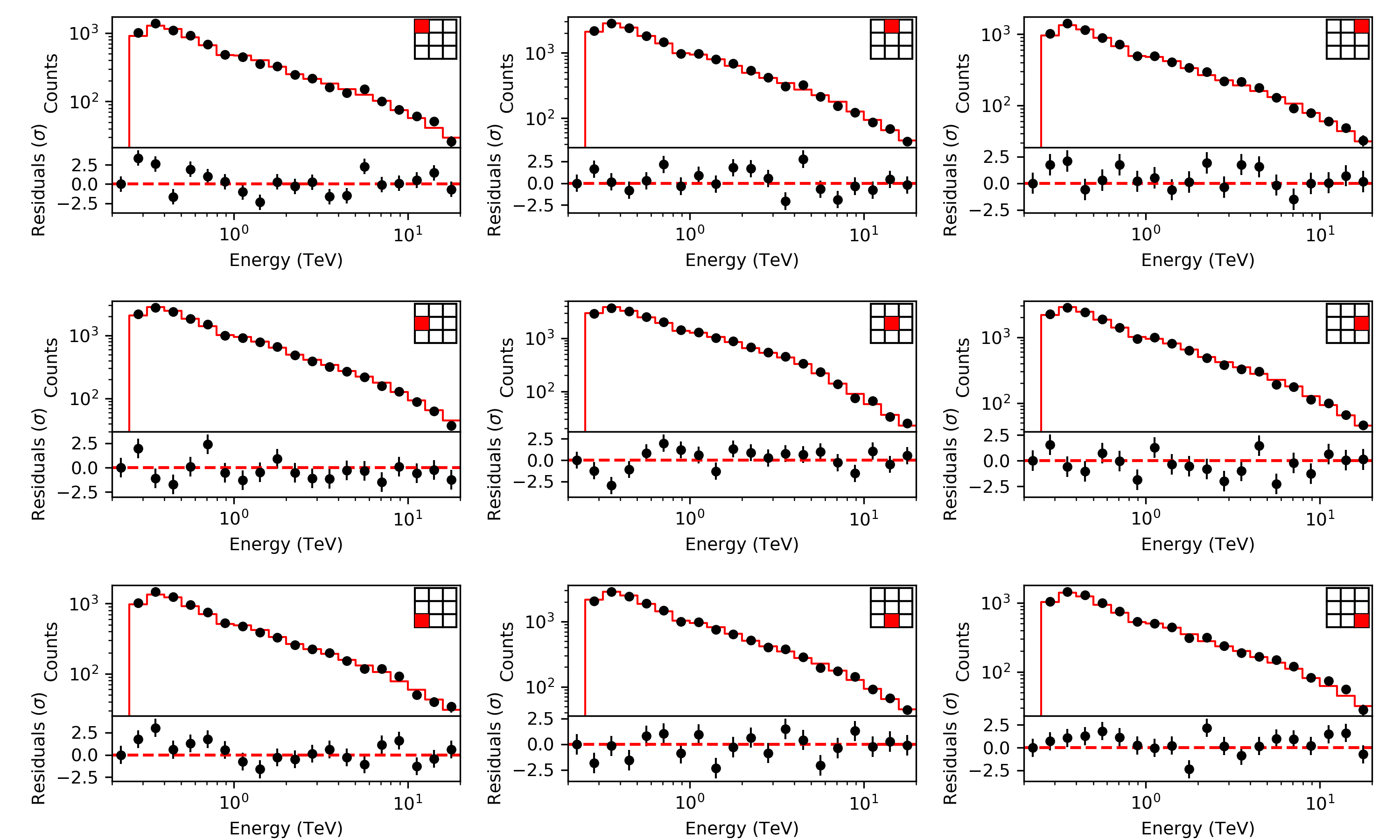}
\caption{
Counts spectra and residuals for nine spatial sub-regions of the field of view for all 45 empty-field
observations stacked in the field-of-view coordinate system.
The location of each sub-region in the field of view is indicated by the red box in the
$3\times3$ grid that is displayed in each panel.
\label{fig:off_sectors}
}
\end{figure*}

\begin{figure*}[!t]
\centering
\includegraphics[width=\textwidth]{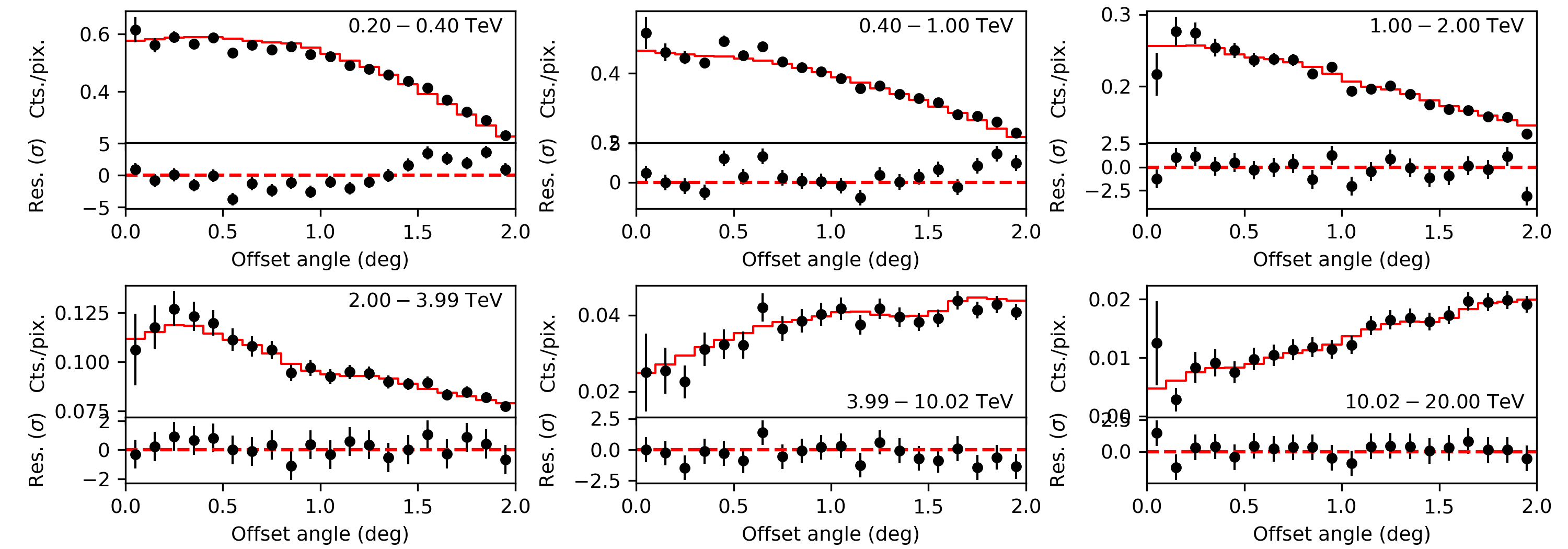}
\caption{
Radial counts profiles and residuals as a function of offset angle $\theta$ for six energy bands between
$0.2$ and $20$ TeV for all 45 empty-field observations stacked in the field-of-view coordinate
system.
The error bars on the counts are computed using the square root of the counts.
\label{fig:off_profiles}
}
\end{figure*}

To ensure the convergence of the maximum likelihood fitting algorithm for this specific
background model, initial values need to be determined for all ten parameters that are reasonably
close to their final values.
The tricky part is the determination of initial values for the background rates $B_i$ since
they vary by more than two orders of magnitude between the lowest and highest energies,
and their values differ significantly between observations.
We found that the following two-step process provides robust results for each individual
observation.
In a first step, a simplified background model is fitted to the data using a maximum likelihood algorithm,
where $B_\mathrm{spectral}(E')$ is assumed to follow a simple power law, reducing the number of
free spectral parameters from eight to two.
The resulting power law is then evaluated at each node energy to determine a first guess
of the values of $B_i$.
In a second step, the full background model is then fitted to the data to refine the $B_i$ values.
The resulting ten parameter values are then stored as the initial values of the background model
parameters.

We implemented the algorithm that prepares this and other background models for each individual
observation in the {\tt csbkgmodel} script.
The script, provided a lookup table for the spatial component, performs all necessary steps and
produces a model definition file that can be used as background model for the analysis of the
H.E.S.S.~data.

We fitted this background model to each empty-field observation using the unbinned maximum
likelihood algorithm implemented in {\tt ctlike} (see Appendix \ref{sec:ctlike} for a description of the
algorithm).
After fitting we performed a visual inspection of the fit residuals for each observation, as well
as for all 45 empty-field observations stacked together in the field-of-view coordinate system.
Figure \ref{fig:off} shows the spectral residuals, integrated over the field of view, and the spatial
residuals, integrated over energy, for the stacked observations.
Figure \ref{fig:off_sectors} shows the spectral residuals of the stacked observations for the field of view
divided into a grid of $3\times3$ sub-regions of equal size to verify that there are no significant spatial
variations in the residuals over the field of view.
Figure \ref{fig:off_profiles} shows the radial counts profiles and residuals as a function of offset angle
$\theta$ for six energy bands.
Overall the residuals look reasonably flat, and, although sometimes the scatter is larger
than expected from purely statistical fluctuations, no strong trends or biases are apparent.
As we will show later in the analysis of MSH 15--52 and RX~J1713.7--3946, where the background
model is most relevant due to the faintness of the gamma-ray source emission, our analysis results
will turn out to not be significantly affected by this scatter.

To ensure that the model can be reliably applied to observations different from those used in the lookup
table generation, we furthermore checked the residuals for five subsets of ten stacked empty-field
observations that were randomly selected.
We generated a background model for each of the ten observations using a lookup table that
was based on the remaining 35 empty-field observations, and hence the selected empty-field
observations are statistically independent of the background lookup table.
Figures \ref{fig:off_stacked_residuals_set1} -- \ref{fig:off_stacked_residuals_profiles_set5} in
Appendix \ref{sec:residuals} show the corresponding residual plots.
In all cases, the spectral, spatial, radial and sub-region residuals look acceptably flat, although some
deviations beyond the level expected from statistical fluctuations is once more clearly visible.

Therefore, in the rest of the paper we adopted the {\tt csbkgmodel} background model each time a
background model was required.
We ran the script on the observation definition file of each source to produce a model definition file that
defines an initial model for the distribution of the background events.
The script was run with the runwise option so that an independent background model component is
generated for each observation.
In this initial iteration of the background-model fitting we did not include a model for the gamma-ray source.
This yields a reasonable approximation of the background model parameters because the events 
recorded over the field of view are in general dominated by background.
This implies, however, that the initial model overestimates the background if gamma-ray emission is
present. 
In the next sections the background model will be refined for each source by fitting it to the data together
with models for the gamma-ray signal.
We checked by inspecting the residuals that the final model for source and background provided a good
representation of the data.

\subsection{Crab nebula observations}
\label{sec:crabobs}

\begin{figure}[!t]
\centering
\includegraphics[width=8.8cm]{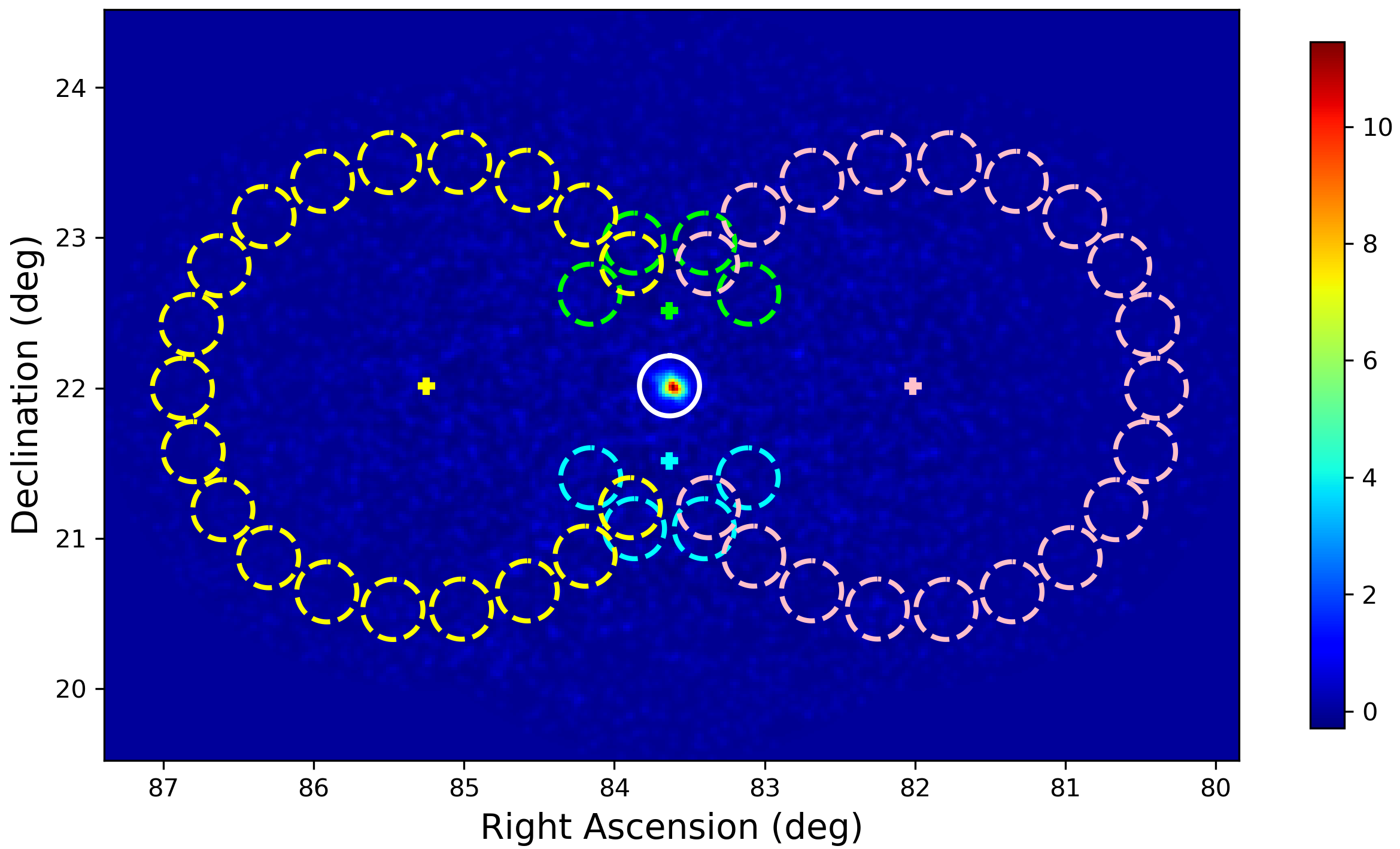}
\caption{
Background-subtracted counts map of the Crab nebula observations for the energy
band 670 GeV -- 30 TeV.
The map was computed for a $0.02\degree\times0.02\degree$ binning and was smoothed with
a Gaussian kernel of $\sigma=0.02\degree$ to reduce statistical noise.
The colour bar represents the number of excess counts per bin.
The white circle indicates the On region selected for On-Off analysis, coloured dashed circles
are the corresponding Off regions and plus symbols the pointing directions, where each colour
corresponds to one of the four observations.
Coordinates are for the epoch J2000.
\label{fig:crab_skymap}
}
\end{figure}

The Crab nebula was the first very-high energy gamma-ray source detected \citep{weekes1989}
and is one of the brightest objects in the TeV sky.
According to previous H.E.S.S.~analyses, the source spectrum is well described by an
exponentially cut-off power law \citep{aharonian2006a} and the source is spatially resolved
\citep{holler2017}.
Figure \ref{fig:crab_skymap} shows a background-subtracted counts map of the Crab nebula
observations, using the background model that was fitted to the data in the unbinned maximum
likelihood analysis.
The On and Off regions used for the classical analysis are also indicated (see below).

We started our analysis by fitting the Crab nebula observations using the unbinned maximum
likelihood algorithm implemented in {\tt ctlike}.
Event energies between 670 GeV and 30 TeV were considered.
Energy dispersion, which is the probability density of measuring an event energy of $E'$ if the
true energy is $E$, was accounted for, which was done for all analyses presented in this paper.
We used a point-source spatial model for the Crab nebula to allow for comparison of our results
with literature values.
The position parameters were left free in the model fit, and we tested several spectral models,
including
a power law (PL) $I(E) = k_0 (E/E_0)^{-\Gamma}$,
a power law with exponential cut-off (EPL) $I(E) = k_0 (E/E_0)^{-\Gamma} \exp(-E/E_c)$, and
a curved power-law model (CPL) $I(E) = k_0 (E/E_0)^{-\Gamma+\beta \ln(E/E_0)}$, also
known as log-parabola model.
The reference energy $E_0$ for all models was set to 1 TeV.

The fit results are summarised in Table \ref{tab:crab_spectral_results}.
We measured the detection significance of the source model using the so-called Test Statistic (TS)
which is defined as
\begin{equation}
\mathrm{TS} = 2 \ln L(M_s+M_b) - 2 \ln L(M_b)
\label{eq:ts}
\end{equation}
\citep{mattox1996},
where $\ln L(M_s+M_b)$ is the log-likelihood value obtained when fitting the source and the background
models together to the data, and $\ln L(M_b)$ is the log-likelihood value obtained when fitting only the
background model to the data.
Under the hypothesis that the background model $M_b$ provides a satisfactory fit of the data, TS
follows a $\chi^2_n$ distribution with $n$ degrees of freedom, where $n$ is the number of 
free parameters in the source model component. Therefore,
\begin{equation}
p = \int_\mathrm{TS}^{+\infty} \chi^2_n(x) \:\: \mathrm{d}x
\end{equation}
gives the chance probability (p-value) that the log-likelihood improves by TS/2 when adding the source
model $M_s$ due to statistical fluctuations only \citep{cash1979}.

\begin{table}[!t]
\small
\caption{Fit results from the unbinned analysis for the Crab nebula observations for different spectral
models assuming a point-source spatial model.
}
\label{tab:crab_spectral_results}
\centering
\setlength\tabcolsep{4.5pt}
\begin{tabular}{l c c c c c}
\hline\hline
Model & TS & $N_{\rm src}$ & $k_0$ & $\Gamma$ & $E_c$ or $\beta$ \\
\hline
PL & $2021.3$ & $684$ & $4.5 \pm 0.2$ & $2.64 \pm 0.07$ & - \\ 
EPL & $2030.6$ & $686$ & $5.0 \pm 0.4$ & $2.22 \pm 0.18$ & $7.4 \pm 3.2$ \\ 
CPL & $2029.9$ & $686$ & $4.5 \pm 0.2$ & $2.24 \pm 0.16$ &  $-0.23 \pm 0.09$ \\ 
\hline
EPL\tablefootmark{a} & n.c. & $4283$ & $3.84 \pm 0.09$ & $2.41 \pm 0.04$ & $15.1 \pm 2.8$ \\ 
CPL\tablefootmark{b} & n.c. & n.c. & $4.47 \pm 0.29$ & $2.23 \pm 0.18$ &  $-0.16 \pm 0.10$ \\ 
\hline
\end{tabular}
\tablefoot{
   TS is the Test Statistic and $N_{\rm src}$ is the number of events attributed to the source model.
   $k_0$ is given in units of $10^{-11}$ \ftev, $E_c$ in units of TeV and $\beta$ is dimensionless.
   \tablefoottext{a}{Values are from \citet{aharonian2006a} for their dataset III which covers the
                             period during which the observations of the H.E.S.S.~public data release were
                             taken.}
   \tablefoottext{b}{Values are from \citet{nigro2019}.}
   ``n.c.'' signals that the information was not communicated in the publications.
}
\end{table}

The largest TS value is obtained for the exponentially cut-off power law, and the TS difference of $9.3$
compared to the power-law model corresponds to a significance level of $3.1\sigma$ for the
detection of the cut off.
The curved power law gives a similar improvement, and statistically the model can not be distinguished
from the exponentially cut-off power law.
We can compare these results to the values published by \citet{aharonian2006a} for their
dataset III, which covers the period during which the observations of the H.E.S.S.~public data
release were taken, but which span a much longer live time of 10.6 hours compared to the 1.75
hours analysed in this paper.
For dataset III, \citet{aharonian2006a} quote $4283$ excess counts from the Crab nebula for
standard selection cuts, which corresponds to $404$ counts per hour live time, giving an
expected number of $707$ excess counts for our dataset.
This estimate is in excellent agreement with our $N_{\rm src}$ values in
Table~\ref{tab:crab_spectral_results}.

For dataset III, \citet{aharonian2006a} obtain
$k_0=(3.84 \pm 0.09) \times 10^{-11}$ \ftev,
$\Gamma = 2.41 \pm 0.04$ and
$E_c = 15.1 \pm 2.8$ TeV
for a power law with exponential cut-off.
Fitting the data with a source model where we fixed the spectral parameters to the values
of \citet{aharonian2006a} resulted in a TS value of $2007.2$ that is lower by $23.4$ with respect to
our best fitting exponentially cut-off power law.
This TS difference corresponds to a significance level of $4.4\sigma$, suggesting that our result
differs significantly from the one found by \citet{aharonian2006a}.
However, \citet{nigro2019} analysed the same dataset that we did using conventional IACT
analysis techniques, and for a curved power-law model they obtain\footnote{
  The value of $\beta$ has been scaled to account for the different definition of the curved power-law
  function in \citet{nigro2019}.}
$k_0=(4.47 \pm 0.29) \times 10^{-11}$ \ftev,
$\Gamma = 2.39 \pm 0.18$ and
$\beta = - 0.16 \pm 0.10$, which is compatible within statistical errors with our results.
Fitting a curved power-law model using their spectral parameters resulted in a TS value of
$2028.7$ that is smaller by only $1.2$ with respect to our best-fitting value, corresponding to a 
significance level of $1.1\sigma$, which confirms that our analysis is consistent with theirs.
This suggests that the difference with respect to \citet{aharonian2006a} should be attributed to
differences between their dataset III and the dataset included in the H.E.S.S.~public data release.
Firstly, the live time of their dataset is six times longer than the live time of the data analysed in
this work; hence their sample is statistically different from ours.
Secondly, the event reconstruction and background reduction software that was used to prepare
the data differs, implying that we do not really analyse the same events.
And thirdly, the instrument response functions that were used to recover the physical source
parameters are different, which may also explain differences in the fitted spectral
parameters.
This illustrates the limitations of the comparison of the ctools results with published literature
values, and emphasises the need for reference results for the H.E.S.S.~public data
release.

\begin{figure}[!t]
\centering
\includegraphics[width=8.8cm]{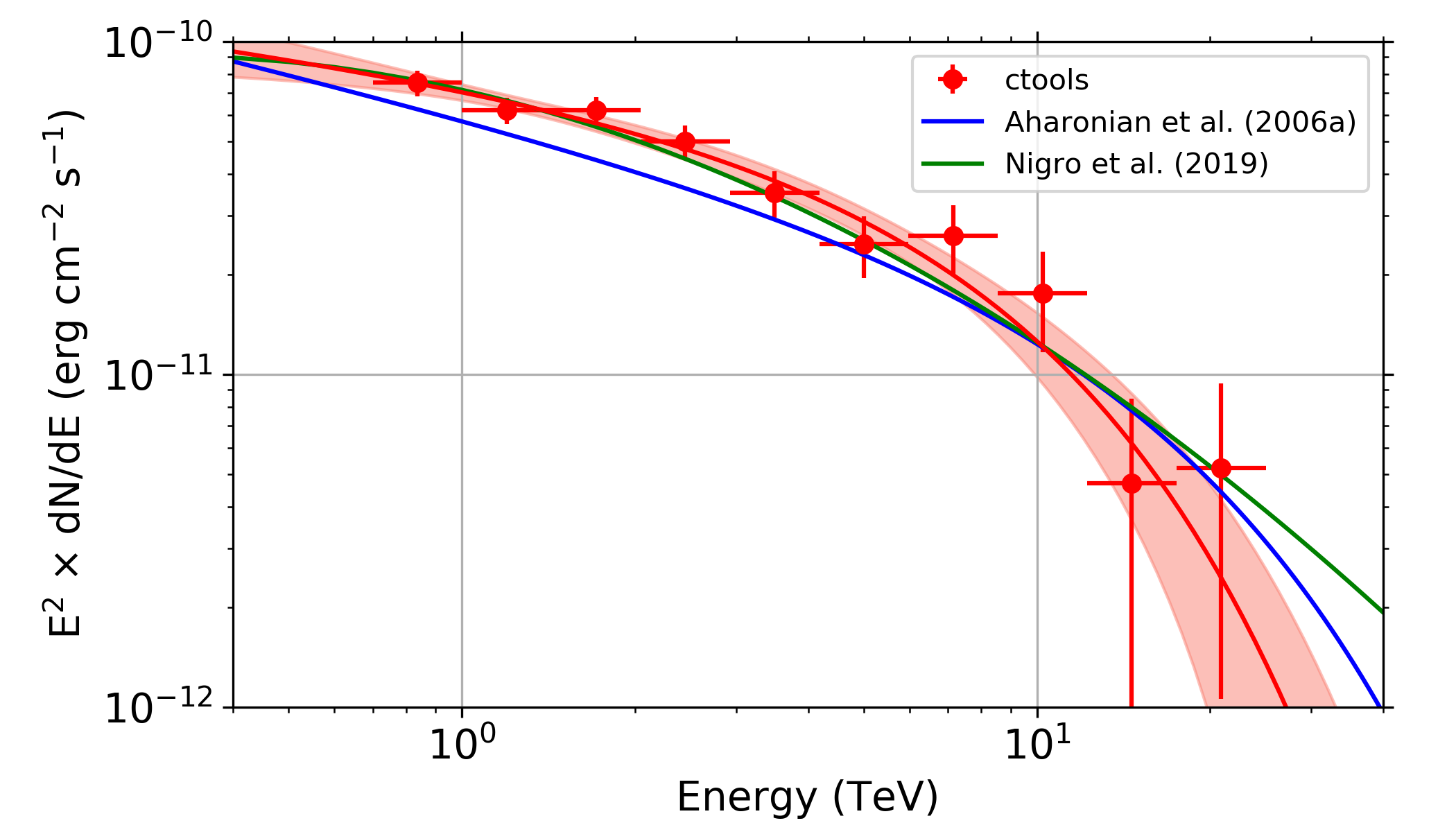}
\caption{
Crab nebula SED derived using {\tt csspec}.
Red data represent the unbinned {\tt csspec} analysis,
the red curve is the curved power-law model fitted by {\tt ctlike}, and the light red band is the 68\%
confidence level uncertainty band of the model fit that was determined using {\tt ctbutterfly}.
The blue curve is the power law with exponential cut-off obtained by \citet{aharonian2006a} for dataset 
III, and the green curve is the curved power-law obtained by \citet{nigro2019}.
\label{fig:crab_sed}
}
\end{figure}

\begin{table*}[!t]
\caption{Fit results for the Crab nebula observations under variations of the analysis method
assuming a point source with an exponentially cut-off power-law spectrum.
}
\label{tab:crab_spectral_method}
\centering
\setlength\tabcolsep{4.5pt}
\begin{tabular}{l c c c c c}
\hline\hline
Analysis method & TS & $N_{\rm src}$ & $k_0$ & $\Gamma$ & $E_c$ \\
\hline
H.E.S.S. published\tablefootmark{a} & n.c. & $4283$ & $3.84 \pm 0.09$ & $2.41 \pm 0.04$ & $15.1 \pm 2.8$ \\ 
Unbinned & $2030.6$ & $686$ & $5.0 \pm 0.4$ & $2.2 \pm 0.2$ & $7.4 \pm 3.2$ \\ 
Joint binned & $1967.5$ & $664$ & $5.0 \pm 0.4$ & $2.2 \pm 0.2$ & $6.9 \pm 2.9$ \\ 
Stacked binned & $1918.1$ & $651$ & $4.8 \pm 0.4$ & $2.1 \pm 0.2$ & $6.2 \pm 2.5$ \\ 
Joint On-Off (wstat) & $1134.3$ & $562$ & $5.0 \pm 0.4$ & $2.2 \pm 0.2$ & $6.8 \pm 3.4$ \\ 
Joint On-Off (cstat) & $996.1$ & $549$ & $4.8 \pm 0.4$ & $2.2 \pm 0.2$ & $8.0 \pm 4.6$ \\ 
Stacked On-Off (wstat) & $1359.5$ & $575$ & $5.0 \pm 0.4$ & $2.2 \pm 0.2$ & $7.4 \pm 3.6$ \\ 
Stacked On-Off (cstat) & $1102.7$ & $553$ & $4.8 \pm 0.4$ & $2.2 \pm 0.2$ & $7.2 \pm 3.5$ \\ 
\hline
\end{tabular}
\tablefoot{
   TS is the Test Statistic and
   $N_{\rm src}$ is the number of events attributed to the source model within the analysis region,
   which is the full field of view for the 3D analyses and the On region for the On-Off analyses.
   $k_0$ is given in units of $10^{-11}$ \ftev\ and $E_c$ in units of TeV.
   \tablefoottext{a}{Values are from \citet{aharonian2006a} for their dataset III which covers the
                             period during which the observations of the H.E.S.S.~public data release were
                             taken.}
   ``n.c.'' signals that the information was not communicated in the publication.
}
\end{table*}

We then used {\tt csspec} in unbinned analysis mode to derive the spectral energy distribution
(SED) of the Crab nebula.
{\tt csspec} performs maximum likelihood model fits for a predefined set of energy bins, where
model parameters are independently fit for each energy bin.
Since most of the energy nodes of the spectral background model component would actually lie
outside a given energy bin, we replaced the energy nodes $B_i$ by a simple power law for the
{\tt csspec} analysis, where the power law index was determined by fitting the power law over the full
energy range.
The spatial parameters $G_\mathrm{x}$ and $G_\mathrm{y}$ and the spectral power-law
prefactor were then fitted independently for each energy bin.
Comparing the SED to the fitted model curve then allows an assessment of the impact on the analysis
results of the choice of the functional form for $B_\mathrm{spectral}(E')$ and the assumed energy
independence of the spatial gradient.

The SED derived using {\tt csspec} for the Crab nebula observations is shown in Fig.~\ref{fig:crab_sed}.
The figure also shows a butterfly diagram of the curved power-law that we generated using
{\tt ctbutterfly} in unbinned analysis mode.
The SED nicely follows the curved power-law spectrum, demonstrating that the spectral results
are robust with respect to the specific parametrisation of the background model.
For comparison, we also show the power law with exponential cut-off obtained by
\citet{aharonian2006a} for dataset III, and the curved power-law obtained by \citet{nigro2019}.
The published H.E.S.S.~result differs significantly from our best-fitting spectrum, but as explained
above, several reasons could explain this discrepancy.
Our spectrum is however in excellent agreement with the spectrum of \citet{nigro2019} that was
based on the same data, confirming the good agreement between the analysis results.

The results we have presented to this point were obtained using an unbinned maximum likelihood
fitting method.
The ctools offer alternative analysis methods, including
a joint binned analysis, where the events for each observation are filled into a 3D counts cube spanned
by Right Ascension, Declination, and the logarithm of the reconstructed energy, and
a stacked binned analysis, where the events of all observations are combined into a single 3D counts
cube and the effective response for the stacked cube is computed and used.
Collectively we call these methods 3D analyses.

\begin{figure}[!t]
\centering
\includegraphics[width=8.8cm]{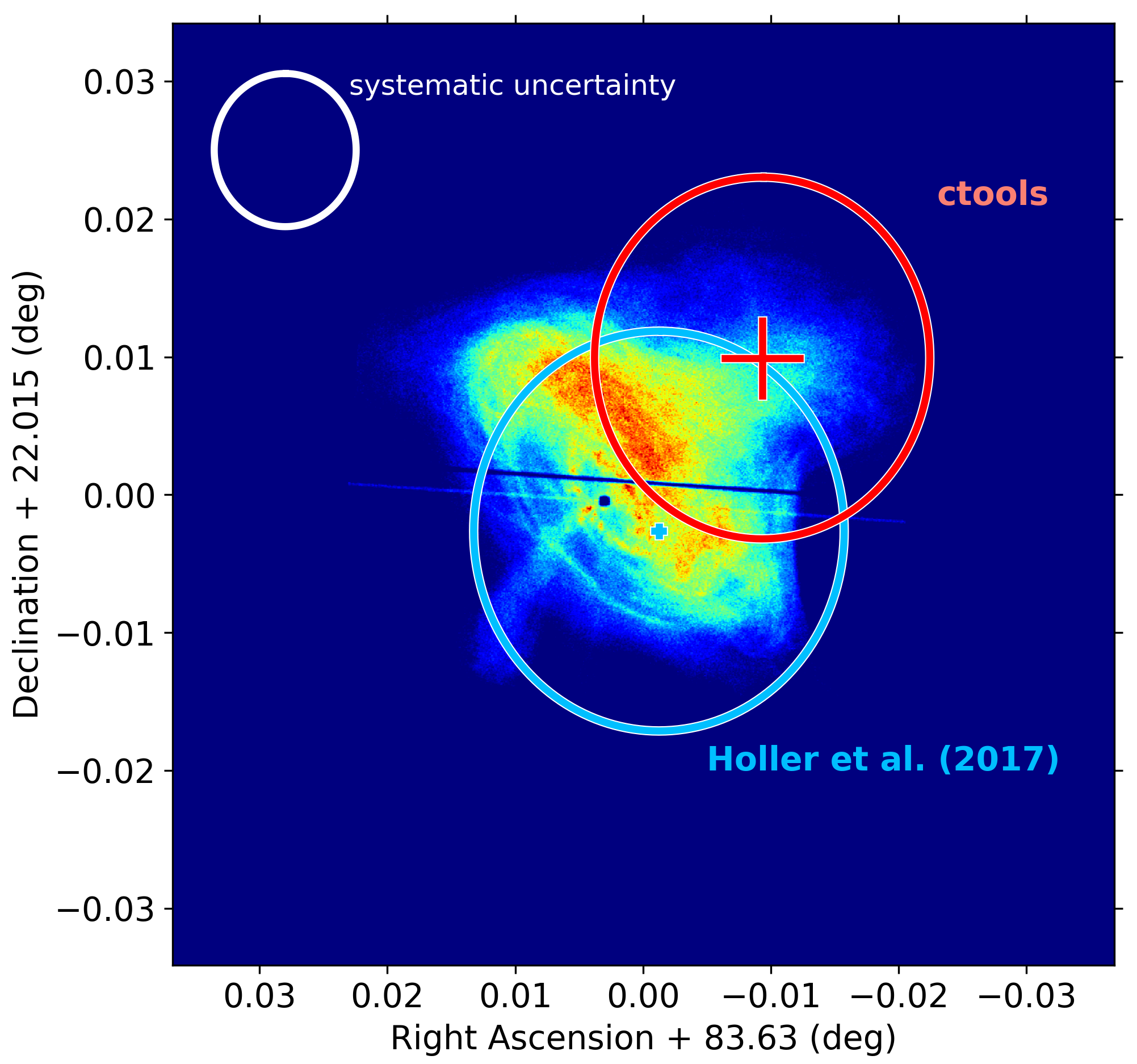}
\caption{
Gamma-ray extension, given as the $1\sigma$ radius of a 2D Gaussian fit, of the Crab nebula as
derived from \citet{holler2017} (blue) and using ctools (red) superimposed on a Chandra 0.3--10 keV
X-ray image of the Crab nebula (credit: NASA/CXC/SAO).
The crosses indicate the statistical uncertainty in the centroid position as quoted by \citet{holler2017}
(blue) and as derived from the Gaussian model fitting using {\tt ctlike} (red).
The systematic positioning accuracy of H.E.S.S. of $20''$ \citep{holler2017} is indicated as a white
circle.
Coordinates are for the epoch J2000.
\label{fig:crab_extension}
}
\end{figure}

\begin{figure*}[!t]
\centering
\includegraphics[width=\textwidth]{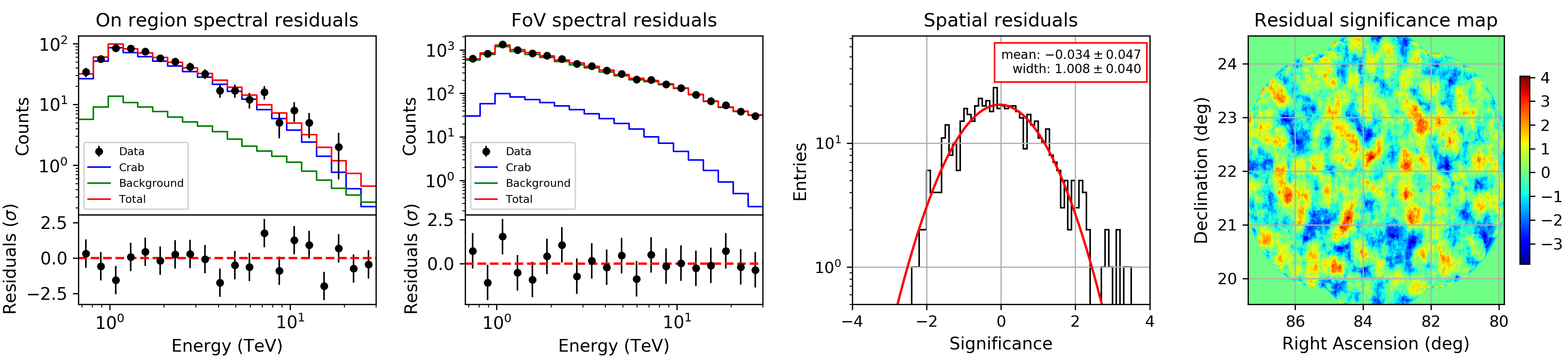}
\caption{
Residuals after fitting the Crab observations using a 2D Gaussian spatial model with an
exponentially cut-off power-law spectrum using an unbinned maximum likelihood fit.
The leftmost panel shows the counts and model spectra and the residuals after subtraction
of the source and background models for the On region.
The second panel shows the spectra and residuals for the entire field of view. 
In both panels, red lines represent the total predicted model counts, blue lines the predicted source
counts and green lines the predicted background counts.
The third panel shows the histogram of significances, determined after summing over the spectral
dimension and by sampling the events into bins of $0.2\degree \times 0.2\degree$.
A Gaussian was fitted to the significance histogram, and the best fitting mean and width of the
Gaussian are given together with the statistical fit errors in the plots.
Perfect residuals would lead to a mean of zero and a width of unity.
The rightmost panel shows a residual significance map summed over all energies for a correlation
radius of $0.2\degree$.
The map is in significance units expressed in Gaussian $\sigma$.
\label{fig:crab_residuals}
}
\end{figure*}

In addition On-Off analyses are also supported, where only the data from an On region are
analysed and the background is estimated from one or several Off regions.
The On-Off analysis, which corresponds to the classical technique for analysing data from IACTs,
exists in several variants.
Firstly, observations can either be analysed jointly or stacked.
For a joint analysis, On-Off spectra from individual observations are kept separately and a joint
maximum likelihood fitting is performed using the effective instrument response for the On
region of each observation.
For a stacked analysis, all events from individual observations are combined into a single On and
a single Off spectrum, and the resulting effective instrument response for the combined observations
is used in the maximum likelihood fitting.

Secondly, maximum likelihood fitting can be either performed using the wstat or the cstat
statistic (see Appendix \ref{sec:wstat_cstat}).
For wstat, it is assumed that the background rate per solid angle is identical in the On and
Off regions, and hence no explicit background model is needed for the analysis.
Such a situation may occur if the On and Off regions are symmetrically located around the
pointing direction, provided that background gradients over the field of view are negligible.
Conversely, for cstat, a background model is used to describe the background rate differences
between On and Off regions that may occur if the regions are not symmetrically located
around the pointing direction, or in the presence of significant background gradients.
For this paper, we used the background model produced by {\tt csbkgmodel} for the cstat
analysis, with individual components for each observation for the joint case, and a single component
for the stacked case.
The On-Off analysis methods are described in more detail in Appendix \ref{sec:onoff_methods}.
We note that for each observation, the cstat analysis has the eight background rates $B_i$ as free
parameters, while the wstat analysis implicitly has one background rate per energy bin as nuisance
parameters.

Table \ref{tab:crab_spectral_method} summarises the Crab nebula spectral fitting results for
a point source with an exponentially cut-off power-law spectrum.
All spatial and spectral source parameters, as well as the parameters of the background model,
were free parameters in the model fits, except for the On-Off analyses, which do not consider
the spatial source and background components and consequently do not allow for an adjustment
of the corresponding parameters.
Spatial bins of $0.02\degree \times 0.02\degree$ and 40 logarithmically spaced energy bins
were used for the binned analysis, with $200 \times 200$ bins around the pointing direction
of each observation for the joint analysis, and $350 \times 250$ bins around the Crab nebula
position (Right Ascension $83.63\degree$ and Declination $22.01\degree$) for the stacked
analysis.
The same number of energy bins was also used for the On-Off analysis, and the On region
was defined as a circle of $0.2\degree$ in radius centred on the Crab nebula position.
The reflected regions method was used to define the Off regions \citep{berge2007}.
The On region and reflected Off regions are shown for illustration in Fig.~\ref{fig:crab_skymap}.

As can be seen from Table \ref{tab:crab_spectral_method}, all analysis methods give compatible
results.
The results for the joint binned analysis are very close to those for the unbinned analysis,
validating the consistent implementation of both analysis methods in the ctools package.
Also the stacked binned analysis gives comparable results, although the
background model for this analysis was generated by stacking the initial background
models obtained for each observation which ignore any source contributions, and
thus formally overestimate the background in case that some source emission is
present.
The nevertheless good agreement with the other analyses can be explained by the fact that,
even for observations of a strong source like the Crab nebula, the field-of-view integrated counts
are largely dominated by background (cf.~Fig.~\ref{fig:crab_residuals}).
In addition, the stacked background model is renormalised during the source fitting,
compensating to a large extent the initial overestimation of the background model.

The On-Off analysis results are close to those obtained with the 3D analyses, providing an
important cross-check between our novel analysis approach with the classical IACT analysis
methods.
The number of source events for the On-Off analyses are about $20\%$ smaller compared to
the 3D analyses, which is in agreement with the fraction of events that are found in the tail of the
point-spread function beyond the On region cut of $0.2\degree$.
We also note that results obtained using wstat are compatible with those obtained using cstat,
suggesting that details of the background model affect only marginally the Crab analysis,
which is understandable since the events in the On region are largely dominated by the
source (cf.~Fig.~\ref{fig:crab_residuals}).
Finally, the stacked On-Off analyses provide results that are compatible with the joint On-Off
analyses, demonstrating once again that the loss of information due to the stacking of individual
observations does not substantially degrade the analysis results.

As next step we investigated the morphology of the Crab nebula.
Using a dedicated analysis, \citet{holler2017} reported recently the measurement of the
extension of the Crab nebula, and although the Instrument Response Functions (IRFs) in the
H.E.S.S.~data release have a lower precision compared to those used by \citet{holler2017}, we wanted
to check whether we were able to reproduce this result with ctools.
We therefore replaced the point-source model by a 2D Gaussian model, similar to the model
used by \citet{holler2017}, and fitted the observations using {\tt ctlike} in unbinned mode.
We obtained a 2D Gaussian extent of $\sigma = 47'' \pm 18''$ using a power law with exponential
cut-off for the spectral component, which is compatible with the value of $\sigma = 52'' \pm 3''$
derived by \citet{holler2017}.
Using a power law or curved power law spectral shape for the source changed the extension
value by at most $3''$.
Figure \ref{fig:crab_extension} illustrates our result in comparison to the result of \citet{holler2017},
superimposed on a Chandra X-ray image of the Crab nebula.
The best-fitting centroid of the 2D Gaussian, which we determined to
$83.621\degree \pm 0.003\degree$ in Right Ascension and
$22.025\degree \pm 0.003\degree$ in Declination (J2000), is slightly offset from the Crab nebula
centre.
We did not find such large offsets between H.E.S.S.~results and ctools localisations for other
sources (see below); hence it is plausible to attribute this offset to systematic uncertainties
in the reconstructed H.E.S.S.~data and/or response functions.

We inspected the spatial and spectral residuals of the fit; corresponding plots are shown in
Fig.~\ref{fig:crab_residuals}.
The spectral residuals are very flat, and also the significance distribution of the residual counts
is very close to the expectations.
The spatial residuals are also relatively flat and appear to be homogeneously distributed
over the field of view.
We also examined the spectral residuals for nine spatial sub-regions, and the radial counts
profiles and residuals as a function of offset angle $\theta$ for six energy bands.
The corresponding plots are shown in Figs.~\ref{fig:crab_residuals_sectors} and \ref{fig:crab_residuals_profiles}
of Appendix \ref{sec:residuals}.
The residuals are relatively flat, and in particular, they do not show evidence for any strong biases
or  trends over the field of view, although fluctuations beyond those expected from pure statistics
exist and small trends are discernable.

\subsection{MSH 15--52 observations}
\label{sec:mshobs}

MSH 15--52 is a radio supernova remnant that houses an X-ray pulsar embedded in a pulsar
wind nebula (PWN).  
The discovery of TeV gamma-ray emission from the MSH 15--52 PWN was reported by
\citet{aharonian2005},
which provided the first image of an extended PWN in this energy range.
The H.E.S.S. observations revealed a complex emission morphology that was fitted by a
two-dimensional Gaussian model with major and minor axes of $6.4' \pm 0.7'$ and
$2.3' \pm 0.5'$ and a position angle of $139\degree \pm 13\degree$.\footnote{
  \citet{aharonian2005} quote a value of $41\degree \pm 13\degree$ with respect to the
  Right Ascension axis, but this value is obviously clockwise from celestial north.
  Throughout this paper, we use the standard astronomical convention of measuring position
  angles counterclockwise from celestial north.}
The H.E.S.S.~data were consistent with a power-law spectrum with an index
$\Gamma = 2.27 \pm 0.03$ and a differential flux at 1 TeV of $(5.7 \pm 0.2) \times 10^{-12}$ \ftev\
\citep{aharonian2005}.

\begin{table*}[!t]
\small
\caption{Fit results for the MSH 15--52 observations for different analysis methods.}
\label{tab:msh_results}
\centering
\begin{tabular}{l c c c c c c c c c}
\hline\hline
Analysis method & TS & $N_{\rm src}$ & R.A. (deg) & Decl. (deg) & $\sigma_{\rm maj}$ & $\sigma_{\rm min}$ & PA (deg) & $k_0$ & $\Gamma$ \\
\hline
H.E.S.S. published\tablefootmark{a} & n.c. & 3481 & $228.529 \pm 0.006$ & $-59.159 \pm 0.003$ & $6.4' \pm 0.7'$ & $2.3' \pm 0.5'$ & $139 \pm 13$ & $5.7 \pm 0.2$ & $2.27 \pm 0.03$ \\
Unbinned & $762.1$ & $1134$ & $228.548 \pm 0.012$ & $-59.174 \pm 0.007$ & $6.3' \pm 0.5'$ & $3.7' \pm 0.4'$ & $147 \pm 6$ & $6.1 \pm 0.3$ & $2.30 \pm 0.06$ \\ 
Joint binned & $781.1$ & $1139$ & $228.552 \pm 0.012$ & $-59.174 \pm 0.007$ & $6.4' \pm 0.5'$ & $4.0' \pm 0.4'$ & $149 \pm 7$ & $6.4 \pm 0.4$ & $2.28 \pm 0.06$ \\ 
Stacked binned & $767.6$ & $1124$ & $228.548 \pm 0.012$ & $-59.172 \pm 0.007$ & $6.6' \pm 0.5'$ & $3.8' \pm 0.4'$ & $148 \pm 6$ & $6.4 \pm 0.4$ & $2.26 \pm 0.06$ \\ 
Joint On-Off (wstat) & $440.9$ & $870$ & $228.548$ & $-59.174$ & $6.3'$ & $3.7'$ & $147$ & $6.5 \pm 0.4$ & $2.35 \pm 0.07$ \\ 
Joint On-Off (cstat) & $527.5$ & $877$ & $228.548$ & $-59.174$ & $6.3'$ & $3.7'$ & $147$ & $6.5 \pm 0.4$ & $2.30 \pm 0.06$ \\ 
Stacked On-Off (wstat) & $530.1$ & $882$ & $228.548$ & $-59.174$ & $6.3'$ & $3.7'$ & $147$ & $6.5 \pm 0.4$ & $2.30 \pm 0.06$ \\ 
Stacked On-Off (cstat) & $548.4$ & $890$ & $228.548$ & $-59.174$ & $6.3'$ & $3.7'$ & $147$ & $6.6 \pm 0.4$ & $2.30 \pm 0.06$ \\ 
\hline
\end{tabular}
\tablefoot{
   TS is the Test Statistic,
   $N_{\rm src}$ is the number of source events attributed to the source within the analysis region,
   which is the full field of view for the 3D analyses and the On region for the On-Off analyses,
   R.A. is the Right Ascension and Decl. the Declination of the fitted source model (J2000),
   $\sigma_{\rm maj}$ and $\sigma_{\rm min}$ are the major and minor axes of the elliptical shape and 
   PA is the position angle, counted counterclockwise from celestial north.
   $k_0$ and $\Gamma$ are the parameters of the power law $I(E) = k_0 (E/E_0)^{-\Gamma}$
   which was used as the spectral component, where $E_0=1$ TeV and $k_0$ is in units of
   $10^{-12}$ \ftev.
   \tablefoottext{a}{Values are from \citet{aharonian2005}.}
   ``n.c.'' signals that the information was not communicated in the publication.
}
\end{table*}

\begin{figure}[!t]
\centering
\includegraphics[width=8.8cm]{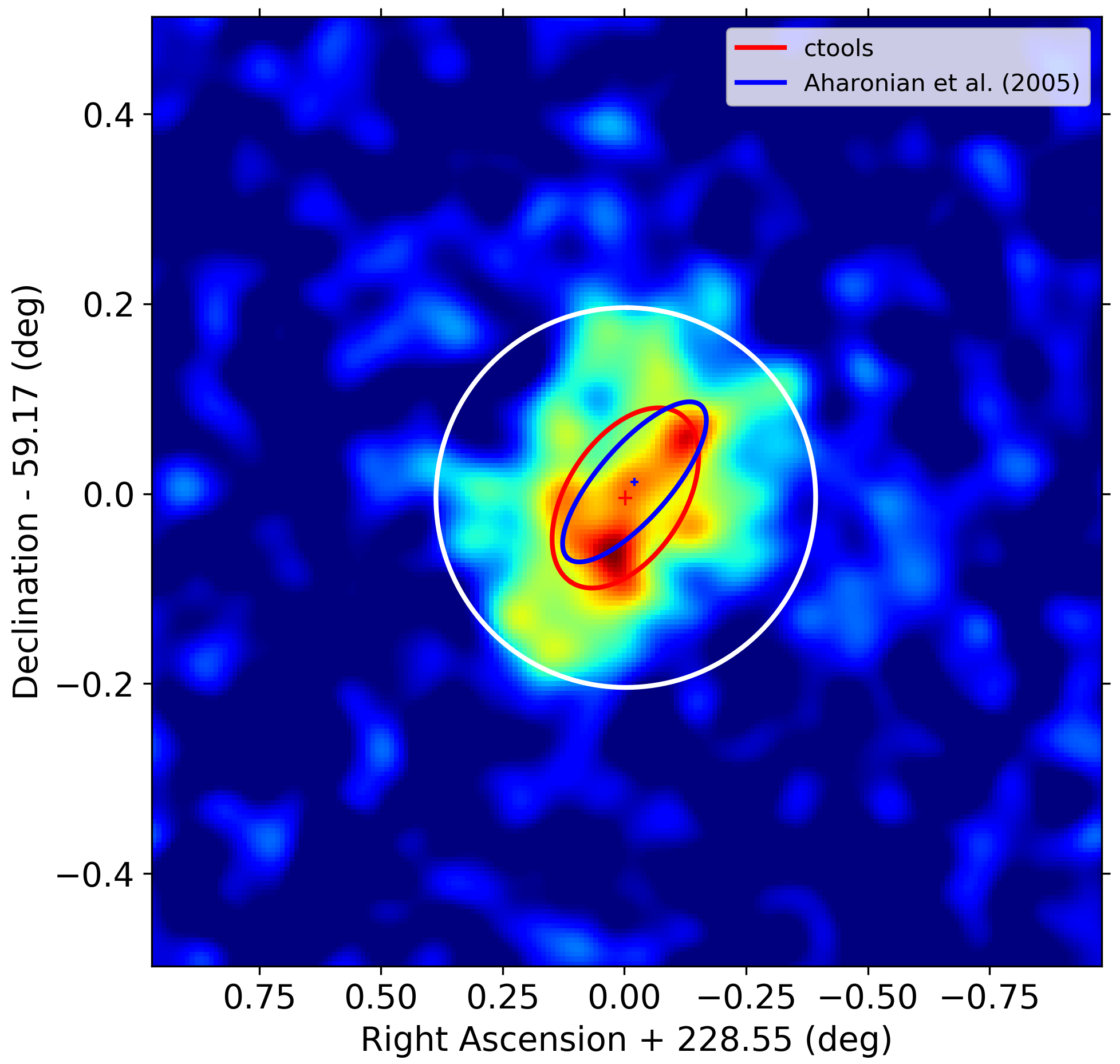}
\caption{
Background-subtracted counts map of the MSH 15--52 observations for the energy
band 381 GeV -- 40 TeV.
The map was computed for a $0.005\degree\times0.005\degree$ binning and was smoothed with a
Gaussian kernel of $\sigma=0.02\degree$ to reduce statistical noise.
The $1\sigma$ contour of the elliptical Gaussian fitted using ctools is indicated as a red ellipse.
The fit results from \citet{aharonian2005} are indicated as a blue ellipse.
The crosses indicate the statistical uncertainty in the centroid positions.
The white circle indicates the On region used in the On-Off analysis.
Coordinates are for the epoch J2000.
\label{fig:msh_skymap}
}
\end{figure}

We therefore analysed the MSH 15--52 observations using a source model composed of an
elliptical 2D Gaussian for the spatial component and a power law for the spectral component.
Events with energies between 381 GeV and 40 TeV were considered.
As for the Crab nebula, we performed unbinned and binned 3D maximum likelihood analyses, and
the different variants of the On-Off analysis that are available in ctools.
The On-Off analyses do not allow the assessment of the spatial source model parameters, hence
we fix them for these analyses to the values obtained in the unbinned analysis.
For the binned analysis, we used spatial bins of $0.02\degree \times 0.02\degree$, with
$200 \times 200$ bins around the pointing direction of each observation for the joint analysis, and
$200 \times 250$ bins around Right Ascension $228.4817\degree$ and Declination $-59.1358\degree$
for the stacked analysis.
The energy range was divided into 40 logarithmically spaced energy bins.
The same number of energy bins was also used for the On-Off analysis, and an On region
of $0.2\degree$ in radius centred on Right Ascension of $228.547\degree$ and 
Declination of $-59.174\degree$ was used.
The Off regions were defined using the reflected regions method.
We note that \citet{aharonian2005} use a larger On region radius of $0.3\degree$ in
their work, yet this larger radius does not allow us to define a sufficient number of Off regions for about half
of the observations for which the offset angle between MSH 15--52 and the pointing direction amounts
to only $0.4\degree$.
We therefore decided to use a smaller On region of $0.2\degree$ (see Fig.~\ref{fig:msh_skymap})
which probably misses some small fraction of the MSH 15--52 events.
Nevertheless, based on the spatial model, {\tt csphagen} computes the event leakage outside the
On region for a given spatial model, and takes it into account when computing the Auxiliary Response
File for the analysis (cf.~Eq.~\ref{eq:arf}).
Consequently, the resulting spectral parameters should reflect the full flux from MSH 15--52
under the assumption that the spatial model adequately describes the gamma-ray emission
morphology.

The results of the {\tt ctlike} maximum likelihood model fits using different analysis methods are
summarised in Table \ref{tab:msh_results}.
\citet{aharonian2005} find 3481 excess counts for 22.1 hours of live time within the circular
On region of $0.3\degree$ which fully encloses the emission, corresponding to 158
excess counts per hour live time.
This rate leads to an estimate of 1307 excess counts for the 8.3 hours of live time included in the
H.E.S.S.~public data release, which is of the same order as the $N_{\rm src}$ values obtained for
the 3D analyses.
Our estimates for the On-Off analyses are lower, typically around $\sim110$ excess counts
per hour live time, suggesting that $\sim30\%$ of the excess counts are lost due
to the choice of a smaller On region in our analysis with respect to \citet{aharonian2005}.

The fitted spatial as well as spectral model parameters are consistent among the different analysis
methods, confirming that ctools also performs correctly for extended source models.
The source extension, and specifically the semi minor axis $\sigma_{\rm min}$ of the elliptical 2D
Gaussian, is somewhat larger than those published by \citet{aharonian2005}.
Figure \ref{fig:msh_skymap} illustrates that  the fit by \citet{aharonian2005} seems to follow more closely
the brightest part of the emission, while the ctools model fit also captures some fainter larger-scale emission.
Also, the prefactor $k_0$ found by \citet{aharonian2005} is lower than the ctools results, suggesting
that some of the larger-scale faint emission may not be attributed to the source by the published H.E.S.S.~analysis.
We also note that the position angle of $147\degree \pm 6\degree$ obtained using the unbinned
ctools analysis is consistent with the value of $150\degree \pm 5\degree$ found in X-rays using
Chandra observations \citep{gaensler2002}.
The published H.E.S.S.~analysis suggests a slightly smaller position angle, but given its large statistical
uncertainty, the value is consistent with the ctools result.

\begin{figure*}[!t]
\centering
\includegraphics[width=\textwidth]{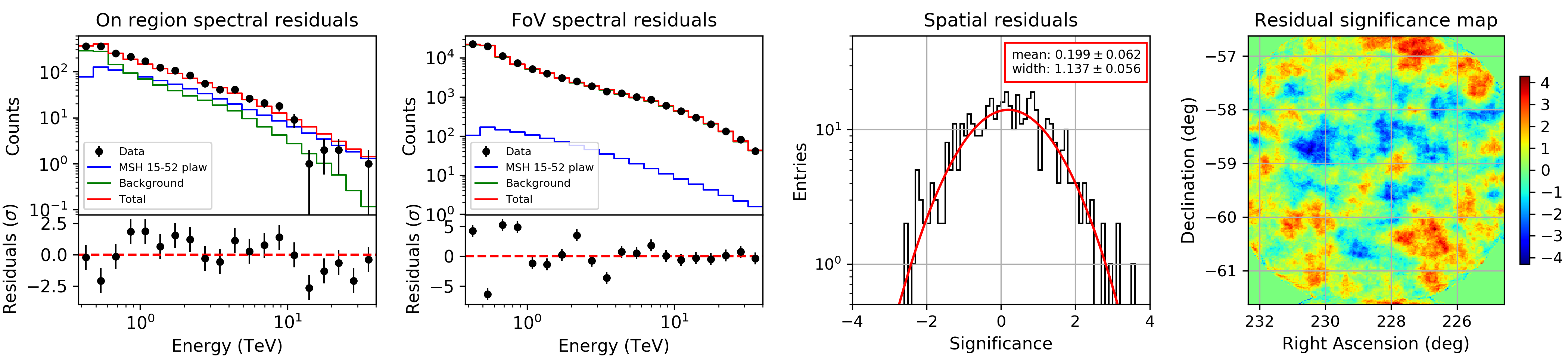}
\caption{
Residuals after fitting the MSH 15--52 observations using an elliptical 2D Gaussian spatial model 
with a power-law spectrum using an unbinned maximum likelihood fit.
See Fig.~\ref{fig:crab_residuals} for a description of the panels.
\label{fig:msh_residuals}
}
\end{figure*}
\begin{figure*}[!t]
\centering
\includegraphics[width=\textwidth]{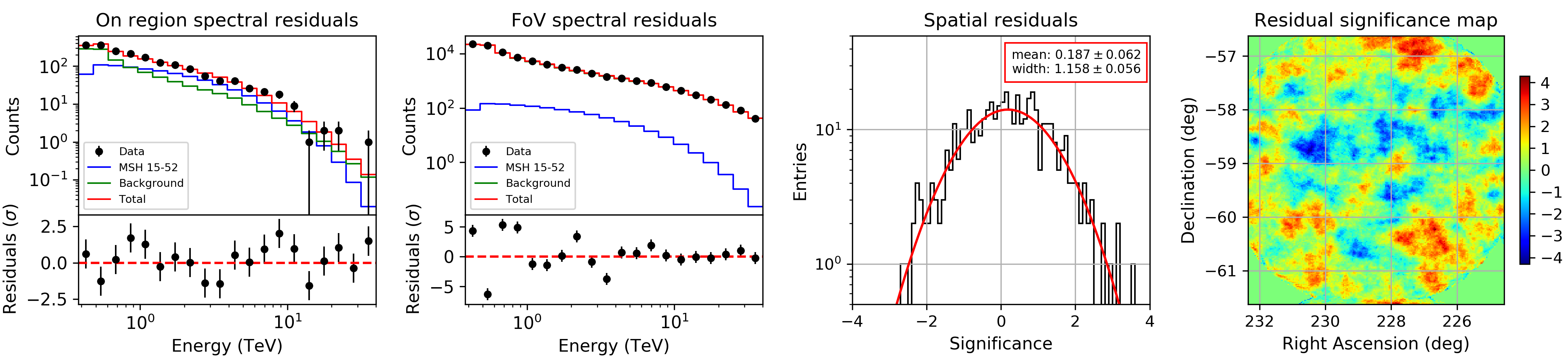}
\caption{
Same as Fig.~\ref{fig:msh_residuals} but for MSH 15--52 fitted using an exponentially cut-off power
law spectrum.
\label{fig:msh_residuals_eplaw}
}
\end{figure*}

The spatial and spectral residuals for the unbinned maximum likelihood fit are displayed in
Fig.~\ref{fig:msh_residuals}.
The spatial residuals are relatively flat and compatible with the expected histogram.
The spectral residuals in the On region are relatively flat, although the counts seem to be
slightly overestimated at low and high energies, and underestimated in between
(but see below).
For the full field of view there are some significant deviations at low energies, but above
about $\sim1$~TeV the residuals are relatively flat.
We also examined the spectral residuals for nine spatial sub-regions, and the radial counts
profiles and residuals as a function of offset angle $\theta$ for six energy bands.
The corresponding plots are shown in Figs.~\ref{fig:msh_residuals_sectors} and \ref{fig:msh_residuals_profiles}
of Appendix \ref{sec:residuals}.
Also here we find significant residuals at low energies in some of the sub-regions.
These residuals are obviously due to the limitations of our background model and
occur preferentially at the edge of the field of view.
To understand the impact of these spectral residuals on the fit results we analysed the data
using variations of the background model or the analysis parameters.
Specifically, we tried background models with ten spectral energy nodes instead
of the eight nodes used by default, which modifies the residual plots at low energies but
leaves the fitted source model parameters unchanged.
We also repeated the analysis for a minimum energy of $800$ GeV, excluding the region showing
significant spectral residuals, with little impact on the fitted source model parameters.
We conclude that our analysis results are robust and not significantly affected by the
spectral residuals that are visible in Figs.~\ref{fig:msh_residuals}, \ref{fig:msh_residuals_sectors} and
\ref{fig:msh_residuals_profiles}.

\begin{figure}[!t]
\centering
\includegraphics[width=8.8cm]{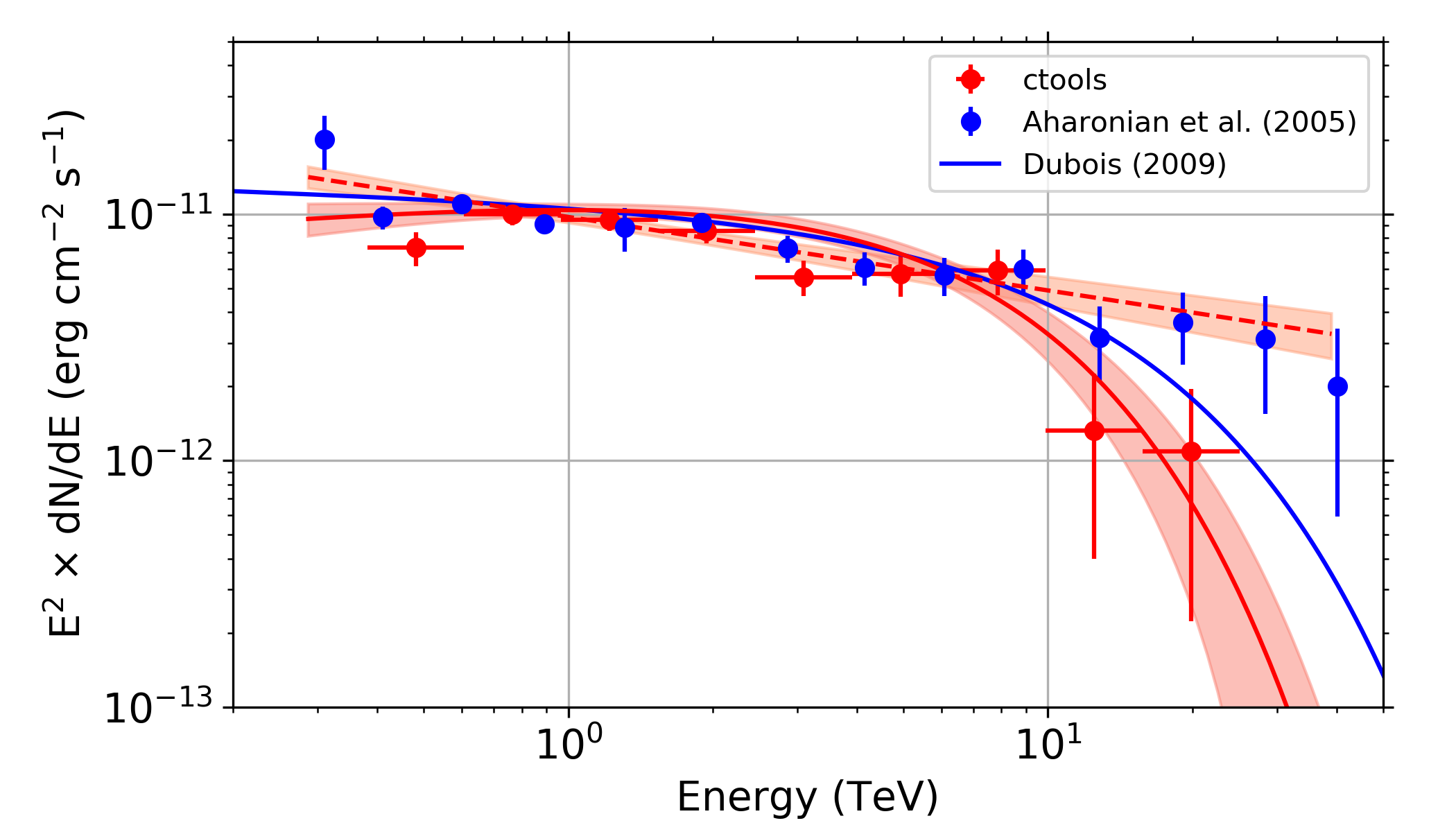}
\caption{
SED of MSH 15--52 derived using {\tt csspec}.
Red data represent the results of the unbinned ctools analyses; blue data are the values from
Figure 3 in \citet{aharonian2005}.
The dashed red line is the fitted power-law spectral model, the full red line is the fitted
exponentially cut-off power law spectral model.
The light red bands are the 68\% confidence level uncertainty bands of the spectral models and
were determined using {\tt ctbutterfly} for unbinned analyses.
The $68\%$ confidence level upper limits are displayed when the statistical error exceeds the 
value of a flux point.
\label{fig:msh_sed}
}
\end{figure}

We then used {\tt csspec} to derive an SED for MSH 15--52 using an unbinned maximum likelihood
analysis.
Spectral points were derived for ten logarithmically spaced energy bins between 381 GeV and
40 TeV.
Similar to the SED generation for the Crab nebula (cf.~section \ref{sec:crabobs}), the spectral
energy nodes $B_i$ were replaced by a simple power law, and the spatial parameters
$G_\mathrm{x}$ and $G_\mathrm{y}$ and the spectral power-law prefactor were fitted
independently for each energy bin.
The result is shown in Fig.~\ref{fig:msh_sed}, where we show for comparison also the spectral
points obtained by \citet{aharonian2005}.
The agreement of the spectral points between our analysis and that of \citet{aharonian2005}
is satisfactory, except for the lowest and the two highest spectral points for which we find slightly
lower fluxes.
Superimposed on Fig.~\ref{fig:msh_sed} is also the spectral power-law model that we determined
in the unbinned maximum likelihood fit, and the corresponding butterfly diagram generated using
{\tt ctbutterfly}.
It turns out that the power-law spectrum clearly overestimates the flux points at low energies.

\begin{table}[!t]
\small
\caption{Spectral fit results from the unbinned analysis of MSH 15--52 using an exponentially cut-off
power law as spectral model.
}
\label{tab:msh_epl_spectrum}
\centering
\setlength\tabcolsep{4.5pt}
\begin{tabular}{l c c c}
\hline\hline
Analysis & $k_0$ & $\Gamma$ & $E_c$ \\
\hline
ctools (unbinned) & $7.7 \pm 0.6$ & $1.84 \pm 0.13$ & $5.9 \pm 1.7$ \\
\citet{dubois2009} & $7.1 \pm 0.2$ & $2.06 \pm 0.05$ & $11.9 \pm 2.2$ \\
HGPS catalogue\tablefootmark{a} & $6.9 \pm 0.3$ & $2.05 \pm 0.06$ & $19.2 \pm 5.0$ \\
\hline
\end{tabular}
\tablefoot{
   $k_0$ is given in units of $10^{-12}$ \ftev, $E_c$ in units of TeV.
   \tablefoottext{a}{Values are from the \citet{hess2018a}.}
}
\end{table}

\begin{table*}[!ht]
\small
\caption{
Fit results for the RX J1713.7--3946 observations under variations of the spatial model
component.
An exponentially cut-off power law was used for the spectral component.
}
\label{tab:rx_spatial_results}
\centering
\begin{tabular}{l c c c c c c c c c}
\hline\hline
Spatial model & TS & $N_{\rm src}$ & R.A. (deg) & Decl. (deg) & $r$ (deg) & $w$ (deg) & $k_0$ & $\Gamma$ & $E_c$ \\
\hline
H.E.S.S. published\tablefootmark{a} & n.c. & $\sim31000$ & - & - & - & - & $2.3 \pm 0.1$ & $2.06 \pm 0.02$ & $12.9 \pm 1.1$ \\
Disk & $681.4$ & $2118$ & $258.35 \pm 0.01$ & $-39.76 \pm 0.01$ & $0.52 \pm 0.01$ & - & $1.9 \pm 0.1$ &  $1.94 \pm 0.09$ & $9.5 \pm 2.9$ \\
Gaussian & $627.7$ & $2333$ & $258.22 \pm 0.02$ & $-39.75 \pm 0.02$ & $0.31 \pm 0.01$ & - & $2.1 \pm 0.2$ & $2.01 \pm 0.09$ & $10.4 \pm 3.5 $ \\
Shell & $678.0$ & $2100$ & $258.35 \pm 0.02$ & $-39.76 \pm 0.01$ & $0.31 \pm 0.02$ & $0.24 \pm 0.03$ & $2.0 \pm 0.1$ & $1.88 \pm 0.10$ & $7.7 \pm 2.4 $ \\
Template & $762.1$ & $1937$ & - & - & - & - & $1.7 \pm 0.1$ & $1.88 \pm 0.08$ & $10.8 \pm 2.9$ \\
Template ($F_x^{0.48}$) & $784.6$ & $2148$ & - & - & - & - & $2.0 \pm 0.1$ & $1.85 \pm 0.08$ & $8.8 \pm 2.2$ \\
\hline
\end{tabular}
\tablefoot{
   TS is the Test Statistic and
   $N_{\rm src}$ is the number of source events attributed to the source.
   R.A. is the Right Ascension and Decl. the Declination of the fitted source model (J2000).
   For the disk model, $r$ is the radial extension of the disk, for the Gaussian model, $r$ is the Gaussian
   $\sigma$, and for the shell model, $r$ is the inner radius and $w$ is the shell width.
   $k_0$ is given in units of $10^{-11}$ \ftev\ and $E_c$ in units of TeV.
   \tablefoottext{a}{Values are from the \citet{hess2018c}.}
   ``n.c.'' signals that the information was not communicated in the publication.
}
\end{table*}

Interestingly, \citet{dubois2009} showed that the spectrum of MSH 15--52 deviates significantly from
a simple power law, and concluded that the H.E.S.S.~data are well fitted by an exponentially cut-off
power law.
Also in the HGPS catalogue, MSH 15--52 is fitted using an exponentially cut-off power law.
Replacing the power law by an exponentially cut-off power law in our unbinned maximum likelihood
analysis leads to a TS value of $780.9$ that is larger by $18.8$ with respect to the power law,
corresponding to a detection of the exponential cut-off at a significance level of $4.3\sigma$.
The spatial parameters found by the fit remain very close to those found for a power-law spectrum.
For illustration, we superimpose the resulting spectral law together with the corresponding butterfly
diagram on the spectral points in Fig.~\ref{fig:msh_sed}, where we also show for comparison the
spectral law determined by \citet{dubois2009}.
The fit results for the exponentially cut-off power-law spectral model are given in
Table \ref{tab:msh_epl_spectrum} where we compare them to the values found by
\citet{dubois2009} and the \citet{hess2018a}.
Our results are very close to those of \citet{dubois2009} and the \citet{hess2018a}, although we
found a slightly smaller cut-off energy compared to their analyses.
Finally, we show the spatial and spectral residuals after fitting the exponentially cut-off power
law in Fig.~\ref{fig:msh_residuals_eplaw}.
The slight overestimation of the counts at low and high energies, and the underestimation in
between, that we observed for the power-law spectrum in the On region
(cf.~Fig.~\ref{fig:msh_residuals}), has now disappeared.

\subsection{RX J1713.7--3946 observations}
\label{sec:rxobs}

RX J1713.7--3946 is a shell-type Galactic supernova remnant and one of the largest and
brightest sources of TeV gamma rays.
Previous H.E.S.S. analyses \citep{aharonian2006b, aharonian2007, aharonian2011, hess2018c}
revealed a complex emission morphology extending over about $1\degree$ in diameter.
To allow comparison of the ctools results with results reported in these works, we selected the
energy range 300 GeV -- 50 TeV for our analysis.

As a first step we focused on the modelling of the source morphology.
Several extended spatial models were fitted to the data, including 2D disk, 2D Gaussian and 2D
shell models as well as a spatial template map that was obtained from X-ray observations of
XMM-Newton \citep{acero2009}.
Since \citet{acero2009} suggested a non-linear flux correlation between X rays and gamma rays,
we also tested modified versions of the X-ray map where all pixels were taken to a given power
$\alpha$, where $\alpha=0.41$ corresponds to the value suggested by \citet{acero2009}.
According to \citet{aharonian2006b} the spectrum of RX J1713.7--3946 is curved and deviates
significantly from a power law.
The \citet{hess2018c} demonstrated that the spectrum of RX J1713.7--3946 is well described by
an exponentially cut-off power law and we hence used the same spectral model in our analysis
to allow comparison to that work.
We used the {\tt ctlike} tool to fit the source model together with the background model to the data
using an unbinned maximum likelihood analysis.
All spatial and spectral source parameters, as well as the parameters of the background model,
were free parameters in the model fit.

The results of these model fits are summarised in Table \ref{tab:rx_spatial_results}.
Among all tested spatial model components, the template map provides the largest statistical
significance.
Using the modified template maps increases the statistical significance even more, with a
best fitting value of $\alpha=0.48 \pm 0.10$, consistent with the value suggested by \citet{acero2009}.
We find slightly smaller prefactors $k_0$, harder spectral indices $\Gamma$ and smaller
cut-off energies $E_c$ compared to the \citet{hess2018c}.

\begin{figure}[!t]
\centering
\includegraphics[width=8.8cm]{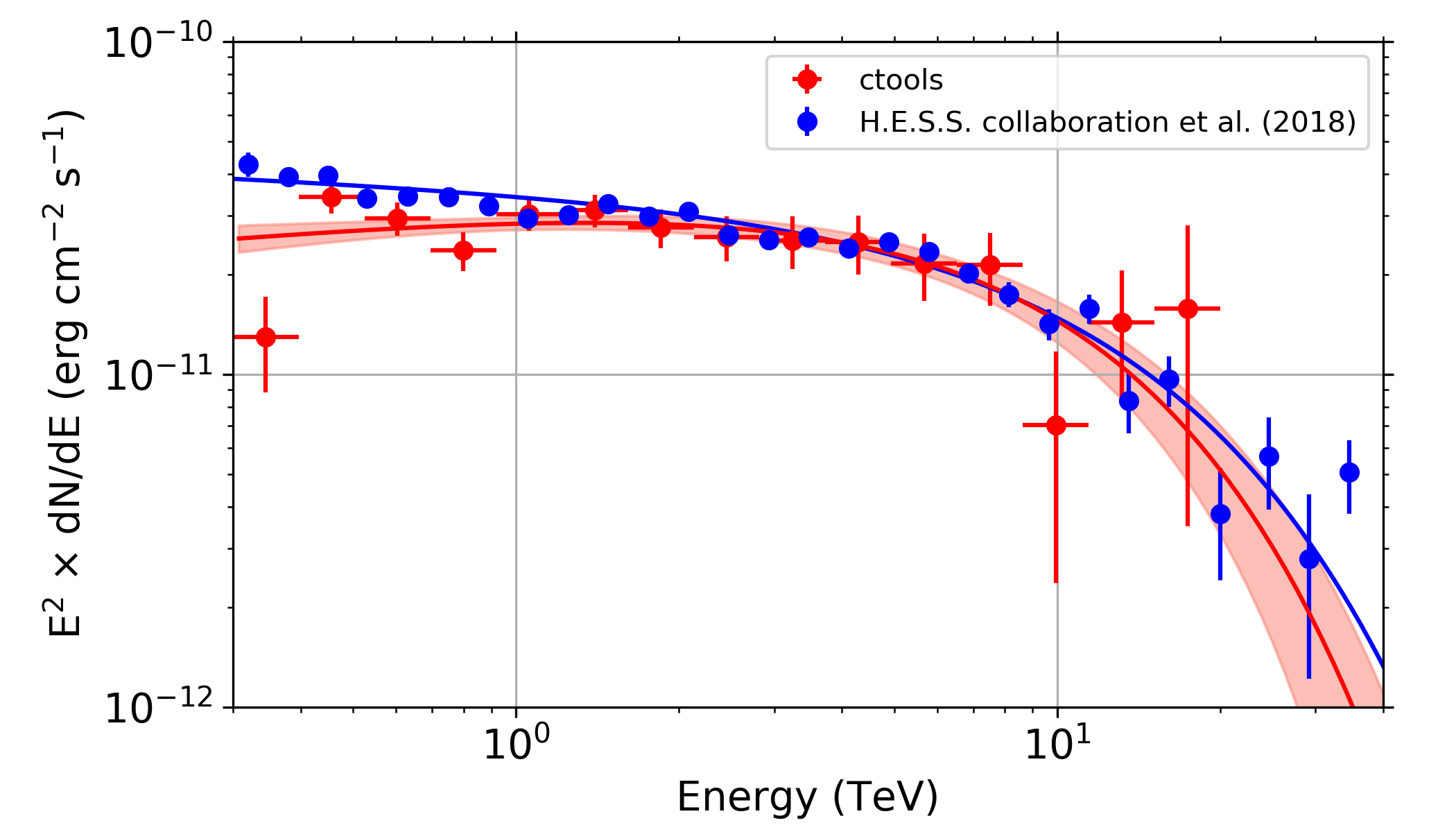}
\caption{
SED of RX J1713.7--3946 derived using {\tt csspec}.
Red data represent the unbinned ctools analysis, blue data are the values from Table F.1
of \citet{hess2018c}.
$68\%$ confidence level upper limits are displayed when the statistical error exceeds the 
value of a flux point.
The red line is the fitted exponentially cut-off power-law spectral model, the light red band is
the 68\% confidence level uncertainty band of the spectral model and was determined using
{\tt ctbutterfly}.
The blue line is the exponentially cut-off power-law spectral model determined by
\citet{hess2018c}.
\label{fig:rx_sed}
}
\end{figure}

\begin{figure*}[!t]
\centering
\includegraphics[width=\textwidth]{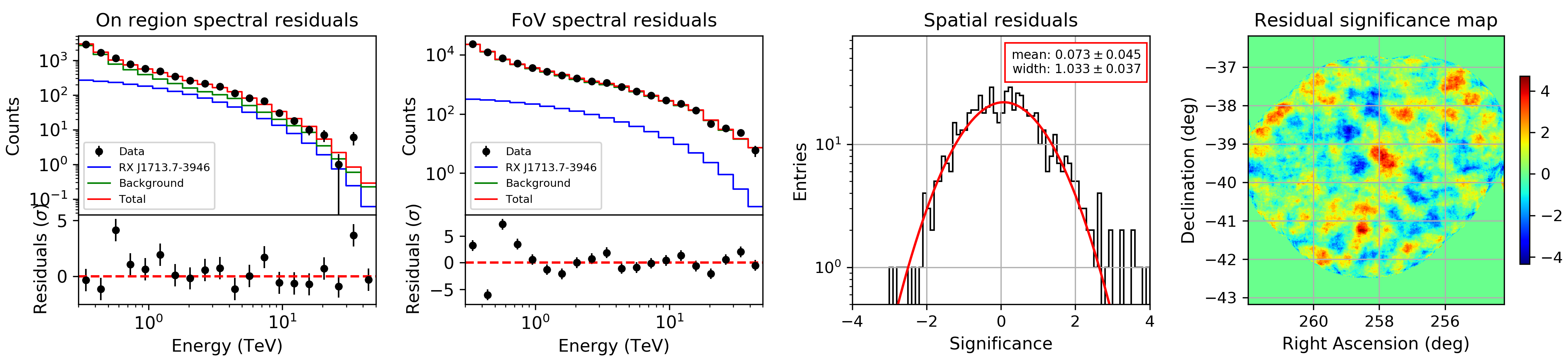}
\caption{
Residuals after fitting the RX J1713.7--3946 observations using the $F_x^{0.48}$ spatial template
with an exponentially cut-off power-law spectral component using an unbinned maximum likelihood fit.
See Fig.~\ref{fig:crab_residuals} for a description of the panels.
\label{fig:rx_residuals}
}
\end{figure*}

\begin{figure*}[!t]
\centering
\includegraphics[height=7.3cm]{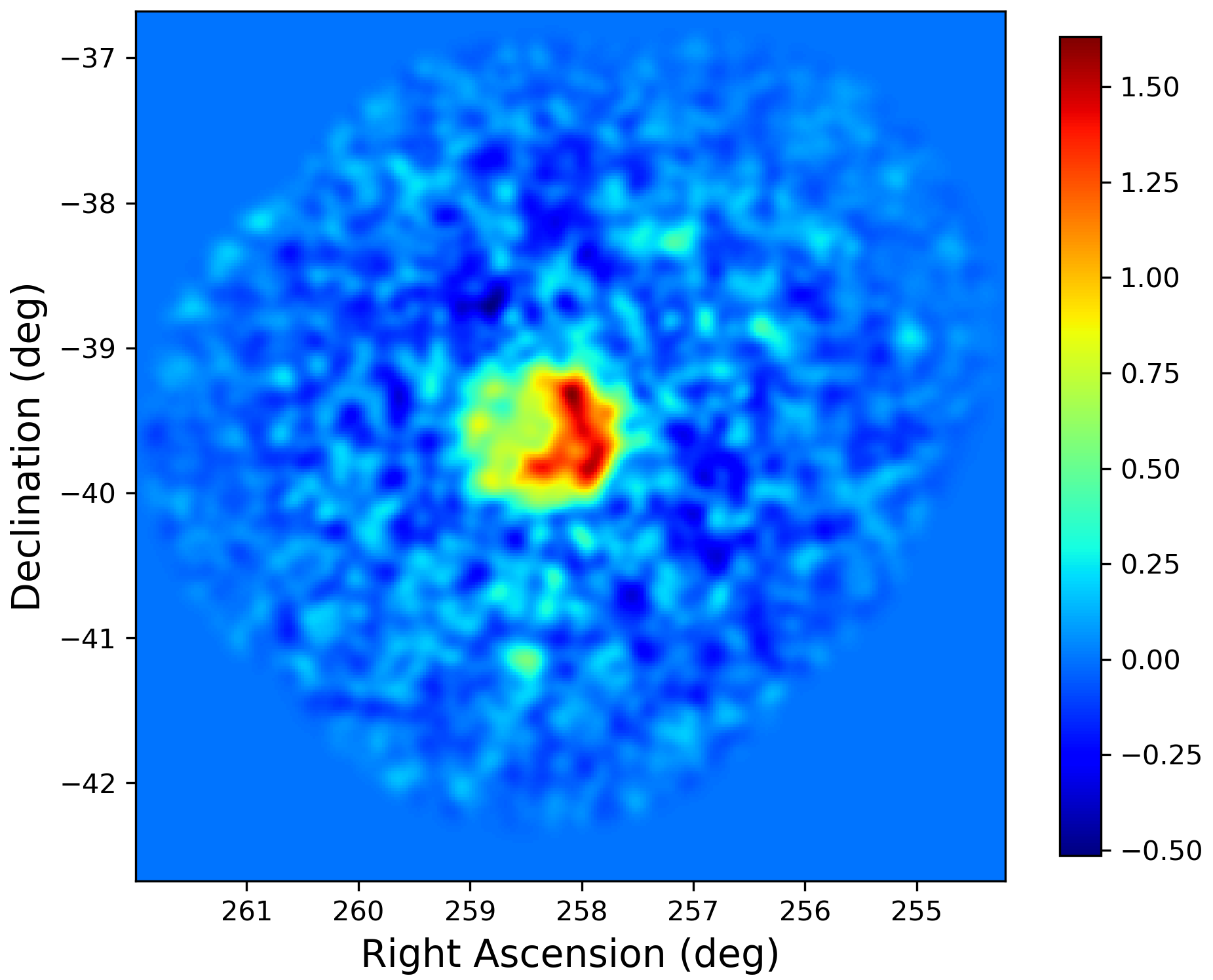}
\includegraphics[height=7.3cm]{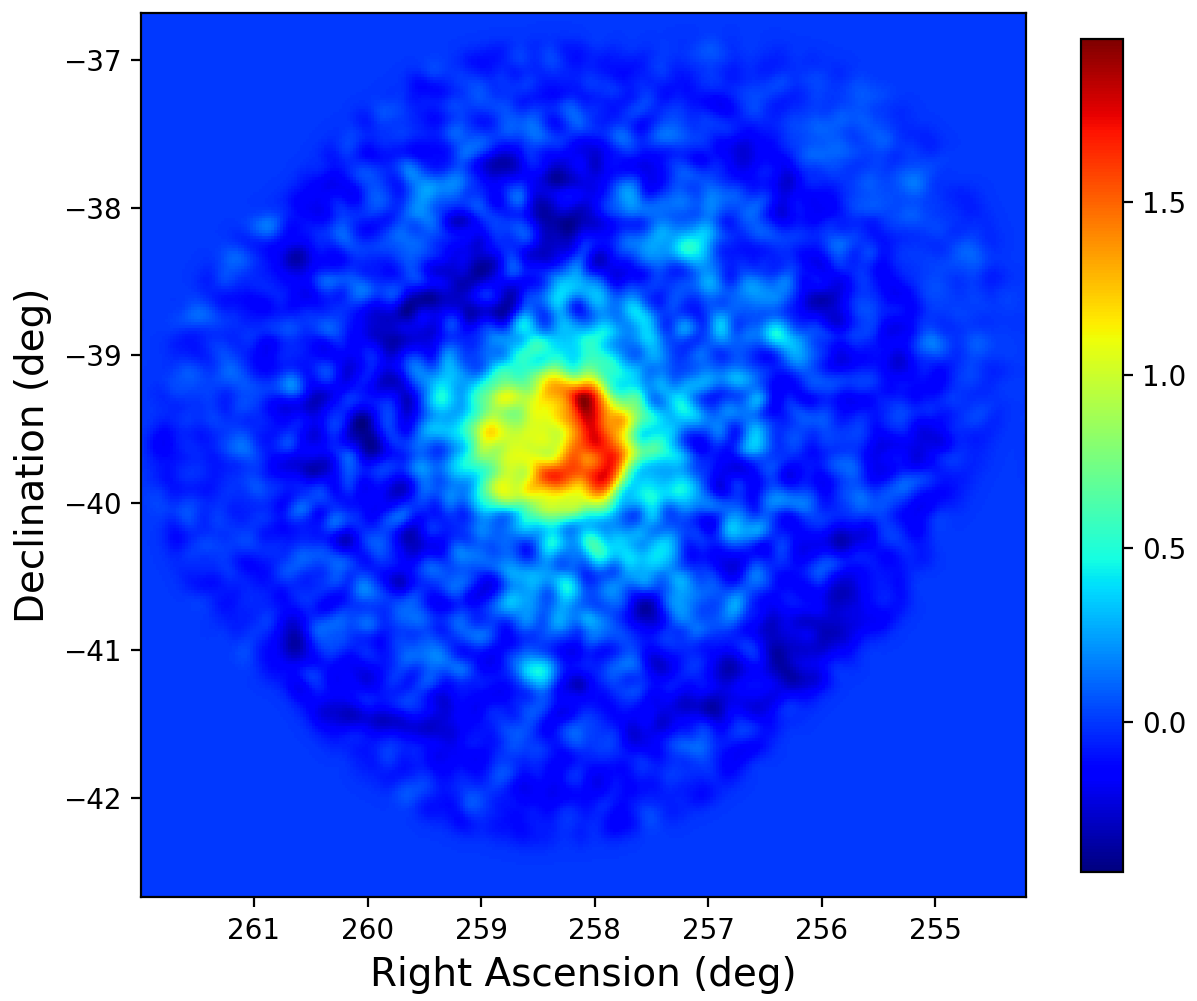}
\caption{
Left: Background-subtracted counts map of the RX J1713.7--3946 observations for the energy
band 300 GeV -- 50 TeV. The background was estimated using the model generated
by {\tt csbkgmodel} with background parameters estimated using an unbinned maximum likelihood
analysis using a $F_x^{0.48}$ spatial template with an exponentially cut-off power-law spectral
component.
Right: Background-subtracted sky map for the same data generated using {\tt ctskymap} with
the ring-background method for a correlation radius of $0.01\degree$.
Both maps were computed for a $0.02\degree\times0.02\degree$ binning and were smoothed
with a Gaussian kernel of $\sigma=0.06\degree$ to reduce statistical noise.
The colour bar represents the number of excess counts per $0.02\degree\times0.02\degree$ bin.
Coordinates are for the epoch J2000.
\label{fig:rx_skymap}
}
\end{figure*}

Adopting the $F_x^{0.48}$ spatial template we then determined the SED of RX J1713.7--3946
using {\tt csspec} with an unbinned maximum likelihood analysis.
We chose 15 energy bins spanning $0.3-20$ TeV for this purpose, and as before, we replaced the
background model by one where $G_\mathrm{x}$, $G_\mathrm{y}$ and the spectral
power-law prefactor were fitted independently for each energy bin.
We also computed the butterfly diagram for the exponentially cut-off power-law spectrum using
{\tt ctbutterfly}, again with an unbinned maximum likelihood analysis.
The results are shown in Fig.~\ref{fig:rx_sed}, which also shows the spectral points derived by
the \citet{hess2018c} from a larger dataset.\footnote{
  The live time of the observations used for the spectral analysis in \citet{hess2018c} amounts to
  116 hours which compares to the live time of 6.3 hours used in this paper (about $\sim18$ times 
  less).
}
The agreement between the ctools SED and the butterfly diagram is pretty adequate, demonstrating
again that the spectral results that we obtained using ctools are robust with respect to details
of the background model parametrisation.
The agreement between the ctools and H.E.S.S.~SED is also quite satisfactory, but the ctools data
point for the lowest energy ($300-397$ GeV) is significantly below the corresponding data points
found by the \citet{hess2018c}, which probably also explains why we found a flatter spectral index
and smaller prefactor in our analysis.
We note that this energy bin is the one that is most sensitive to imperfections in the background
model because it has the smallest signal-to-background ratio, as is evidenced in the left panel of
Fig.~\ref{fig:rx_residuals}.
This spectral point changes well beyond the statistical error for the different analysis methods
considered. 
On the other hand, our lowest spectral point is in agreement with the \fermi-LAT analysis of
RX J1713.7--3946 \citep[cf.~Fig.~5 in the][]{hess2018c} and the flatter spectral index also provides
a better spectral connection to the GeV data points.

The spatial and spectral residuals for the unbinned maximum likelihood fit of the $F_x^{0.48}$ spatial
template are displayed in Fig.~\ref{fig:rx_residuals}.
We also examined the spectral residuals for nine spatial sub-regions, and the radial counts
profiles and residuals as a function of offset angle $\theta$ for six energy bands.
The corresponding plots are shown in Figs.~\ref{fig:rx_residuals_sectors} and \ref{fig:rx_residuals_profiles}
of Appendix \ref{sec:residuals}.
For the full field of view, the spectral residuals show some significant deviations from zero below
$\la800$~GeV, but within the region of RX J1713.7--3946 the spectral residuals are relatively flat,
except for two outliers near $600$~GeV and $30$~TeV.
While the $30$~TeV outlier is beyond the energy range of our SED (and for which less than one event
is expected from RX J1713.7--3946 in the energy bin), the SED at $600$~GeV looks rather smooth,
suggesting that the outlier does not significantly impact our fitting result.

\begin{table*}
\caption{
Fit results for the RX J1713.7--3946 observations for different analysis methods.
The template map scaled to $F_x^{0.48}$ was used as spatial model component, and an exponentially
cut-off power law was used for the spectral component.
}
\label{tab:rx_results}
\centering
\begin{tabular}{l c c c c c}
\hline\hline
Analysis method & TS & $N_{\rm src}$ & $k_0$ & $\Gamma$ & $E_c$ \\
\hline
H.E.S.S. published\tablefootmark{a} & n.c. & $\sim31000$ & $2.3 \pm 0.1$ & $2.06 \pm 0.02$ & $12.9 \pm 1.1$ \\
Unbinned & $784.6$ & $2148$ & $2.0 \pm 0.1$ & $1.85 \pm 0.08$ & $8.8 \pm 2.2$ \\
Joint binned & $852.9$ & $2243$ & $2.1 \pm 0.1$ & $1.86 \pm 0.08$ & $9.1 \pm 2.3$ \\ 
Stacked binned & $806.2$ & $2119$ & $1.9 \pm 0.1$ & $1.85 \pm 0.08$ & $9.4 \pm 2.4$ \\
Joint On-Off (wstat)\tablefootmark{b} & $551.6$ & $2563$ & $3.2 \pm 0.4$ & $1.90 \pm 0.15$ & $4.9 \pm 2.3$ \\
Joint On-Off (wstat)\tablefootmark{c} & $336.5$ & $1929$ & $2.8 \pm 0.4$ & $1.51 \pm 0.20$ & $3.5 \pm 1.3$ \\
Joint On-Off (cstat) & $353.9$ & $2034$ & $2.1 \pm 0.1$ & $1.90 \pm 0.10$ & $12.3 \pm 3.8$ \\
Stacked On-Off (wstat)\tablefootmark{b} & $558.4$ & $2612$ & $2.8 \pm 0.2$ & $2.04 \pm 0.11$ & $9.0 \pm 4.2$ \\
Stacked On-Off (wstat)\tablefootmark{c} & $341.9$ & $1977$ & $2.3 \pm 0.2$ & $1.75 \pm 0.14$ & $6.9 \pm 2.5$ \\
Stacked On-Off (cstat) & $351.2$ & $2009$ & $2.3 \pm 0.2$ & $1.79 \pm 0.13$ & $7.9 \pm 2.8$ \\
\hline
\end{tabular}
\tablefoot{
   TS is the Test Statistic and $N_{\rm src}$ is
   the number of source events attributed to the source within the analysis region, which is the full field
   of view for the 3D analyses and the On region for the On-Off analyses.
   $k_0$, $\Gamma$, and $E_c$ are the parameters of the exponentially cut-off power law, where
   $k_0$ is given in units of $10^{-11}$ \ftev\ and $E_c$ in units of TeV.
   \tablefoottext{a}{Values are from the \citet{hess2018c}.}
   \tablefoottext{b}{For this result the scale factors $\alpha_{k,i}$ were used, which are based on the assumption that
                             the background rate per solid angle is the same in the On and the Off regions.}
   \tablefoottext{c}{For this result the scale factors $\alpha_{k,i}(M_b)$ that were computed for the cstat analysis
                             were used, which take into account differences in the expected background rates between On and 
                             Off regions.}
    ``n.c.'' signals that the information was not communicated in the publication.
}
\end{table*}

Figure \ref{fig:rx_residuals} demonstrates that the observed events within the source region are
dominated by background events, which is different from the situation for the other three sources
studied in this paper.
The RX J1713.7--3946 results are thus most sensitive to inadequacies of our background
model.
We therefore examined the robustness of the $F_x^{0.48}$ spatial template fit result under variations
of the background model or the analysis parameters.
Specifically, we tried background models from which we remove the gradient component
Eq.~\ref{eq:bkg_gradient}, which reduces the accuracy of the model, or where we increased the
number of energy nodes, which increases the accuracy of the model but which can lead to
convergence problems in the fit.
We also increased the energy threshold from 300 GeV to 800 GeV.
None of these modifications changed the fit result significantly, demonstrating that our result is
robust with respect to analysis details.

We also note that there are some excess values in the histogram of spatial residuals that
correspond to an excess feature situated in the western part of the supernova remnant, illustrating that
the X-ray template does not fully describe the emission morphology of the gamma-ray
emission.
We note that this excess coincides with the zones 3 and 4 defined in \citet{hess2018c}
for which the gamma-ray emission is found to reach beyond the extent of the X-ray emission.
Our analysis is thus consistent with this finding.

As the next step, we generated background-subtracted counts maps of the RX J1713.7--3946 observations
to image the gamma-ray emission from the supernova remnant using two different methods.
Firstly, the background model that was fitted to the data using the $F_x^{0.48}$ spatial
template with an exponentially cut-off power-law spectrum was subtracted from the data, and the resulting
residual map was slightly smoothed using a Gaussian kernel of $\sigma=0.06\degree$.
Secondly, {\tt ctskymap} was used to produce a background-subtracted counts map using the so-called
ring method, which estimates the background counts from a ring around the source, excluding areas
where significant gamma-ray emission is detected.
We selected an inner and outer ring radius of $0.8\degree$ and $1\degree$, respectively,
and specified an exclusion region of $0.6\degree$ in radius centred on Right Ascension of
$258.1125\degree$ and Declination of $-39.6867\degree$.
The correlation radius was set to $0.01\degree$.

The resulting sky maps are shown in Fig.~\ref{fig:rx_skymap}.
The source morphologies in both maps look very similar, and also quantitatively, the number of
excess counts per sky map pixel in both maps is comparable.
Around the excess emission attributed to RX J1713.7--3946 the sky maps look relatively flat,
and in particular, no large-scale gradients are discernable.

Finally, we compared the spectral fitting results for different analysis methods.
We used again the $F_x^{0.48}$ spatial template together with an exponentially cut-off power law for
the spectral component, and we repeated the analysis using the joint and stacked binned
methods, as well as the variants of the On-Off analysis.
For the binned analysis, we used spatial bins of $0.02\degree \times 0.02\degree$, with
$200 \times 200$ bins around the pointing direction of each observation for the joint analysis, and
$300 \times 300$ bins around Right Ascension $258.1125\degree$ and Declination $-39.6867\degree$
for the stacked analysis.
The energy range was divided into 40 logarithmically spaced energy bins.

The same number of energy bins was also used for the On-Off analysis, and an On region
of $0.6\degree$ in radius centred on Right Ascension of $258.1125\degree$ and 
Declination of $-39.6867\degree$ was used.
In this case we cannot use the standard implementation of the reflected-region method owing to the large size
of the source, which makes impossible finding suitable reflected regions within the field of view.
Therefore, the Off region was defined as a ring with inner and outer radius of $0.8\degree$ and $1\degree$,
respectively, around the same centre as for the On region.
Since this choice likely leads to different background rates per solid angle in the On and Off regions,
the classical wstat analysis may lead to inaccurate results as it supposes the same background rates per
solid angle in the On and Off regions.
We nevertheless kept the classical wstat analysis as a reference, but also explored
an alternative wstat analysis, where the scale factors $\alpha_{k,i}$ were
replaced by the model-dependent scale factors $\alpha_{k,i}(M_b)$ that were used for the cstat analysis.
In that way, spatial variations of the background rates per solid angle are taken into account, but the
background model is not explicitly used in the spectral analysis, and the background rates are
treated as nuisance parameters based on the On and Off region counts for each energy bin
(cf.~Appendix \ref{sec:wstat_cstat}).

Table \ref{tab:rx_results} summarises the spectral fitting results obtained with the different analysis
methods.
Several notable features deserve some discussion.
Firstly, the 3D analyses result in larger TS values compared to the On-Off analysis, which is in
line with the fact that the 3D analyses use more information to constrain the background model, leading
to less statistical uncertainty on the source model.
Secondly, the joint wstat analyses lead to relatively small cut-off energy values.
As noted in Appendix \ref{sec:wstat_cstat}, wstat is known to be inaccurate if energy bins have zero
Off counts, which is the case above $\sim5-10$~TeV for the joint analyses of the RX J1713.7--3946 
observations.
Reducing the number of energy bins from 40 to 20 for the joint wstat analyses pushes the problems
towards higher energies, resulting in somewhat larger cut-off energy values, steeper spectral indices
and smaller prefactors.
Nevertheless, even with a more severe rebinning the problem does not fully disappear, and the joint
wstat results should therefore be interpreted with caution.
Stacking the observations results in much larger statistics per energy bin, and the stacked results, in particular
when the model-dependent scale factors $\alpha_{k,i}(M_b)$ are used, are close to the results
obtained with the 3D analyses.

Thirdly, the joint and stacked On-Off analyses using wstat result in spectral prefactors $k_0$
that are $40-60\%$ larger than for the unbinned analysis.
This is because the estimation of the background from a ring around the supernova
remnant leads to an underestimation of the background rates, since the ring is on average
observed at larger offset angles than the source itself, and background rates drop
with increasing offset angle.
That is, the hypothesis underlying the use of wstat, which is that the background rate per solid angle
is the same in the On and Off regions, is not met.
Consequently, the underestimation of the background rate leads to an overestimation of the source
flux.
Taking the background rate variation into account using the alternative wstat analysis
that uses the model-dependent scale factors $\alpha_{k,i}(M_b)$ reduces the prefactors and
makes the results more compatible with those of the unbinned analysis.

This bias can also be overcome by alternative background estimation techniques used in the publications 
by the H.E.S.S.~Collaboration, such as the use of independent Off observations \citep{aharonian2007},
or by applying the reflected-region method independently to several subregions within the source
\citep{hess2018c}.
This is beyond the scope of this validation study and not investigated here.
We note, however, that the 3D analysis and the On-Off analysis based on cstat offer viable alternatives
to obtain accurate results also for this very extended source, providing prefactors that are much
closer to the values found for the unbinned analysis or published by the H.E.S.S. collaboration.

\subsection{PKS 2155--304 observations}
\label{sec:pksobs}

\begin{figure}[!t]
\centering
\includegraphics[width=8.8cm]{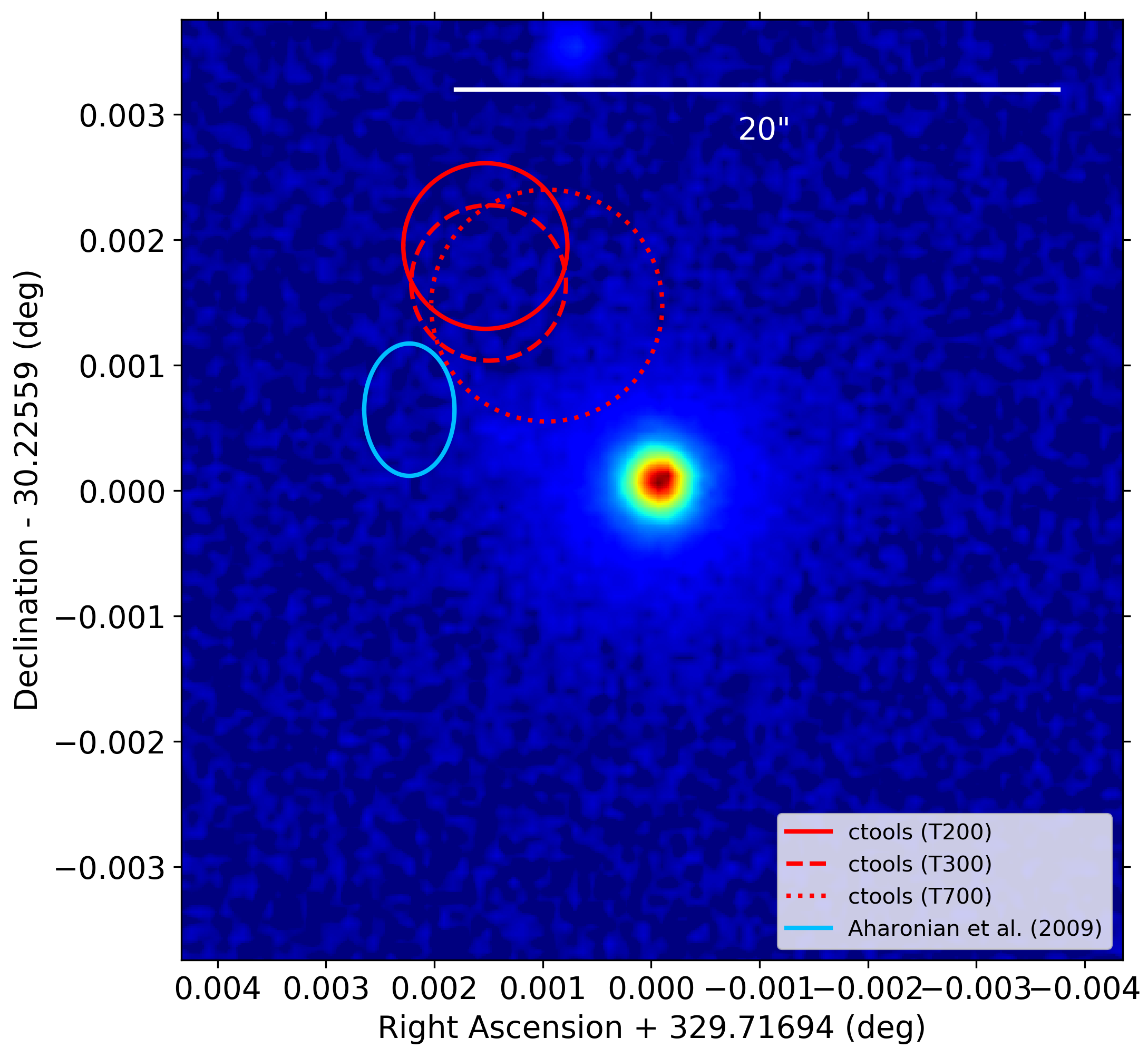}
\caption{
68\% confidence level error circles of the TeV source position derived using {\tt ctlike} (red) and
obtained in the H.E.S.S. analysis \citep[][blue]{aharonian2009}, overlaid on over a PanSTARRS
y-band image of PKS 2155--304.
The white bar indicates the systematic uncertainty for H.E.S.S.~localisation of $20\arcsec$ quoted in
\citet{aharonian2009}.
Coordinates are for the epoch J2000.
\label{fig:pks_position}
}
\end{figure}

PKS 2155--304 is one of the brightest and most studied blazars in the southern hemisphere,
and was first detected at TeV energies by the Durham Mark 6 telescope \citep{chadwick1999}.
PKS 2155--304 exhibits strong TeV variability, and the H.E.S.S.~public data release includes
observations taken during summer 2006 when the blazar was undergoing a major outburst,
with ultrafast TeV flux variability at hour time scales \citep{aharonian2009}.

\begin{figure*}[!t]
\centering
\includegraphics[width=\textwidth]{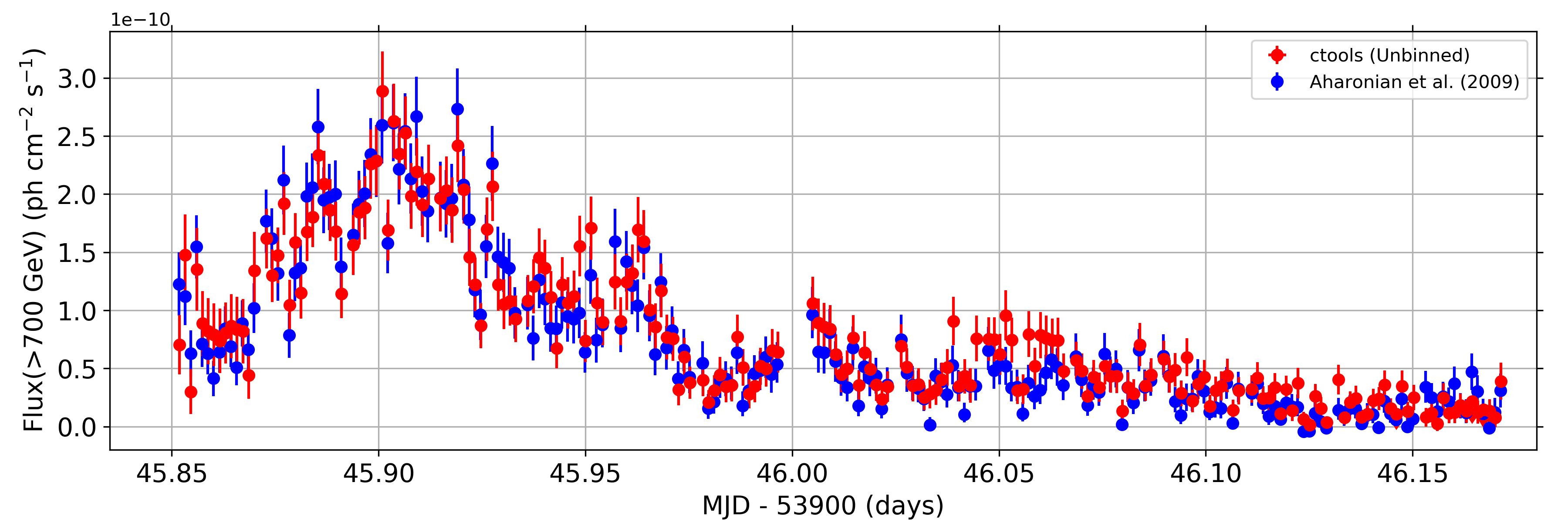}
\includegraphics[width=\textwidth]{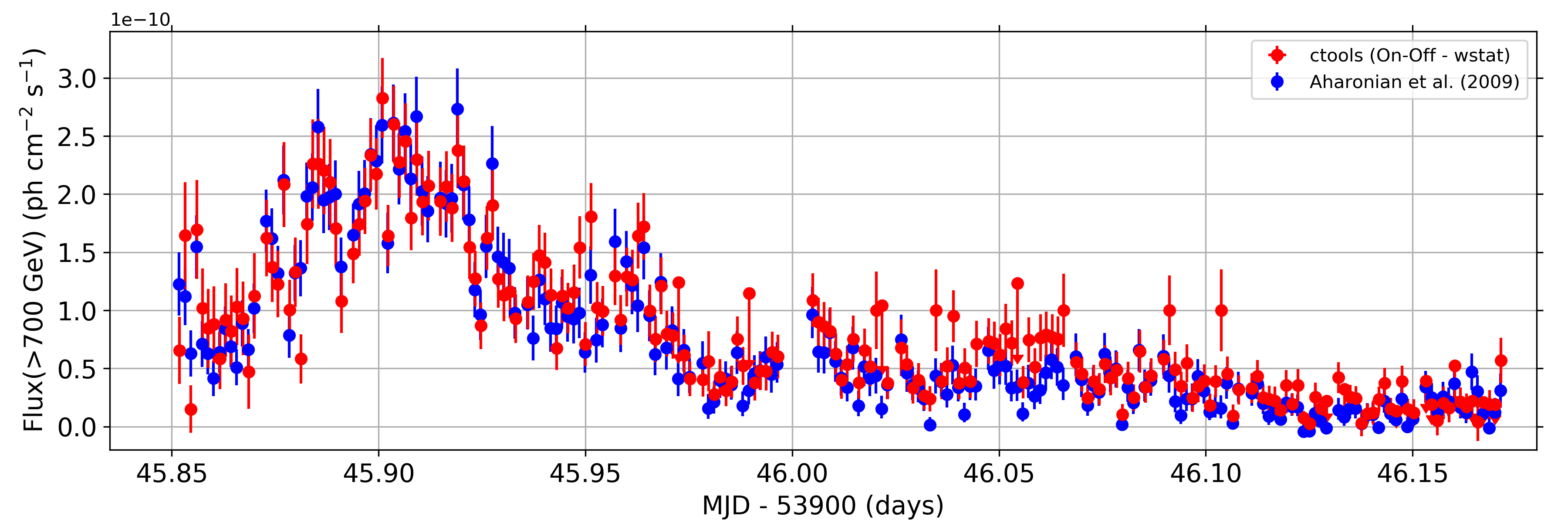}
\caption{
Light curves of PKS 2155--304 derived using {\tt cslightcrv} (red points) compared to the light curve
obtained by the H.E.S.S.~Collaboration \citep[][blue points]{aharonian2009}.
The ctools light curve in the top panel was derived using an unbinned maximum likelihood analysis, the
one in the bottom panel using an On-Off analysis with background estimates taken from
Off regions.
\label{fig:pks_lightcurve}
}
\end{figure*}

The dataset provided in the H.E.S.S.~public data release corresponds to the same observations as those 
studied by \citet{aharonian2009}, although the event reconstruction software and background discrimination
are different from those used for that paper.
\citet{aharonian2009} quote a best-fit position of the TeV source of
$329.7192\degree \pm 0.0004\degree$ in Right Ascension and
$-30.2249\degree \pm 0.0005\degree$ in Declination, which is $7\arcsec$ offset from the true source
position at Right Ascension of $329.7169\degree$ and Declination of $-30.2256\degree$
\citep{ma1998}.
The systematic position uncertainty is $20\arcsec$ \citep{aharonian2009}, considerably larger than the offset
from the true source position.
We performed an unbinned maximum likelihood analysis using {\tt ctlike} for the datasets defined
as T200, T300, and T700 in \citet{aharonian2009}\footnote{
  The datasets vary by the energy threshold and the time intervals that are covered
  \citep[see Table 1 of][]{aharonian2009}. The energy thresholds are 200 GeV, 300 GeV, and 700 GeV for
  T200, T300, and T700, respectively. The maximum event energy used in our analysis was
  10 TeV.}
to determine the best-fitting TeV source position.
The source was modelled as a point source with a curved power-law spectrum, and the
background model was generated using {\tt csbkgmodel}.

Figure \ref{fig:pks_position} shows the results of the analysis.
The best-fit {\tt ctlike} source position is offset by about $5\arcsec$ from the quoted H.E.S.S.~position,
a distance slightly larger than the statistical location uncertainties, but considerably smaller than the
systematic uncertainty of $20\arcsec$ quoted in \citet{aharonian2009}.
As illustrated by the different red circles, the choice of the dataset impacts slightly the fitted
source position, but within uncertainties the fitted positions are consistent.
Using an exponentially cut-off power law or a curved power law as spectral component
does not change the fitted source position significantly.
The difference from the quoted H.E.S.S.~position may be due to a difference in reconstruction
software between \citet{aharonian2009} and the H.E.S.S.~public data release, as well as by
a different spatial fitting technique (we used a 3D maximum likelihood fit while the H.E.S.S.~results
are generally obtained using a 2D technique that sums over an interval of reconstructed
energies).
We also note that the offset of the TeV source by $\sim7\arcsec$ from the true source position is
considerably smaller than the offset observed for the Crab nebula (see section \ref{sec:crabobs}),
suggesting that the Crab nebula offset is not related to an intrinsic problem in ctools.

Since PKS 2155--304 is highly variable during the period covered by the H.E.S.S.~flare dataset,
we used {\tt cslightcrv} to derive light curves of PKS 2155--304 and compared them to the one
shown in \citet{aharonian2009}.
In particular, we used the same time binning that was employed for Fig.~1 in
\citet{aharonian2009}\footnote{
  Each of the 15 observations was split into 14 intervals of equal length, resulting in $\sim2$ minute
  long time intervals}
and we used the event selection T700 that covers the full time interval of the observations.
PKS 2155--304 was modelled as a point source, located at the true position
(Right Ascension $329.7169\degree$ and Declination $-30.2256\degree$),
with a power-law spectral component.
The index of the power-law component was fixed to a fiducial value of $\Gamma=3.4$ (see below).

\begin{figure*}[!t]
\centering
\includegraphics[width=8.8cm]{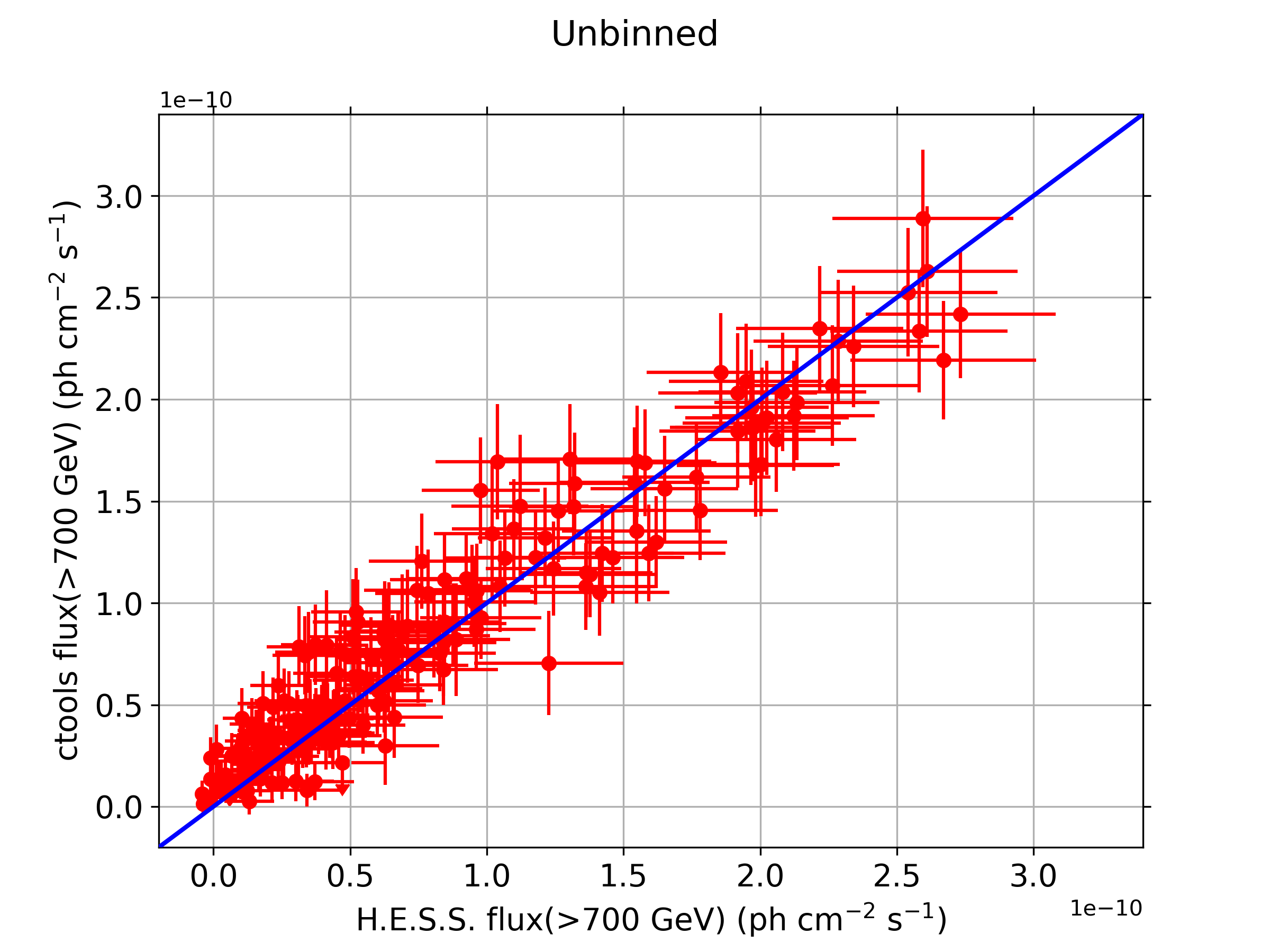}
\includegraphics[width=8.8cm]{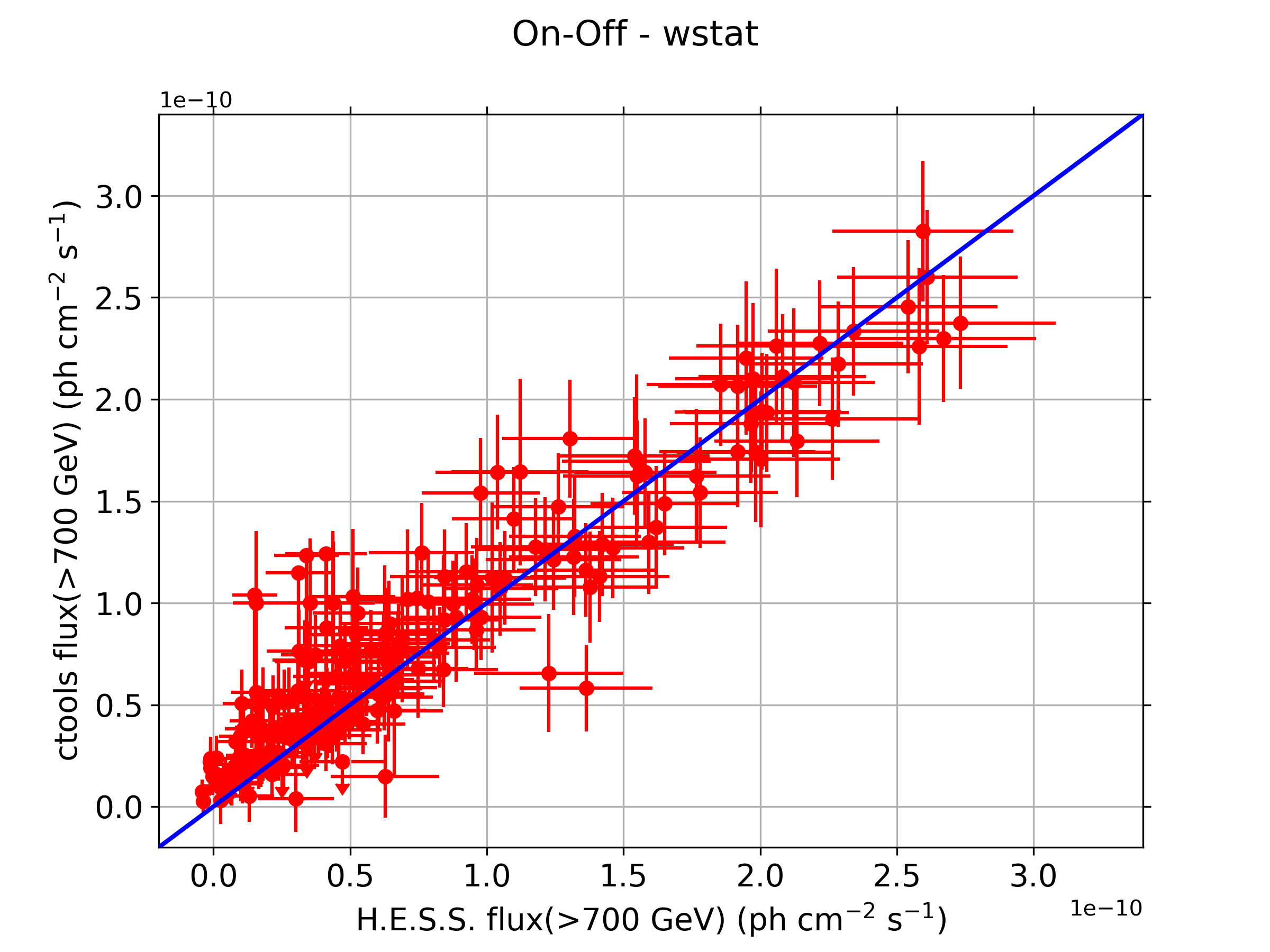}
\caption{
Correlation of PKS 2155--304 light curve fluxes obtained by \citet{aharonian2009} (horizontal axis) with
fluxes obtained by ctools (vertical axis).
The left plot shows results for the unbinned maximum likelihood analysis, and the right plot shows
results obtained for an On-Off analysis with background estimates taken from Off regions.
The blue line indicates equal fluxes.
\label{fig:pks_correlation}
}
\end{figure*}

We used {\tt cslightcrv} in two different modes.
First, {\tt cslightcrv} was run using an unbinned maximum likelihood analysis, where all events
within a given $\sim2$ minute long time interval were fitted with the source model on top of
the background model.
As usual, we used a background model that was generated using {\tt csbkgmodel} for the
unbinned analysis.
Second, {\tt cslightcrv} was run in On-Off mode, where events were selected from a circular On
region of $0.2\degree$ in radius, centred on the true source position, and the background
was estimated from reflected Off regions, assuming that the background rate per solid angle in
the On region is the same as in the Off regions.
Consequently, the wstat statistic was used for the fitting.
Ten logarithmically spaced energy bins between 700 GeV and 10 TeV were used for the On-Off
analysis.

The resulting light curves are shown in Fig.~\ref{fig:pks_lightcurve} where they are compared to
the light curve obtained by \citet{aharonian2009} (see the bottom panel of their Fig.~1).
Overall the agreement is excellent between the ctools results and the H.E.S.S.~analysis.
All features of the light curve obtained from the H.E.S.S.~Collaboration analysis are reproduced
with our ctools analyses, and our flux points overlap with the H.E.S.S.~Collaboration flux points
within statistical uncertainties.
To demonstrate this, we show in Fig.~\ref{fig:pks_correlation} correlation plots between the ctools
flux points with the flux points obtained by \citet{aharonian2009}.
The flux points correlate nicely between the published H.E.S.S.~and the ctools analyses, as
demonstrated by the fact that the flux points intersect with the diagonal line that indicates equal
flux measurements.
This demonstrates that ctools enables an accurate study of the time variability of gamma-ray
sources, and that the standard H.E.S.S.~background model for 3D analysis, as implemented in
{\tt csbkgmodel}, also works satisfactorily for short, minute-long time intervals.

\begin{figure}[!t]
\centering
\includegraphics[width=8.8cm]{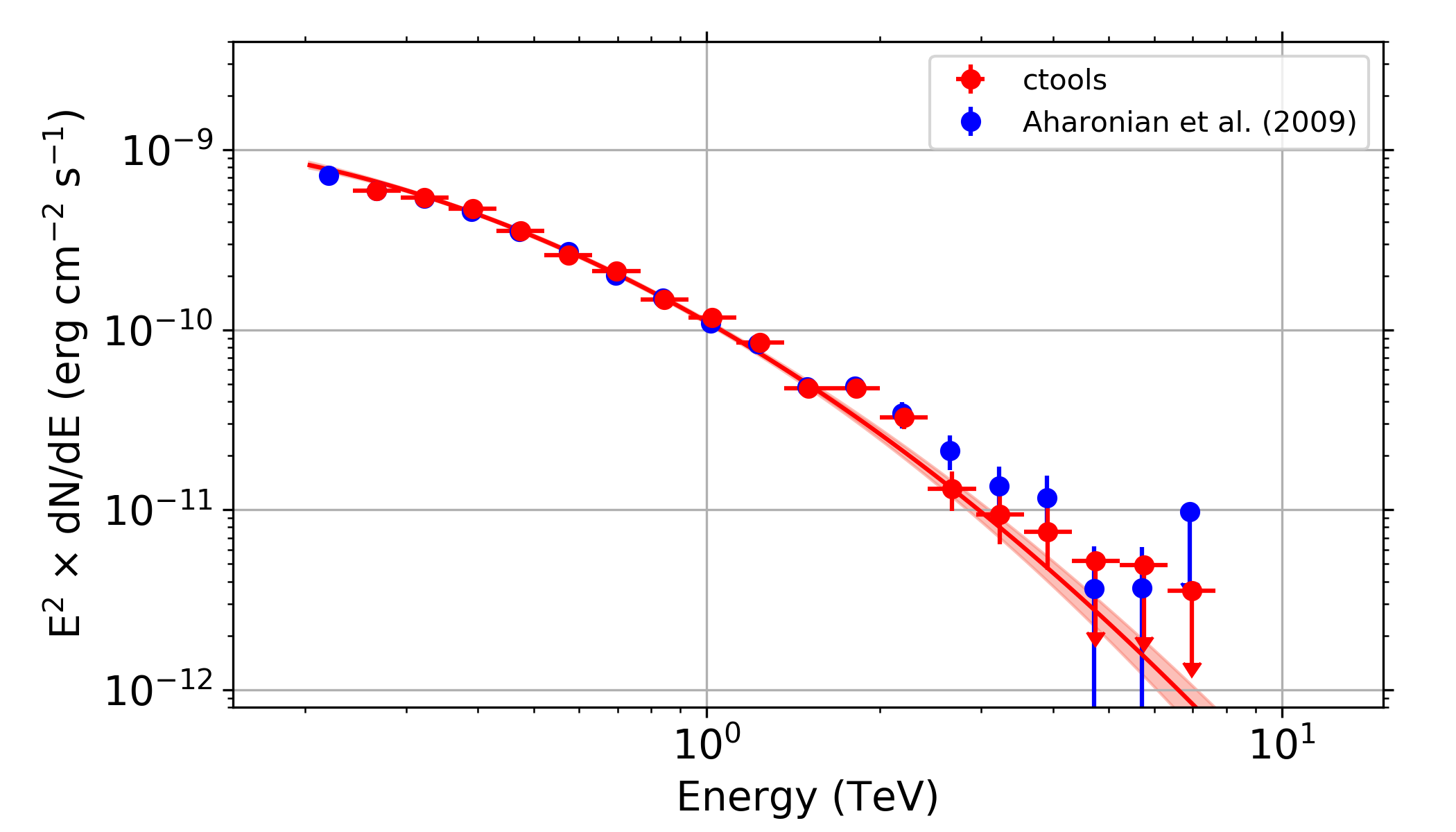}
\caption{
SED of PKS 2155--304 derived using {\tt csspec} for the T200 dataset.
Red data represent the unbinned ctools analysis, blue data are the values from Figure 9 in \citet{aharonian2009}.
The red line is the fitted curved power-law spectral model, the light red band is the 68\% confidence level
uncertainty band of the spectral model and was determined using {\tt ctbutterfly}.
$68\%$ confidence level upper limits are displayed when the statistical error exceeds the 
value of a flux point.
\label{fig:pks_sed}
}
\end{figure}

\begin{figure*}[!t]
\centering
\includegraphics[width=\textwidth]{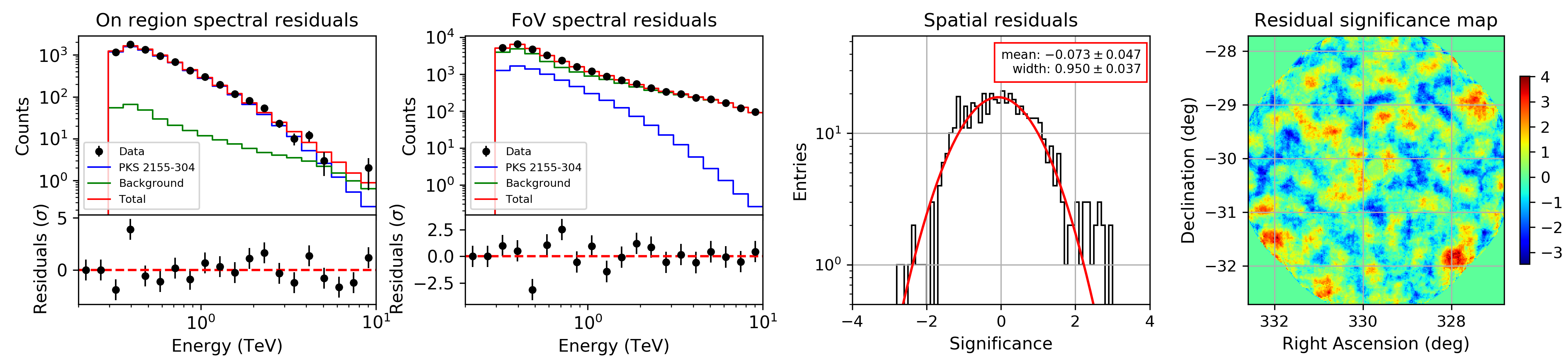}
\caption{
Residuals after fitting the PKS 2155--304 observations using a point source with a curved power-law
spectrum using an unbinned maximum likelihood fit.
See Fig.~\ref{fig:crab_residuals} for a description of the panels.
\label{fig:pks_residuals}
}
\end{figure*}

We then performed a spectral analysis of the PKS 2155--304 observations using the T200
dataset, which covers the largest energy interval.
Firstly, we determined the SED of PKS 2155--304 using {\tt csspec} in the usual manner by
fitting the background parameters independently for each spectral point.
We then derived a butterfly diagram using {\tt ctbutterfly} for a power-law, an exponentially cut-off
power-law, and a curved power-law spectral component using an unbinned maximum likelihood
analysis.
The best fit is obtained using the curved power law, followed by the exponentially cut-off power law
and the power law (cf.~Table \ref{tab:pks_spectral_results}).
Figure \ref{fig:pks_sed} shows the SED on top of the butterfly diagram of the curved power law,
superimposed on the spectrum derived by \citet{aharonian2009} for the same dataset.
Overall the agreement between the ctools analysis and the H.E.S.S.~result is excellent.
Only near $\sim3$~TeV the H.E.S.S.~analysis suggests slightly larger flux values, but within
error bars, both SEDs are consistent.

\begin{table*}[!t]
\caption{Fit results for PKS 2155--304 for dataset T200 under variations of the spectral model
component and analysis method assuming a point-source spatial model.}
\label{tab:pks_spectral_results}
\centering
\begin{tabular}{l c c c c c c c}
\hline\hline
PL & TS & $N_{\rm src}$ & $k_0$ & $\Gamma$ & & F$_{0.3-3 {\rm TeV}}$ \\
 & & counts & \ftev\ & & & \eunit\ \\
\hline
H.E.S.S. published\tablefootmark{a} & n.c. & n.c. & $(7.5 \pm 0.1) \times 10^{-11}$ & $3.25 \pm 0.01$ & - & $4.1 \times 10^{-10}$ \\
Unbinned & $46992.5$ & $8922$ & $(6.7 \pm 0.2) \times 10^{-11}$ & $3.45 \pm 0.02$ & - & $4.1 \times 10^{-10}$ \\
Joint binned & $43417.4$ & $8116$ & $(6.7\pm 0.2) \times 10^{-11}$ & $3.47 \pm 0.03$ & - & $4.1 \times 10^{-10}$ \\
Stacked binned & $43118.4$ & $8071$ & $(6.5 \pm 0.2) \times 10^{-11}$ & $3.48 \pm 0.03$ & - & $4.2 \times 10^{-10}$ \\
Joint On-Off (wstat) & $17799.2$ & $7877$ & $(6.5 \pm 0.2) \times 10^{-11}$ & $3.49 \pm 0.03$ & - & $4.1 \times 10^{-10}$ \\
Joint On-Off (cstat) & $16244.4$ & $7488$ & $(6.5 \pm 0.2) \times 10^{-11}$ & $3.49 \pm 0.03$ & - & $4.1 \times 10^{-10}$ \\
Stacked On-Off (wstat) & $18366.7$ & $7887$ & $(6.6 \pm 0.2) \times 10^{-11}$ & $3.48 \pm 0.03$ & - & $4.1 \times 10^{-10}$ \\
Stacked On-Off (cstat) & $16866.9$ & $7511$ & $(6.5 \pm 0.2) \times 10^{-11}$ & $3.49 \pm 0.03$ & - & $4.1 \times 10^{-10}$ \\
\hline\hline
EPL & TS & $N_{\rm src}$ & $k_0$ & $\Gamma$ & $E_c$ & F$_{0.3-3 {\rm TeV}}$ \\
 & & counts & \ftev\ & & TeV & \eunit\ \\
\hline
H.E.S.S. published\tablefootmark{a} & n.c. & n.c. & $(2.1 \pm 0.2) \times 10^{-10}$ & $2.65 \pm 0.06$ & $1.0 \pm 0.1$ & $4.3 \times 10^{-10}$ \\
Unbinned & $47087.2$ & $8902$ & $(1.7 \pm 0.2) \times 10^{-10}$ & $2.85 \pm 0.08$ & $1.1 \pm 0.2$ & $4.2 \times 10^{-10}$ \\
Joint binned & $43507.5$ & $8097$ & $(1.8 \pm 0.2) \times 10^{-10}$ & $2.82 \pm 0.08$ & $1.1 \pm 0.2$ & $4.2 \times 10^{-10}$ \\ 
Stacked binned & $43197.2$ & $8056$ & $(1.6 \pm 0.2) \times 10^{-10}$ & $2.89 \pm 0.08$ & $1.2 \pm 0.2$ & $4.2 \times 10^{-10}$ \\
Joint On-Off (wstat) & $17870.4$ & $7872$ & $(1.6 \pm 0.2) \times 10^{-10}$ & $2.91 \pm 0.09$ & $1.2 \pm 0.2$ & $4.2 \times 10^{-10}$ \\
Joint On-Off (cstat) & $16321.4$ & $7480$ & $(1.7 \pm 0.2) \times 10^{-10}$ & $2.87 \pm 0.09$ & $1.1 \pm 0.2$ & $4.2 \times 10^{-10}$ \\
Stacked On-Off (wstat) & $18437.9$ & $7882$ & $(1.6 \pm 0.2) \times 10^{-10}$ & $2.90 \pm 0.09$ & $1.2 \pm 0.2$ & $4.2 \times 10^{-10}$ \\ 
Stacked On-Off (cstat) & $16940.7$ & $7503$ & $(1.7 \pm 0.2) \times 10^{-10}$ & $2.88 \pm 0.09$ & $1.2 \pm 0.2$ & $4.2 \times 10^{-10}$ \\ 
\hline\hline
CPL & TS & $N_{\rm src}$ & $k_0$ & $\Gamma$ & $\beta$ & F$_{0.3-3 {\rm TeV}}$ \\
 & & counts & \ftev\ & & & \eunit\ \\
\hline
H.E.S.S. published\tablefootmark{a} & n.c. & n.c. & $(7.5 \pm 0.2) \times 10^{-11}$ & $3.69 \pm 0.05$ & $-0.78 \pm 0.07$ & $4.3 \times 10^{-10}$ \\
Unbinned & $47091.7$ & $8900$ & $(6.9 \pm 0.2) \times 10^{-11}$ & $3.83 \pm 0.05$ & $-0.35 \pm 0.04$ & $4.3 \times 10^{-10}$ \\ 
Joint binned & $43512.4$ & $8097$ & $(7.0 \pm 0.2) \times 10^{-11}$ & $3.83 \pm 0.06$ & $-0.37 \pm 0.05$ & $4.3 \times 10^{-10}$ \\ 
Stacked binned & $43199.5$ & $8054$ & $(6.8 \pm 0.2) \times 10^{-11}$ & $3.82 \pm 0.05$ & $-0.35 \pm 0.05$ & $4.2 \times 10^{-10}$ \\ 
Joint On-Off (wstat) & $17873.3$ & $7877$ & $(6.8 \pm 0.2) \times 10^{-11}$ & $3.82 \pm 0.06$ & $-0.34 \pm 0.05$ & $4.2 \times 10^{-10}$ \\ 
Joint On-Off (cstat) & $16322.0$ & $7481$ & $(6.7 \pm 0.2) \times 10^{-11}$ & $3.85 \pm 0.06$ & $-0.36 \pm 0.05$ & $4.2 \times 10^{-10}$ \\ 
Stacked On-Off (wstat) & $18442.2$ & $7887$ & $(6.9 \pm 0.2) \times 10^{-11}$ & $3.81 \pm 0.06$ & $-0.34 \pm 0.05$ & $4.2 \times 10^{-10}$ \\ 
Stacked On-Off (cstat) & $16944.1$ & $7505$ & $(6.8 \pm 0.2) \times 10^{-11}$ & $3.83 \pm 0.06$ & $-0.35 \pm 0.05$ & $4.2 \times 10^{-10}$ \\ 
\hline
\end{tabular}
\tablefoot{
   TS is the Test Statistic and $N_{\rm src}$ is the number of events attributed to the source within
   the analysis region, which is the full field of view for the 3D analyses and the On
   region for the On-Off analyses,
   $k_0$, $\Gamma$, $E_c$, and $\beta$ are the spectral parameters, where $k_0$ is evaluated at
   1 TeV.
   F$_{0.3-3 {\rm TeV}}$ is the resulting energy flux of PKS 2155--304 within 300 GeV and 3 TeV.
   \tablefoottext{a}{Values are from Table 1 of \citet{aharonian2009}.}
   ``n.c.'' signals that the information was not communicated in the publication.
}
\end{table*}

Secondly, we performed {\tt ctlike} model fits using a power law, an exponentially cut-off power-law,
and a curved power-law spectral component for the different analysis methods available in
ctools, and compared them to results given in Table 1 of \citet{aharonian2009}.
For the binned analysis, we used spatial bins of $0.02\degree \times 0.02\degree$, with
$200 \times 200$ bins around the pointing direction of each observation for the joint analysis, and
$250 \times 250$ bins around Right Ascension $329.7169\degree$ and Declination $-30.2256\degree$
for the stacked analysis.
The energy range was divided into 40 logarithmically spaced energy bins.
The same number of energy bins was also used for the On-Off analysis.
An On region of $0.2\degree$ in radius centred on Right Ascension of $329.7169\degree$ and 
Declination of $-30.2256\degree$ was adopted; Off regions were defined using the reflected regions
method.
The results are summarised in Table \ref{tab:pks_spectral_results}.
All spatial and spectral source parameters, as well as the parameters of the background model,
were free parameters in the model fits (except for the On-Off analysis, which does not allow an
adjustment of the spatial source and background parameters).

We confirm for all analysis methods that the largest TS values are obtained for a curved power law,
followed by the exponentially cut-off power law and the power law.
Among the different analysis methods, spectral model parameters are very close and within the
statistical parameter uncertainties.
With respect to the published H.E.S.S.~results, the ctools results indicate a steeper
spectrum and smaller spectral normalisation at 1 TeV, but the energy flux within the
300 GeV -- 3 TeV energy range is comparable.
A steeper ctools spectrum is consistent with the fact that the high-energy ctools spectral points are
below the ones determined by \citet{aharonian2009}, as evidenced in Fig.~\ref{fig:pks_sed}.
We do not know the precise origin of this difference, but it plausibly can be attributed to
differences in the analysis details or reconstruction methods used.

Finally, the spatial and spectral residuals for the unbinned maximum likelihood fit of the curved
power law spectrum are displayed in Fig.~\ref{fig:pks_residuals}.
Spectral residuals for nine sub-sectors and radial counts profiles and residuals as a function of offset
angle $\theta$ are shown in Figs.~\ref{fig:pks_residuals_sectors} and \ref{fig:pks_residuals_profiles} of
Appendix \ref{sec:residuals}.
The spectral residuals are relatively flat, yet the significance distribution of the residual counts
shows some positive excess values, and residual features around PKS 2155--304 are also
visible in the residual map.
These features are probably attributable to the strong variability of the source, which is not properly
taken into account for the spectral and residual analyses that assume a constant source flux.
In general, the point-spread function will vary in time, as a consequence of the changing zenith angle of
the pointing direction during the observation period.
For the PKS 2155--304 observations, the zenith angle changed between $7.2\degree$ and
$27.1\degree$, resulting in a variation of the $68\%$ containment radius at 1 TeV and an offset
angle of $0.5\degree$ between $4.7'$ and $5.4'$, respectively.
Assuming a constant source flux will give equal weight to the point-spread function of each
observation, while for a varying source flux, the observations during which the source flares
will have greater weight.
It is hence not surprising that some spatial residuals remain, 
which are mainly situated at the edge of the field of view and hence
should not significantly affect the results for the source.
In addition, the excess values seen in Fig.~\ref{fig:pks_residuals} are relatively modest.

\section{Joint analysis of H.E.S.S.~and \fermi-LAT data}
\label{sec:joint-fermi}

\begin{figure*}[!t]
\centering
\begin{tabular}{cc}
\includegraphics[width=0.5\textwidth]{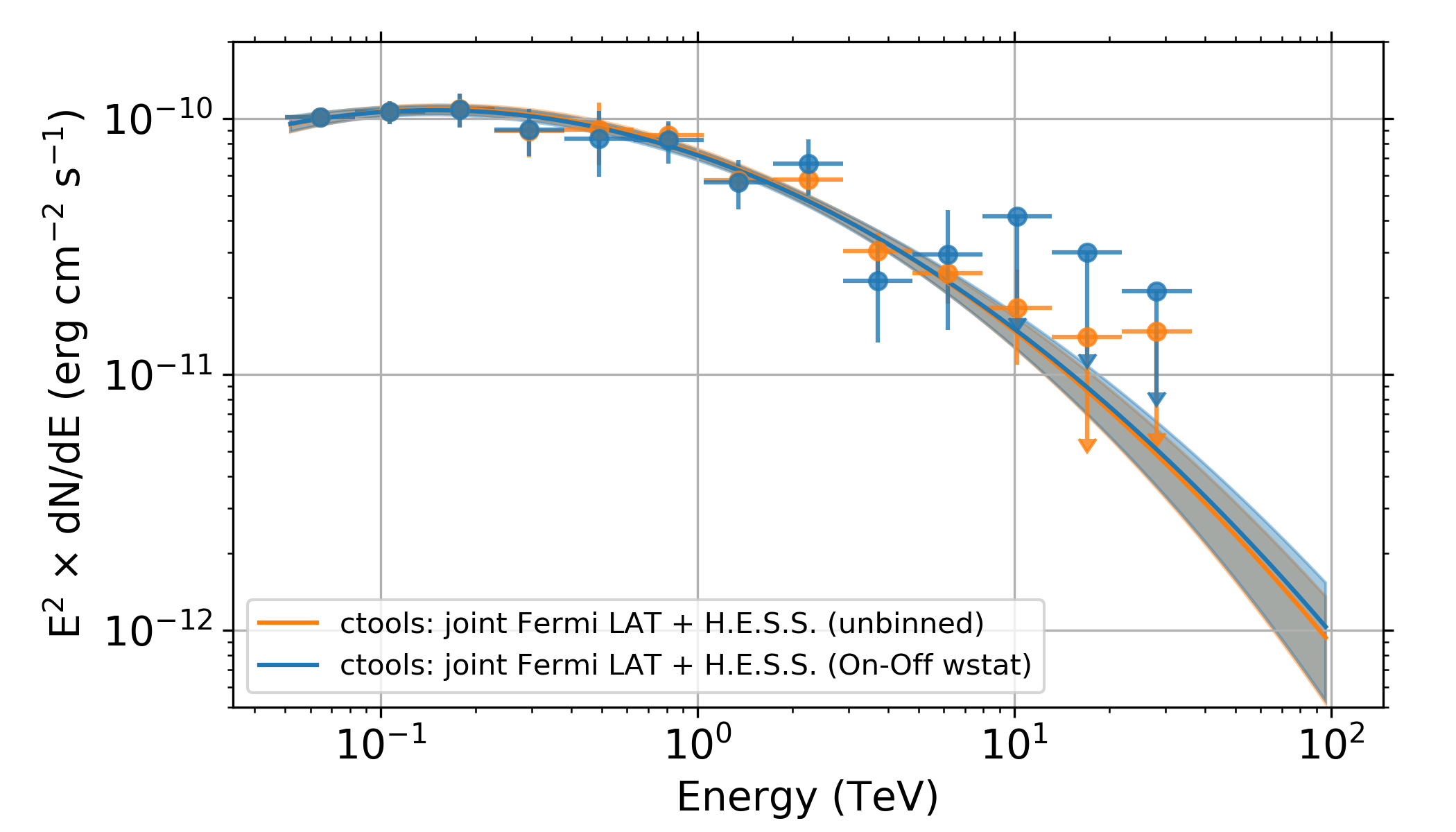} &
\includegraphics[width=0.5\textwidth]{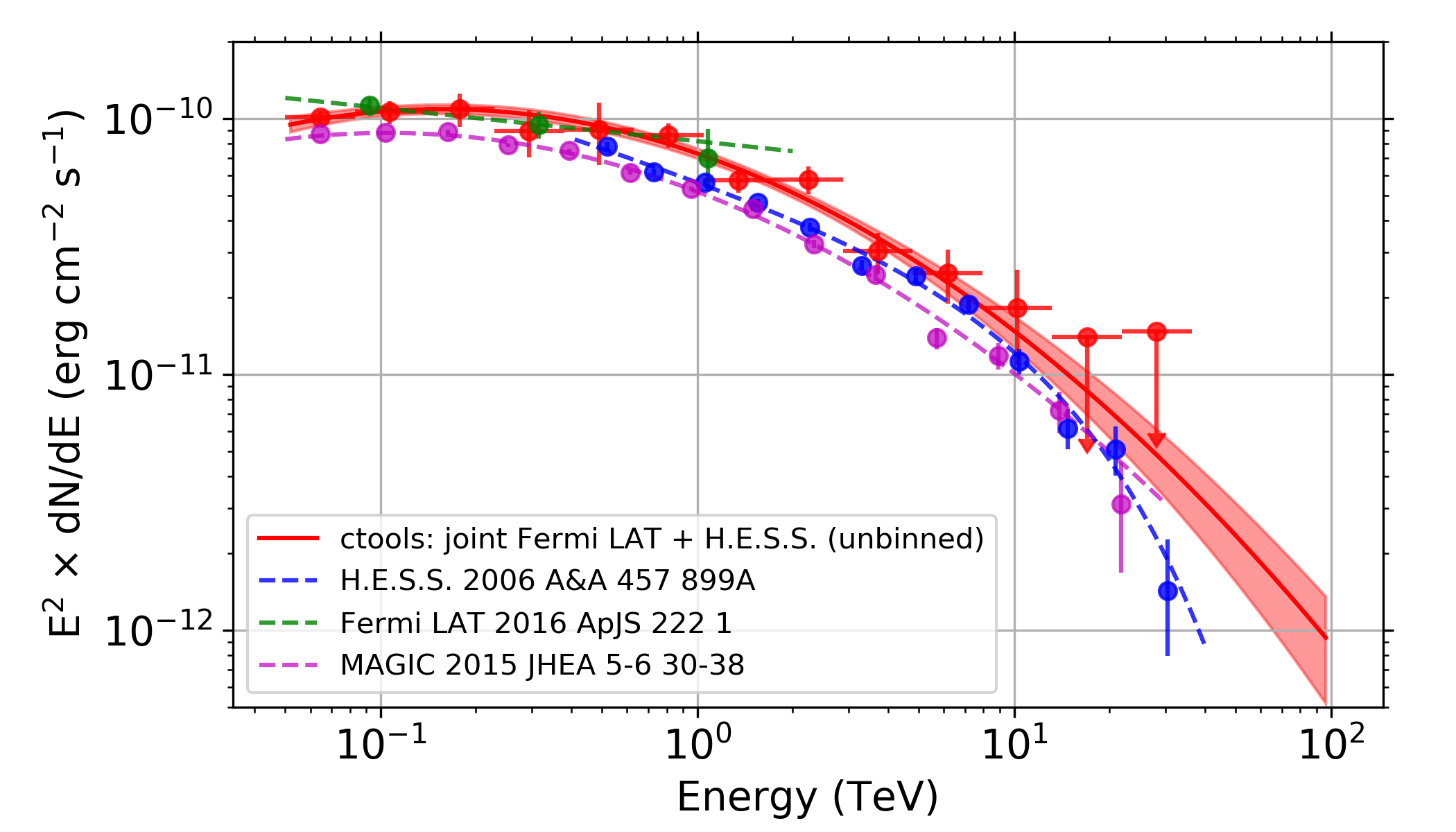} \\
\end{tabular}
\caption{
Left: SED of the Crab nebula derived from the joint analysis of \fermi-LAT and H.E.S.S.~data above 50~GeV.
H.E.S.S.~data were analysed using two different methods, unbinned and On-Off with wstat statistics (see
Section~\ref{sec:crabobs}).
Lines represent the best-fit curved power-law spectral models over the entire energy range, and the shaded bands 
represent the 68\% confidence-level uncertainty bands of the spectral models determined using {\tt ctbutterfly}.
Upper limits from the On-Off analysis at the highest energies are above the maximum flux shown in the plot.
Right: Best-fit curved power-law model and SED derived from our joint analysis of LAT and H.E.S.S.~data (with
unbinned analysis of H.E.S.S.~data), compared to published results from H.E.S.S.~only \citep{aharonian2006a},
\fermi-LAT only above 50~GeV \citep{2FHL}, and MAGIC \citep{MAGICCrab2015}. 
\label{fig:jointfermi}
}
\end{figure*} 

The ctools also are capable of analysing IACT data jointly with data from other gamma-ray instruments.
The H.E.S.S.~public data release enables testing this feature for the first time with real IACT data.
Here we illustrate, as a test case, a joint spectral analysis of the Crab nebula using H.E.S.S.~observations
(Section~\ref{sec:crabobs}) along with data from the \fermi Large Area Telescope \citep[LAT,][]{atwood2009}.

\subsection{Preparation of \fermi-LAT data}

The \fermi LAT is a space-borne pair-tracking gamma-ray imaging telescope detecting photons in the
energy range from 20~MeV to greater than 1~TeV.
It has been operating since 2008 and its data are publicly available.\footnote{
  \url{https://fermi.gsfc.nasa.gov/ssc/data/}}
For this analysis we retrieved all candidate photon events measured with the LAT over about 124 months
from the beginning of the scientific operations (MET\footnote{
  \fermi Mission Elapsed Time, that is seconds since the reference time of January 1, 2001, at midnight in the
  Coordinated Universal Time (UTC) system.}
from 239\,557\,417~s to 565\,315\,205~s), in a circle centred at
Right Ascension of $83.633\degree$,
Declination of $22.015\degree$,
and a radius of $2.3\degree$, and belonging to the P8R3 SOURCE event class \citep{atwood2013,bruel2018}.

We prepared the LAT data using fermitools 1.0.1\footnote{
  \url{https://fermi.gsfc.nasa.gov/ssc/data/analysis/software/}}
and used the \texttt{P8R3\_SOURCE\_V2} instrument response functions.
We selected candidate photons with energies between 50~GeV and 1~TeV.
The lower energy limit is chosen to avoid complications in accounting for the Crab pulsar.
Based on the spectra measured in \citet{MAGICCrab2016}, the contribution of the pulsar above 50~GeV is
$\lesssim4\%$ with respect to the nebula; thus the pulsar can be neglected for the purpose of this validation study.
An accurate determination of the nebula spectrum at lower energies can be obtained by using an event selection
based on the pulsar phase.
However, this is beyond the scope of this paper and not investigated here.
The upper energy limit follows the standard recommendations for the analysis of LAT data.\footnote{
  \url{https://fermi.gsfc.nasa.gov/ssc/data/analysis/LAT_caveats.html}}
We also selected events with measured arrival directions $<105\degree$ from the local zenith to reduce the
contamination from the bright gamma-ray emission from the Earth's atmosphere.

We used \texttt{gtbin} to bin the events in measured arrival direction and energy. 
For the arrival direction we used a $3\degree \times 3\degree$ region centred at
Right Ascension $83.633\degree$,
Declination $22.015\degree$, and with a grid
step of $0.05\degree$ in \textit{Plate Carr\'ee} (CAR) projection in celestial coordinates.
For the energy binning we used 15 logarithmically spaced bins between 50~GeV and 1~TeV.
We combined together all LAT event types (with different PSF and energy-dispersion qualities).
It would be possible to treat event types separately using a joint likelihood in the following \texttt{ctlike} 
analysis, but this is not investigated here.
We calculated the LAT live time as a function of direction in the sky and incidence angle with respect to the LAT
boresight using \texttt{gtltcube}.
We computed the exposure using \texttt{gtexpcube2} over the analysis region, plus a $3\degree$ border to ensure
a proper convolution of the models with the PSF.
Finally we used \texttt{gtsrcmaps} to calculate the spatial convolution with the LAT PSF of the standard LAT
models for the diffuse backgrounds,\footnote{
  \url{https://fermi.gsfc.nasa.gov/ssc/data/access/lat/BackgroundModels.html}.}
namely interstellar emission from the Milky Way (\texttt{gll\_iem\_v07.fits}) and the isotropic background 
(\texttt{iso\_P8R3\_SOURCE\_V2\_v1.txt}).
A model for the Crab nebula was not included in this step: the source of interest will be modelled directly in
\texttt{ctlike} as described in the next paragraph. 

\begin{table*}
\caption{
Best-fit spectral parameters of the Crab nebula at energies $>50$~GeV from the joint fit to the \fermi-LAT 
and H.E.S.S.~data compared to results from MAGIC \citep{MAGICCrab2015} and from the joint analysis
of LAT, MAGIC, VERITAS, FACT, and H.E.S.S. \citep{nigro2019}.
The spectrum is modelled using a curved power law $dN/dE (E) = k_0 (E/E_0)^{-\Gamma + \beta \ln(E/E_0)}$
with a pivot energy $E_0=1$~TeV. $k_0$ is given in units of $10^{-11}$~cm$^{-2}$~s$^{-1}$~TeV$^{-1}$.
\label{tab:fermihess}
}
\centering
\begin{tabular}{lccc}
\hline\hline
& $k_0$ & $\Gamma$ & $\beta$\\
\hline
LAT (binned) + \hess (unbinned)                                                  & $4.6 \pm 0.2$       & $2.43 \pm 0.03$ & $-0.115 \pm 0.016$ \\
LAT (binned) + \hess (stacked On-Off wstat)                                & $4.5 \pm 0.2$       & $2.42 \pm 0.04$ & $-0.111 \pm 0.018$ \\
LAT (binned) + \hess (unbinned, response scale)                        & $4.3 \pm 0.5$       & $2.45 \pm 0.05$ & $-0.113 \pm 0.017$ \\
LAT (binned) + \hess (stacked On-Off wstat, response scale)      & $4.3 \pm 0.5$       & $2.43 \pm 0.05$ & $-0.110 \pm 0.018$ \\
MAGIC\tablefootmark{a}                                                               & $3.23 \pm 0.03$   & $2.47 \pm 0.01$ & $-0.104 \pm 0.004$ \\
LAT + MAGIC + VERITAS + FACT + H.E.S.S. \tablefootmark{b} & $3.85 \pm 0.11$	& $2.51 \pm 0.03$ & $-0.104 \pm 0.009$ \\
\hline
\end{tabular}
\tablefoot{
   The values of $\beta$ from the literature have been scaled to take into account the different definitions of
   the curved power law.
   \tablefoottext{a}{Values are from \citet{MAGICCrab2015}.}
   \tablefoottext{b}{Values are from \citet{nigro2019}.}
}
\end{table*}

\subsection{Spectrum of the Crab nebula from \fermi-LAT and H.E.S.S.~data}

We used \texttt{ctlike} to perform a joint fit of the LAT data, prepared as described in the previous section,
along with H.E.S.S.~data treated as described in Section~\ref{sec:crabobs}.
We note that for LAT data \texttt{ctlike} at present does not account for energy dispersion.
We chose only two methods to analyse H.E.S.S.~data here as two extreme representative cases, namely the
unbinned analysis and the stacked On-Off analysis with wstat statistic.
We modelled the spectrum of the Crab nebula as a curved power law, 
$dN/dE (E) = k_0 (E/E_0)^{-\Gamma + \beta \ln(E/E_0)}$, with a pivot energy $E_0=1$~TeV, to compare with
the best fit to MAGIC data above 50~GeV presented in \citet{MAGICCrab2015}.
For LAT data we included in the likelihood analysis the models for the diffuse backgrounds described in the
previous section.
Owing to the limited size of the analysis region and low counting statistics we fixed the normalisation of the
isotropic background template to one, and we left as free parameter only the normalisation of the Galactic
interstellar model.
For the latter we verified that the fitted normalisation is always compatible within statistical uncertainties with one.
Diffuse gamma-ray emission can be ignored in the analysis of H.E.S.S.~data because its intensities are
expected to be small with respect to the residual cosmic-ray background and the Crab nebula itself owing to the
higher energy threshold.
Conversely, for H.E.S.S.~we included the model previously described for the residual cosmic-ray background in 
the unbinned analysis.
In the analysed region, no other gamma-ray sources are detected above 10~GeV \citep{ajello2017}.

A general validation  of ctools for the analysis of LAT data has not been performed yet.
It is beyond the scope of this paper, and left for future work.
Nevertheless, we have verified that if we analyse the LAT dataset alone by using \texttt{ctlike} the best-fit 
parameter values and errors are in excellent agreement with the results obtained using \texttt{gtlike} (the 
likelihood tool of the fermitools suite). 

Table~\ref{tab:fermihess} summarises the results from the two different analyses of the LAT and
H.E.S.S.~data, which are in excellent agreement.
Specifically, the second analysis, that is a binned analysis for the LAT and a stacked On-Off wstat analysis
for \hess, corresponds to the well-established standard analysis method for each instrument, and gives
comparable results as a binned analysis for the LAT combined with an unbinned analysis for \hess.
We have thus demonstrated that the unbinned analysis for H.E.S.S.~data can be effectively used in joint
analyses as well.
We have also shown that it is possible to combine analysis techniques tailored to each instrument in a joint
analysis, going beyond the approach illustrated in \citet{nigro2019} that analyse LAT data by unconventionally
using the On-Off method to match the standard approach for IACT data.

The spectral parameters $\Gamma$ and $\beta$ are in good agreement with the results obtained from
the independent observations with MAGIC over an equivalent energy range $> 50$~GeV \citep{MAGICCrab2015},
and from the joint analysis of LAT, MAGIC, VERITAS, FACT, and H.E.S.S. data in \citet{nigro2019}.
The normalisation $k_0$ from our analysis is larger by 30\% than what was reported by MAGIC, that, given the
completely different datasets, is reasonably consistent with systematics related to flux measurements by IACT
telescopes (see for example the discussion on flux systematics in \citealt{hess2018a}, or the variations between values 
of $k_0$ from different instruments in \citealt{nigro2019}).

We then used \texttt{csspec} to derive the SED of the Crab nebula over 15 bins in energy from 50~GeV 
to 100~TeV.
The resulting SEDs are shown in Figure~\ref{fig:jointfermi}.
The SEDs show once again that there is an excellent agreement between results obtained using the 
unbinned and On-Off analysis of H.E.S.S.~data in this case.
The findings are also consistent with published results \citep{aharonian2006a, MAGICCrab2015, 2FHL}.
We note that the higher fluxes found in our analysis compared to \citet{MAGICCrab2015} are also
inferred from \fermi-LAT data alone below 2~TeV in \citet{2FHL}.

We checked for differences in the absolute flux level between H.E.S.S.~and the LAT.
On one hand, the relevant systematic uncertainties for the LAT are the uncertainties in the effective area, 
that for our dataset\footnote{
  \url{https://fermi.gsfc.nasa.gov/ssc/data/analysis/LAT_caveats.html}}
without accounting for energy dispersion amount to 5\% between 50~GeV and 100~GeV, and then 
increase linearly as a function of logarithm of energy to reach 15\% at 1 TeV, and the uncertainties in
the absolute energy scale, that amount to $-3.0\% \pm 0.4\% \, (\mathrm{statistical}) \pm 2.0\% \, 
(\mathrm{systematic})$ \citep{Fermi-e-2017}. 
On the other hand, for H.E.S.S.~the systematic uncertainties in the flux measurement are 30\% \citep{hess2018a}. 
Therefore, for simplicity, we neglected the smaller uncertainties for the LAT, that is we fixed the LAT
effective area to nominal, and we introduced for H.E.S.S.~a scaling factor that multiplies the effective
area.
We included the scale parameter as an additional free parameter in the fit to the data, and we found that the
best-fit values are $1.09 \pm 0.16$ for the unbinned analysis, and $1.04 \pm 0.15$ for the wstat analysis,
respectively.
Therefore, the absolute flux levels measured by the LAT and H.E.S.S.~are consistent within 
statistical uncertainties for the datasets and models considered.
The scale factors obtained are well within the level of systematic uncertainties reported for the two
instruments.
In Table~\ref{tab:fermihess} we show the analysis results also for the case with the scale factor as 
free parameter: all the parameter values are in good agreement with those obtained without applying the
response scaling.
We note that with the effective area scaling factor as a free parameter the uncertainty in the prefactor
$k_0$ is significantly increased due to the degeneracy between the two parameters, or, in other words,
due to the fact that once a free scaling factor is introduced for H.E.S.S.~the constraints on the normalisation
are provided by LAT data alone.

\section{Conclusion}
\label{sec:conclusion}

We presented a comprehensive analysis of the first public H.E.S.S.~data release observations
using the ctools astronomical gamma-ray data analysis software package version 1.6.
We introduced a parametrised background model that describes the expected number
of background events as a function of energy and position in the field of view for each
observation.
The background model was defined and validated using the empty-field observations that are included 
in the H.E.S.S.~data release.

We used this model to derive analysis results for the Crab nebula, MSH 15--52, RX J1713.7--3946,
and PKS 2155--304.
Results were obtained using the different analysis methods available in ctools, including
binned and unbinned 3D maximum-likelihood methods, and several variants of the classical
On-Off techniques.
The achieved results were consistent among these different analysis methods.
We compared our results to equivalent H.E.S.S.~results reported in literature that were obtained
by the H.E.S.S.~Collaboration using their internal software.
Overall we found a favourable agreement with the published H.E.S.S.~results.
A detailed quantitative comparison is however not possible, since, in general, the published H.E.S.S.~results
are based on larger datasets than those included in the public H.E.S.S.~data release, and the
H.E.S.S.-internal processing software version used for the publications also differs from the
software version used for the public H.E.S.S.~data release.

This was the first validation of ctools on real Imaging Atmospheric Cherenkov Telescope data, and
to our knowledge, this work is the first successful application of an unbinned 3D maximum
likelihood analysis to extract the characteristics of gamma-ray sources from such data.
This validation represents an important milestone, as it paves the way towards a broader
use of ctools for Imaging Atmospheric Cherenkov Telescope data analysis, and towards its future application
to data from the Cherenkov Telescope Array.
Likely, the parametrised background model used in this work is too simplistic and
needs refinement, in particular for the application of ctools to CTA data.
Yet there is in principle no obstacle to defining more complex models that reliably represent 
the distribution of the background events in the data.
In this paper we made the proof of principle, and we are confident that this principle also holds
for more complex situations.

Finally, we also demonstrated that ctools readily enables a broad-band multi-instrument data analysis
by combining and jointly analysing public data from the two different instruments \fermi-LAT and
H.E.S.S.~to constrain spectral models.
Such a multi-instrument analysis is fundamental for understanding gamma-ray sources, since
emission spectra from a single physical process typically span several orders of magnitude in
energy.
Consequently, broad-band studies are needed to discriminate between emission processes and
to constrain the physical properties of the underlying particle populations.
And with the upcoming CTA, such joint studies will probably be needed to disentangle and study
the emission from the many sources that will often spatially overlap along the line of sight in the
inner regions of our Galaxy.

\begin{acknowledgements}
We would like to acknowledge highly valuable discussions with members of
the CTA Consortium, the H.E.S.S.~Collaboration, and the \fermi-LAT Collaboration.
We are especially indebted to Philippe Bruel and Abelardo Moralejo for their careful reading
of the manuscript and several insightful suggestions.
We would like to thank Luigi Costamante for having provided the published H.E.S.S.~light curve
flux points for PKS 2155--304.
This research made use of ctools, a community-developed analysis package for Imaging Air Cherenkov
Telescope data \citep{ctools2016}.
ctools is based on GammaLib, a community-developed toolbox for the high-level analysis of astronomical
gamma-ray data \citep{gammalib2011}.
This work has made use of the Python 2D plotting library matplotlib \citep{hunter2007}.
This work has been carried out thanks to the support of the OCEVU Labex (ANR-11-LABX-0060)
and the A$^\ast$MIDEX project (ANR-11-IDEX-0001-02) funded by the ``Investissements d'Avenir''
French government programme managed by the ANR.\\

The \fermi-LAT Collaboration acknowledges generous ongoing support
from a number of agencies and institutes that have supported both the
development and the operation of the LAT as well as scientific data analysis.
These include the National Aeronautics and Space Administration and the
Department of Energy in the United States, the Commissariat \`a l'Energie Atomique
and the Centre National de la Recherche Scientifique / Institut National de Physique
Nucl\'eaire et de Physique des Particules in France, the Agenzia Spaziale Italiana
and the Istituto Nazionale di Fisica Nucleare in Italy, the Ministry of Education,
Culture, Sports, Science and Technology (MEXT), High Energy Accelerator Research
Organization (KEK) and Japan Aerospace Exploration Agency (JAXA) in Japan, and
the K.~A.~Wallenberg Foundation, the Swedish Research Council and the
Swedish National Space Board in Sweden.
Additional support for science analysis during the operations phase is gratefully
acknowledged from the Istituto Nazionale di Astrofisica in Italy and the Centre
National d'\'Etudes Spatiales in France. This work performed in part under DOE
Contract DE-AC02-76SF00515.

\end{acknowledgements}

\bibliographystyle{aa} 
\bibliography{references}

\begin{thebibliography}{49}
\expandafter\ifx\csname natexlab\endcsname\relax\def\natexlab#1{#1}\fi

\bibitem[{{Abdollahi} {et~al.}(2017){Abdollahi}, {Ackermann}, {Ajello},
  {Atwood}, {Baldini}, {Barbiellini}, {Bastieri}, {Bellazzini}, {Bloom},
  {Bonino}, {Brandt}, {Bregeon}, {Bruel}, {Buehler}, {Cameron}, {Caputo},
  {Caragiulo}, {Castro}, {Cavazzuti}, {Cecchi}, {Chekhtman}, {Ciprini},
  {Cohen-Tanugi}, {Costanza}, {Cuoco}, {Cutini}, {D'Ammando}, {de Palma},
  {Desiante}, {Digel}, {Di Lalla}, {Di Mauro}, {Di Venere}, {Drell},
  {Drlica-Wagner}, {Favuzzi}, {Focke}, {Funk}, {Fusco}, {Gargano},
  {Gasparrini}, {Giglietto}, {Giordano}, {Giroletti}, {Green}, {Guillemot},
  {Guiriec}, {Harding}, {Jogler}, {J{\'o}hannesson}, {Kamae}, {Kuss}, {La
  Mura}, {Latronico}, {Longo}, {Loparco}, {Lubrano}, {Maldera}, {Malyshev},
  {Manfreda}, {Mazziotta}, {Michelson}, {Mirabal}, {Mitthumsiri}, {Mizuno},
  {Moiseev}, {Monzani}, {Morselli}, {Moskalenko}, {Negro}, {Nuss}, {Orlando},
  {Paneque}, {Perkins}, {Pesce-Rollins}, {Piron}, {Pivato}, {Porter},
  {Principe}, {Rain{\`o}}, {Rando}, {Razzano}, {Reimer}, {Reimer}, {Sgr{\`o}},
  {Simone}, {Siskind}, {Spada}, {Spandre}, {Spinelli}, {Tajima}, {Thayer},
  {Tibaldo}, {Torres}, {Troja}, {Wood}, {Worley}, {Zaharijas}, {Zimmer}, \&
  {Fermi-LAT Collaboration}}]{Fermi-e-2017}
{Abdollahi}, S., {Ackermann}, M., {Ajello}, M., {et~al.} 2017, \prd, 95, 082007

\bibitem[{{Acero} {et~al.}(2017){Acero}, {Aloisio}, {Amans}, {Amato},
  {Antonelli}, {Aramo}, {Armstrong}, {Arqueros}, {Asano}, {Ashley}, \&
  et~al.}]{acero2017}
{Acero}, F., {Aloisio}, R., {Amans}, J., {et~al.} 2017, \apj, 840, 74

\bibitem[{{Acero} {et~al.}(2009){Acero}, {Ballet}, {Decourchelle},
  {Lemoine-Goumard}, {Ortega}, {Giacani}, {Dubner}, \&
  {Cassam-Chena{\"i}}}]{acero2009}
{Acero}, F., {Ballet}, J., {Decourchelle}, A., {et~al.} 2009, \aap, 505, 157

\bibitem[{{Ackermann} {et~al.}(2016){Ackermann}, {Ajello}, {Atwood}, {Baldini},
  {Ballet}, {Barbiellini}, {Bastieri}, {Becerra Gonzalez}, {Bellazzini},
  {Bissaldi}, {Blandford}, {Bloom}, {Bonino}, {Bottacini}, {Brandt}, {Bregeon},
  {Bruel}, {Buehler}, {Buson}, {Caliandro}, {Cameron}, {Caputo}, {Caragiulo},
  {Caraveo}, {Cavazzuti}, {Cecchi}, {Charles}, {Chekhtman}, {Cheung}, {Chiang},
  {Chiaro}, {Ciprini}, {Cohen}, {Cohen-Tanugi}, {Cominsky}, {Conrad}, {Cuoco},
  {Cutini}, {D'Ammando}, {de Angelis}, {de Palma}, {Desiante}, {Di Mauro}, {Di
  Venere}, {Dom{\'{\i}}nguez}, {Drell}, {Favuzzi}, {Fegan}, {Ferrara}, {Focke},
  {Fortin}, {Franckowiak}, {Fukazawa}, {Funk}, {Furniss}, {Fusco}, {Gargano},
  {Gasparrini}, {Giglietto}, {Giommi}, {Giordano}, {Giroletti}, {Glanzman},
  {Godfrey}, {Grenier}, {Grondin}, {Guillemot}, {Guiriec}, {Harding}, {Hays},
  {Hewitt}, {Hill}, {Horan}, {Iafrate}, {Hartmann}, {Jogler},
  {J{\'o}hannesson}, {Johnson}, {Kamae}, {Kataoka}, {Kn{\"o}dlseder}, {Kuss},
  {La Mura}, {Larsson}, {Latronico}, {Lemoine-Goumard}, {Li}, {Li}, {Longo},
  {Loparco}, {Lott}, {Lovellette}, {Lubrano}, {Madejski}, {Maldera},
  {Manfreda}, {Mayer}, {Mazziotta}, {Michelson}, {Mirabal}, {Mitthumsiri},
  {Mizuno}, {Moiseev}, {Monzani}, {Morselli}, {Moskalenko}, {Murgia}, {Nuss},
  {Ohsugi}, {Omodei}, {Orienti}, {Orlando}, {Ormes}, {Paneque}, {Perkins},
  {Pesce-Rollins}, {Petrosian}, {Piron}, {Pivato}, {Porter}, {Rain{\`o}},
  {Rando}, {Razzano}, {Razzaque}, {Reimer}, {Reimer}, {Reposeur}, {Romani},
  {S{\'a}nchez-Conde}, {Saz Parkinson}, {Schmid}, {Schulz}, {Sgr{\`o}},
  {Siskind}, {Spada}, {Spandre}, {Spinelli}, {Suson}, {Tajima}, {Takahashi},
  {Takahashi}, {Takahashi}, {Thayer}, {Thompson}, {Tibaldo}, {Torres}, {Tosti},
  {Troja}, {Vianello}, {Wood}, {Wood}, {Yassine}, {Zaharijas}, \&
  {Zimmer}}]{2FHL}
{Ackermann}, M., {Ajello}, M., {Atwood}, W.~B., {et~al.} 2016, \apjs, 222, 5

\bibitem[{{Aharonian} {et~al.}(2009){Aharonian}, {Akhperjanian}, {Anton},
  {Barres de Almeida}, {Bazer-Bachi}, {Becherini}, {Behera}, {Benbow},
  {Bernl{\"o}hr}, {Boisson}, {Bochow}, {Borrel}, {Brion}, {Brucker}, {Brun},
  {B{\"u}hler}, {Bulik}, {B{\"u}sching}, {Boutelier}, {Chadwick},
  {Charbonnier}, {Chaves}, {Cheesebrough}, {Chounet}, {Clapson}, {Coignet},
  {Costamante}, {Dalton}, {Daniel}, {Davids}, {Degrange}, {Deil}, {Dickinson},
  {Djannati-Ata{\"i}}, {Domainko}, {O'C.~Drury}, {Dubois}, {Dubus}, {Dyks},
  {Dyrda}, {Egberts}, {Emmanoulopoulos}, {Espigat}, {Farnier}, {Feinstein},
  {Fiasson}, {F{\"o}rster}, {Fontaine}, {F{\"u}{\ss}ling}, {Gabici}, {Gallant},
  {G{\'e}rard}, {Giebels}, {Glicenstein}, {Gl{\"u}ck}, {Goret}, {G{\"o}hring},
  {Hauser}, {Hauser}, {Heinz}, {Heinzelmann}, {Henri}, {Hermann}, {Hinton},
  {Hoffmann}, {Hofmann}, {Holleran}, {Hoppe}, {Horns}, {Jacholkowska}, {de
  Jager}, {Jahn}, {Jung}, {Katarzy{\'n}ski}, {Katz}, {Kaufmann}, {Kendziorra},
  {Kerschhaggl}, {Khangulyan}, {Kh{\'e}lifi}, {Keogh}, {Klu{\'z}niak},
  {Kneiske}, {Komin}, {Kosack}, {Lamanna}, {Lenain}, {Lohse}, {Marandon},
  {Martin}, {Martineau-Huynh}, {Marcowith}, {Maurin}, {McComb}, {Medina},
  {Moderski}, {Monard}, {Moulin}, {Naumann-Godo}, {de Naurois}, {Nedbal},
  {Nekrassov}, {Niemiec}, {Nolan}, {Ohm}, {Olive}, {de O{\~n}a Wilhelmi},
  {Orford}, {Ostrowski}, {Panter}, {Paz Arribas}, {Pedaletti}, {Pelletier},
  {Petrucci}, {Pita}, {P{\"u}hlhofer}, {Punch}, {Quirrenbach}, {Raubenheimer},
  {Raue}, {Rayner}, {Renaud}, {Rieger}, {Ripken}, {Rob}, {Rosier-Lees},
  {Rowell}, {Rudak}, {Rulten}, {Ruppel}, {Sahakian}, {Santangelo},
  {Schlickeiser}, {Sch{\"o}ck}, {Schr{\"o}der}, {Schwanke}, {Schwarzburg},
  {Schwemmer}, {Shalchi}, {Sikora}, {Skilton}, {Sol}, {Spangler}, {Stawarz},
  {Steenkamp}, {Stegmann}, {Superina}, {Szostek}, {Tam}, {Tavernet}, {Terrier},
  {Tibolla}, {Tluczykont}, {van Eldik}, {Vasileiadis}, {Venter}, {Venter},
  {Vialle}, {Vincent}, {Vivier}, {V{\"o}lk}, {Volpe}, {Wagner}, {Ward},
  {Zdziarski}, \& {Zech}}]{aharonian2009}
{Aharonian}, F., {Akhperjanian}, A.~G., {Anton}, G., {et~al.} 2009, \aap, 502,
  749

\bibitem[{{Aharonian} {et~al.}(2005){Aharonian}, {Akhperjanian}, {Aye},
  {Bazer-Bachi}, {Beilicke}, {Benbow}, {Berge}, {Berghaus}, {Bernl{\"o}hr},
  {Boisson}, {Bolz}, {Braun}, {Breitling}, {Brown}, {Bussons Gordo},
  {Chadwick}, {Chounet}, {Cornils}, {Costamante}, {Degrange},
  {Djannati-Ata{\"i}}, {O'C.~Drury}, {Dubus}, {Emmanoulopoulos}, {Espigat},
  {Feinstein}, {Fleury}, {Fontaine}, {Fuchs}, {Funk}, {Gallant}, {Giebels},
  {Gillessen}, {Glicenstein}, {Goret}, {Hadjichristidis}, {Hauser},
  {Heinzelmann}, {Henri}, {Hermann}, {Hinton}, {Hofmann}, {Holleran}, {Horns},
  {de Jager}, {Kh{\'e}lifi}, {Komin}, {Konopelko}, {Latham}, {Le Gallou},
  {Lemi{\`e}re}, {Lemoine-Goumard}, {Leroy}, {Lohse}, {Martineau-Huynh},
  {Marcowith}, {Masterson}, {McComb}, {de Naurois}, {Nolan}, {Noutsos},
  {Orford}, {Osborne}, {Ouchrif}, {Panter}, {Pelletier}, {Pita},
  {P{\"u}hlhofer}, {Punch}, {Raubenheimer}, {Raue}, {Raux}, {Rayner},
  {Redondo}, {Reimer}, {Reimer}, {Ripken}, {Rob}, {Rolland}, {Rowell},
  {Sahakian}, {Saug{\'e}}, {Schlenker}, {Schlickeiser}, {Schuster}, {Schwanke},
  {Siewert}, {Sol}, {Steenkamp}, {Stegmann}, {Tavernet}, {Terrier},
  {Th{\'e}oret}, {Tluczykont}, {Vasileiadis}, {Venter}, {Vincent}, {V{\"o}lk},
  \& {Wagner}}]{aharonian2005}
{Aharonian}, F., {Akhperjanian}, A.~G., {Aye}, K.-M., {et~al.} 2005, \aap, 435,
  L17

\bibitem[{{Aharonian} {et~al.}(2006{\natexlab{a}}){Aharonian}, {Akhperjanian},
  {Bazer-Bachi}, {Beilicke}, {Benbow}, {Berge}, {Bernl{\"o}hr}, {Boisson},
  {Bolz}, {Borrel}, {Braun}, {Breitling}, {Brown}, {B{\"u}hler},
  {B{\"u}sching}, {Carrigan}, {Chadwick}, {Chounet}, {Cornils}, {Costamante},
  {Degrange}, {Dickinson}, {Djannati-Ata{\"\i}}, {O'C. Drury}, {Dubus},
  {Egberts}, {Emmanoulopoulos}, {Espigat}, {Feinstein}, {Ferrero}, {Fiasson},
  {Fontaine}, {Funk}, {Funk}, {Gallant}, {Giebels}, {Glicenstein}, {Goret},
  {Hadjichristidis}, {Hauser}, {Hauser}, {Heinzelmann}, {Henri}, {Hermann},
  {Hinton}, {Hofmann}, {Holleran}, {Horns}, {Jacholkowska}, {de Jager},
  {Kh{\'e}lifi}, {Komin}, {Konopelko}, {Kosack}, {Latham}, {Le Gallou},
  {Lemi{\`e}re}, {Lemoine-Goumard}, {Lohse}, {Martin}, {Martineau-Huynh},
  {Marcowith}, {Masterson}, {McComb}, {de Naurois}, {Nedbal}, {Nolan},
  {Noutsos}, {Orford}, {Osborne}, {Ouchrif}, {Panter}, {Pelletier}, {Pita},
  {P{\"u}hlhofer}, {Punch}, {Raubenheimer}, {Raue}, {Rayner}, {Reimer},
  {Reimer}, {Ripken}, {Rob}, {Rolland}, {Rowell}, {Sahakian}, {Saug{\'e}},
  {Schlenker}, {Schlickeiser}, {Schwanke}, {Sol}, {Spangler}, {Spanier},
  {Steenkamp}, {Stegmann}, {Superina}, {Tavernet}, {Terrier}, {Th{\'e}oret},
  {Tluczykont}, {van Eldik}, {Vasileiadis}, {Venter}, {Vincent}, {V{\"o}lk},
  {Wagner}, \& {Ward}}]{aharonian2006a}
{Aharonian}, F., {Akhperjanian}, A.~G., {Bazer-Bachi}, A.~R., {et~al.}
  2006{\natexlab{a}}, \aap, 457, 899

\bibitem[{{Aharonian} {et~al.}(2006{\natexlab{b}}){Aharonian}, {Akhperjanian},
  {Bazer-Bachi}, {Beilicke}, {Benbow}, {Berge}, {Bernl{\"o}hr}, {Boisson},
  {Bolz}, {Borrel}, {Braun}, {Breitling}, {Brown}, {Chadwick}, {Chounet},
  {Cornils}, {Costamante}, {Degrange}, {Dickinson}, {Djannati-Ata{\"\i}}, {O'C.
  Drury}, {Dubus}, {Emmanoulopoulos}, {Espigat}, {Feinstein}, {Fontaine},
  {Fuchs}, {Funk}, {Gallant}, {Giebels}, {Glicenstein}, {Goret},
  {Hadjichristidis}, {Hauser}, {Hauser}, {Heinzelmann}, {Henri}, {Hermann},
  {Hinton}, {Hofmann}, {Holleran}, {Horns}, {Jacholkowska}, {de Jager},
  {Kh{\'e}lifi}, {Klages}, {Komin}, {Konopelko}, {Latham}, {Le Gallou},
  {Lemi{\`e}re}, {Lemoine-Goumard}, {Lohse}, {Martin}, {Martineau-Huynh},
  {Marcowith}, {Masterson}, {McComb}, {de Naurois}, {Nedbal}, {Nolan},
  {Noutsos}, {Orford}, {Osborne}, {Ouchrif}, {Panter}, {Pelletier}, {Pita},
  {P{\"u}hlhofer}, {Punch}, {Raubenheimer}, {Raue}, {Rayner}, {Reimer},
  {Reimer}, {Ripken}, {Rob}, {Rolland}, {Rowell}, {Sahakian}, {Saug{\'e}},
  {Schlenker}, {Schlickeiser}, {Schuster}, {Schwanke}, {Siewert}, {Sol},
  {Spangler}, {Steenkamp}, {Stegmann}, {Superina}, {Tavernet}, {Terrier},
  {Th{\'e}oret}, {Tluczykont}, {van Eldik}, {Vasileiadis}, {Venter}, {Vincent},
  {V{\"o}lk}, \& {Wagner}}]{aharonian2006b}
{Aharonian}, F., {Akhperjanian}, A.~G., {Bazer-Bachi}, A.~R., {et~al.}
  2006{\natexlab{b}}, \aap, 449, 223

\bibitem[{{Aharonian} {et~al.}(2007){Aharonian}, {Akhperjanian}, {Bazer-Bachi},
  {Beilicke}, {Benbow}, {Berge}, {Bernl{\"o}hr}, {Boisson}, {Bolz}, {Borrel},
  {Braun}, {Brion}, {Brown}, {B{\"u}hler}, {B{\"u}sching}, {Carrigan},
  {Chadwick}, {Chounet}, {Coignet}, {Cornils}, {Costamante}, {Degrange},
  {Dickinson}, {Djannati-Ata{\"\i}}, {O'C. Drury}, {Dubus}, {Egberts},
  {Emmanoulopoulos}, {Espigat}, {Feinstein}, {Ferrero}, {Fiasson}, {Fontaine},
  {Funk}, {Funk}, {F{\"u}{\ss}ling}, {Gallant}, {Giebels}, {Glicenstein},
  {Gl{\"u}ck}, {Goret}, {Hadjichristidis}, {Hauser}, {Hauser}, {Heinzelmann},
  {Henri}, {Hermann}, {Hinton}, {Hoffmann}, {Hofmann}, {Holleran}, {Hoppe},
  {Horns}, {Jacholkowska}, {de Jager}, {Kendziorra}, {Kerschhaggl},
  {Kh{\'e}lifi}, {Komin}, {Konopelko}, {Kosack}, {Lamanna}, {Latham}, {Le
  Gallou}, {Lemi{\`e}re}, {Lemoine-Goumard}, {Lohse}, {Martin},
  {Martineau-Huynh}, {Marcowith}, {Masterson}, {Maurin}, {McComb}, {Moulin},
  {de Naurois}, {Nedbal}, {Nolan}, {Noutsos}, {Olive}, {Orford}, {Osborne},
  {Panter}, {Pelletier}, {Pita}, {P{\"u}hlhofer}, {Punch}, {Ranchon},
  {Raubenheimer}, {Raue}, {Rayner}, {Reimer}, {Reimer}, {Ripken}, {Rob},
  {Rolland}, {Rosier-Lees}, {Rowell}, {Sahakian}, {Santangelo}, {Saug{\'e}},
  {Schlenker}, {Schlickeiser}, {Schr{\"o}der}, {Schwanke}, {Schwarzburg},
  {Schwemmer}, {Shalchi}, {Sol}, {Spangler}, {Spanier}, {Steenkamp},
  {Stegmann}, {Superina}, {Tam}, {Tavernet}, {Terrier}, {Tluczykont}, {van
  Eldik}, {Vasileiadis}, {Venter}, {Vialle}, {Vincent}, {V{\"o}lk}, {Wagner},
  \& {Ward}}]{aharonian2007}
{Aharonian}, F., {Akhperjanian}, A.~G., {Bazer-Bachi}, A.~R., {et~al.} 2007,
  \aap, 464, 235

\bibitem[{{Aharonian} {et~al.}(2011){Aharonian}, {Akhperjanian}, {Bazer-Bachi},
  {Beilicke}, {Benbow}, {Berge}, {Bernl{\"o}hr}, {Boisson}, {Bolz}, {Borrel},
  {Braun}, {Brion}, {Brown}, {B{\"u}hler}, {B{\"u}sching}, {Carrigan},
  {Chadwick}, {Chounet}, {Coignet}, {Cornils}, {Costamante}, {Degrange},
  {Dickinson}, {Djannati-Ata{\"i}}, {Drury}, {Dubus}, {Egberts},
  {Emmanoulopoulos}, {Espigat}, {Feinstein}, {Ferrero}, {Fiasson}, {Fontaine},
  {Funk}, {Funk}, {F{\"u}{\ss}ling}, {Gallant}, {Giebels}, {Glicenstein},
  {Gl{\"u}ck}, {Goret}, {Hadjichristidis}, {Hauser}, {Hauser}, {Heinzelmann},
  {Henri}, {Hermann}, {Hinton}, {Hoffmann}, {Hofmann}, {Holleran}, {Hoppe},
  {Horns}, {Jacholkowska}, {de Jager}, {Kendziorra}, {Kerschhaggl},
  {Kh{\'e}lifi}, {Komin}, {Konopelko}, {Kosack}, {Lamanna}, {Latham}, {Le
  Gallou}, {Lemi{\`e}re}, {Lemoine-Goumard}, {Lohse}, {Martin},
  {Martineau-Huynh}, {Marcowith}, {Masterson}, {Maurin}, {McComb}, {Moulin},
  {de Naurois}, {Nedbal}, {Nolan}, {Noutsos}, {Olive}, {Orford}, {Osborne},
  {Panter}, {Pelletier}, {Pita}, {P{\"u}hlhofer}, {Punch}, {Ranchon},
  {Raubenheimer}, {Raue}, {Rayner}, {Reimer}, {Reimer}, {Ripken}, {Rob},
  {Rolland}, {Rosier-Lees}, {Rowell}, {Sahakian}, {Santangelo}, {Saug{\'e}},
  {Schlenker}, {Schlickeiser}, {Schr{\"o}der}, {Schwanke}, {Schwarzburg},
  {Schwemmer}, {Shalchi}, {Sol}, {Spangler}, {Spanier}, {Steenkamp},
  {Stegmann}, {Superina}, {Tam}, {Tavernet}, {Terrier}, {Tluczykont}, {van
  Eldik}, {Vasileiadis}, {Venter}, {Vialle}, {Vincent}, {V{\"o}lk}, {Wagner},
  {Ward}, \& {H.E.S.S.~Collaboration}}]{aharonian2011}
{Aharonian}, F., {Akhperjanian}, A.~G., {Bazer-Bachi}, A.~R., {et~al.} 2011,
  \aap, 531, C1

\bibitem[{{Ajello} {et~al.}(2017){Ajello}, {Atwood}, {Baldini}, {Ballet},
  {Barbiellini}, {Bastieri}, {Bellazzini}, {Bissaldi}, {Blandford}, {Bloom},
  {Bonino}, {Bregeon}, {Britto}, {Bruel}, {Buehler}, {Buson}, {Cameron},
  {Caputo}, {Caragiulo}, {Caraveo}, {Cavazzuti}, {Cecchi}, {Charles},
  {Chekhtman}, {Cheung}, {Chiaro}, {Ciprini}, {Cohen}, {Costantin}, {Costanza},
  {Cuoco}, {Cutini}, {D'Ammando}, {de Palma}, {Desiante}, {Digel}, {Di Lalla},
  {Di Mauro}, {Di Venere}, {Dom{\'{\i}}nguez}, {Drell}, {Dumora}, {Favuzzi},
  {Fegan}, {Ferrara}, {Fortin}, {Franckowiak}, {Fukazawa}, {Funk}, {Fusco},
  {Gargano}, {Gasparrini}, {Giglietto}, {Giommi}, {Giordano}, {Giroletti},
  {Glanzman}, {Green}, {Grenier}, {Grondin}, {Grove}, {Guillemot}, {Guiriec},
  {Harding}, {Hays}, {Hewitt}, {Horan}, {J{\'o}hannesson}, {Kensei}, {Kuss},
  {La Mura}, {Larsson}, {Latronico}, {Lemoine-Goumard}, {Li}, {Longo},
  {Loparco}, {Lott}, {Lubrano}, {Magill}, {Maldera}, {Manfreda}, {Mazziotta},
  {McEnery}, {Meyer}, {Michelson}, {Mirabal}, {Mitthumsiri}, {Mizuno},
  {Moiseev}, {Monzani}, {Morselli}, {Moskalenko}, {Negro}, {Nuss}, {Ohsugi},
  {Omodei}, {Orienti}, {Orlando}, {Palatiello}, {Paliya}, {Paneque}, {Perkins},
  {Persic}, {Pesce-Rollins}, {Piron}, {Porter}, {Principe}, {Rain{\`o}},
  {Rando}, {Razzano}, {Razzaque}, {Reimer}, {Reimer}, {Reposeur}, {Saz
  Parkinson}, {Sgr{\`o}}, {Simone}, {Siskind}, {Spada}, {Spandre}, {Spinelli},
  {Stawarz}, {Suson}, {Takahashi}, {Tak}, {Thayer}, {Thayer}, {Thompson},
  {Torres}, {Torresi}, {Troja}, {Vianello}, {Wood}, \& {Wood}}]{ajello2017}
{Ajello}, M., {Atwood}, W.~B., {Baldini}, L., {et~al.} 2017, \apjs, 232, 18

\bibitem[{{Aleksi{\'c}} {et~al.}(2015){Aleksi{\'c}}, {Ansoldi}, {Antonelli},
  {Antoranz}, {Babic}, {Bangale}, {Barrio}, {Becerra Gonz{\'a}lez}, {Bednarek},
  {Bernardini}, {Biasuzzi}, {Biland}, {Blanch}, {Bonnefoy}, {Bonnoli},
  {Borracci}, {Bretz}, {Carmona}, {Carosi}, {Colin}, {Colombo}, {Contreras},
  {Cortina}, {Covino}, {Da Vela}, {Dazzi}, {De Angelis}, {De Caneva}, {De
  Lotto}, {de O{\~n}a Wilhelmi}, {Delgado Mendez}, {Doert}, {Dominis Prester},
  {Dorner}, {Doro}, {Einecke}, {Eisenacher}, {Elsaesser}, {Fonseca}, {Font},
  {Frantzen}, {Fruck}, {Galindo}, {Garc{\'{\i}}a L{\'o}pez}, {Garczarczyk},
  {Garrido Terrats}, {Gaug}, {Godinovi{\'c}}, {Gonz{\'a}lez Mu{\~n}oz},
  {Gozzini}, {Hadasch}, {Hanabata}, {Hayashida}, {Herrera}, {Hildebrand},
  {Hose}, {Hrupec}, {Idec}, {Kadenius}, {Kellermann}, {Kodani}, {Konno},
  {Krause}, {Kubo}, {Kushida}, {La Barbera}, {Lelas}, {Lewandowska},
  {Lindfors}, {Lombardi}, {L{\'o}pez}, {L{\'o}pez-Coto}, {L{\'o}pez-Oramas},
  {Lorenz}, {Lozano}, {Makariev}, {Mallot}, {Maneva}, {Mankuzhiyil},
  {Mannheim}, {Maraschi}, {Marcote}, {Mariotti}, {Mart{\'{\i}}nez}, {Mazin},
  {Menzel}, {Miranda}, {Mirzoyan}, {Moralejo}, {Munar-Adrover}, {Nakajima},
  {Niedzwiecki}, {Nilsson}, {Nishijima}, {Noda}, {Nowak}, {Orito},
  {Overkemping}, {Paiano}, {Palatiello}, {Paneque}, {Paoletti}, {Paredes},
  {Paredes-Fortuny}, {Persic}, {Prada Moroni}, {Prandini}, {Preziuso},
  {Puljak}, {Reinthal}, {Rhode}, {Rib{\'o}}, {Rico}, {Rodriguez Garcia},
  {R{\"u}gamer}, {Saggion}, {Saito}, {Saito}, {Satalecka}, {Scalzotto},
  {Scapin}, {Schultz}, {Schweizer}, {Shore}, {Sillanp{\"a}{\"a}}, {Sitarek},
  {Snidaric}, {Sobczynska}, {Spanier}, {Stamatescu}, {Stamerra}, {Steinbring},
  {Storz}, {Strzys}, {Takalo}, {Takami}, {Tavecchio}, {Temnikov}, {Terzi{\'c}},
  {Tescaro}, {Teshima}, {Thaele}, {Tibolla}, {Torres}, {Toyama}, {Treves},
  {Uellenbeck}, {Vogler}, {Wagner}, {Zanin}, {Horns}, {Mart{\'{\i}}n}, \&
  {Meyer}}]{MAGICCrab2015}
{Aleksi{\'c}}, J., {Ansoldi}, S., {Antonelli}, L.~A., {et~al.} 2015, Journal of
  High Energy Astrophysics, 5, 30

\bibitem[{{Ansoldi} {et~al.}(2016){Ansoldi}, {Antonelli}, {Antoranz}, {Babic},
  {Bangale}, {Barres de Almeida}, {Barrio}, {Becerra Gonz{\'a}lez}, {Bednarek},
  {Bernardini}, {Biasuzzi}, {Biland}, {Blanch}, {Bonnefoy}, {Bonnoli},
  {Borracci}, {Bretz}, {Carmona}, {Carosi}, {Colin}, {Colombo}, {Contreras},
  {Cortina}, {Covino}, {Da Vela}, {Dazzi}, {De Angelis}, {De Caneva}, {De
  Lotto}, {de O{\~n}a Wilhelmi}, {Delgado Mendez}, {Di Pierro}, {Dominis
  Prester}, {Dorner}, {Doro}, {Einecke}, {Eisenacher Glawion}, {Elsaesser},
  {Fern{\'a}ndez-Barral}, {Fidalgo}, {Fonseca}, {Font}, {Frantzen}, {Fruck},
  {Galindo}, {Garc{\'{\i}}a L{\'o}pez}, {Garczarczyk}, {Garrido Terrats},
  {Gaug}, {Godinovi{\'c}}, {Gonz{\'a}lez Mu{\~n}oz}, {Gozzini}, {Hanabata},
  {Hayashida}, {Herrera}, {Hirotani}, {Hose}, {Hrupec}, {Hughes}, {Idec},
  {Kellermann}, {Knoetig}, {Kodani}, {Konno}, {Krause}, {Kubo}, {Kushida}, {La
  Barbera}, {Lelas}, {Lewandowska}, {Lindfors}, {Lombardi}, {Longo},
  {L{\'o}pez}, {L{\'o}pez-Coto}, {L{\'o}pez-Oramas}, {Lorenz}, {Makariev},
  {Mallot}, {Maneva}, {Mannheim}, {Maraschi}, {Marcote}, {Mariotti},
  {Mart{\'{\i}}nez}, {Mazin}, {Menzel}, {Miranda}, {Mirzoyan}, {Moralejo},
  {Munar-Adrover}, {Nakajima}, {Neustroev}, {Niedzwiecki}, {Nevas Rosillo},
  {Nilsson}, {Nishijima}, {Noda}, {Orito}, {Overkemping}, {Paiano},
  {Palatiello}, {Paneque}, {Paoletti}, {Paredes}, {Paredes-Fortuny}, {Persic},
  {Poutanen}, {Prada Moroni}, {Prandini}, {Puljak}, {Reinthal}, {Rhode},
  {Rib{\'o}}, {Rico}, {Rodriguez Garcia}, {Saito}, {Saito}, {Satalecka},
  {Scalzotto}, {Scapin}, {Schultz}, {Schweizer}, {Shore}, {Sillanp{\"a}{\"a}},
  {Sitarek}, {Snidaric}, {Sobczynska}, {Stamerra}, {Steinbring}, {Strzys},
  {Takalo}, {Takami}, {Tavecchio}, {Temnikov}, {Terzi{\'c}}, {Tescaro},
  {Teshima}, {Thaele}, {Torres}, {Toyama}, {Treves}, {Ward}, {Will}, \&
  {Zanin}}]{MAGICCrab2016}
{Ansoldi}, S., {Antonelli}, L.~A., {Antoranz}, P., {et~al.} 2016, \aap, 585,
  A133

\bibitem[{{Atwood} {et~al.}(2013){Atwood}, {Albert}, {Baldini}, {Tinivella},
  {Bregeon}, {Pesce-Rollins}, {Sgr{\`o}}, {Bruel}, {Charles}, {Drlica-Wagner},
  {Franckowiak}, {Jogler}, {Rochester}, {Usher}, {Wood}, {Cohen-Tanugi}, \&
  {S.~Zimmer for the Fermi-LAT Collaboration}}]{atwood2013}
{Atwood}, W., {Albert}, A., {Baldini}, L., {et~al.} 2013, in Proc. 4th Fermi
  Symposium (eConf C121028), 8

\bibitem[{{Atwood} {et~al.}(2009){Atwood}, {Abdo}, {Ackermann}, {Althouse},
  {Anderson}, {Axelsson}, {Baldini}, {Ballet}, {Band}, {Barbiellini}, \&
  et~al.}]{atwood2009}
{Atwood}, W.~B., {Abdo}, A.~A., {Ackermann}, M., {et~al.} 2009, \apj, 697, 1071

\bibitem[{{Bal{\'a}zs} {et~al.}(2017){Bal{\'a}zs}, {Conrad}, {Farmer},
  {Jacques}, {Li}, {Meyer}, {Queiroz}, \& {S{\'a}nchez-Conde}}]{balazs2017}
{Bal{\'a}zs}, C., {Conrad}, J., {Farmer}, B., {et~al.} 2017, \prd, 96, 083002

\bibitem[{{Berge} {et~al.}(2007){Berge}, {Funk}, \& {Hinton}}]{berge2007}
{Berge}, D., {Funk}, S., \& {Hinton}, J. 2007, \aap, 466, 1219

\bibitem[{{Bruel} {et~al.}(2018){Bruel}, {Burnett}, {Digel}, {Johannesson},
  {Omodei}, \& {Wood}}]{bruel2018}
{Bruel}, P., {Burnett}, T.~H., {Digel}, S.~W., {et~al.} 2018, arXiv e-prints
  [\eprint[arXiv]{1810.11394}]

\bibitem[{{Burtovoi} {et~al.}(2017){Burtovoi}, {Saito}, {Zampieri}, \&
  {Hassan}}]{burtovoi2017}
{Burtovoi}, A., {Saito}, T.~Y., {Zampieri}, L., \& {Hassan}, T. 2017, \mnras,
  471, 431

\bibitem[{{Cash}(1979)}]{cash1979}
{Cash}, W. 1979, \apj, 228, 939

\bibitem[{{Chadwick} {et~al.}(1999){Chadwick}, {Lyons}, {McComb}, {Orford},
  {Osborne}, {Rayner}, {Shaw}, {Turver}, \& {Wieczorek}}]{chadwick1999}
{Chadwick}, P.~M., {Lyons}, K., {McComb}, T.~J.~L., {et~al.} 1999,
  Astroparticle Physics, 11, 145

\bibitem[{{CTA Consortium}(2019)}]{CTAScienceBook}
{CTA Consortium}. 2019, {Science with the Cherenkov Telescope Array} (World
  Scientific Publishing Co)

\bibitem[{{De Franco} {et~al.}(2017){De Franco}, {Inoue}, {S{\'a}nchez-Conde},
  \& {Cotter}}]{defranco2017}
{De Franco}, A., {Inoue}, Y., {S{\'a}nchez-Conde}, M.~A., \& {Cotter}, G. 2017,
  Astroparticle Physics, 93, 8

\bibitem[{{Dubois}(2009)}]{dubois2009}
{Dubois}, F. 2009, PhD thesis, University of Savoy

\bibitem[{{Gaensler} {et~al.}(2002){Gaensler}, {Arons}, {Kaspi}, {Pivovaroff},
  {Kawai}, \& {Tamura}}]{gaensler2002}
{Gaensler}, B.~M., {Arons}, J., {Kaspi}, V.~M., {et~al.} 2002, \apj, 569, 878

\bibitem[{{H.E.S.S.~Collaboration}(2018)}]{hess2018b}
{H.E.S.S.~Collaboration}. 2018, ArXiv e-prints [\eprint[arXiv]{1810.04516}]

\bibitem[{{H.E.S.S. Collaboration} {et~al.}(2018){H.E.S.S. Collaboration},
  {Abdalla}, {Abramowski}, {Aharonian}, {Benkhali}, {Akhperjanian},
  {Andersson}, {Ang{\"u}ner}, {Arrieta}, {Aubert}, {Backes}, {Balzer},
  {Barnard}, {Becherini}, {Tjus}, {Berge}, {Bernhard}, {Bernl{\"o}hr},
  {Blackwell}, {B{\"o}ttcher}, {Boisson}, {Bolmont}, {Bordas}, {Bregeon},
  {Brun}, {Brun}, {Bryan}, {Bulik}, {Capasso}, {Carr}, {Casanova}, {Cerruti},
  {Chakraborty}, {Chalme-Calvet}, {Chaves}, {Chen}, {Chevalier},
  {Chr{\'e}tien}, {Colafrancesco}, {Cologna}, {Condon}, {Conrad}, {Cui},
  {Davids}, {Decock}, {Degrange}, {Deil}, {Devin}, {deWilt}, {Dirson},
  {Djannati-Ata{\"\i}}, {Domainko}, {Donath}, {Drury}, {Dubus}, {Dutson},
  {Dyks}, {Edwards}, {Egberts}, {Eger}, {Ernenwein}, {Eschbach}, {Farnier},
  {Fegan}, {Fernandes}, {Fiasson}, {Fontaine}, {F{\"o}rster}, {Fukuyama},
  {Funk}, {F{\"u}{\ss}ling}, {Gabici}, {Gajdus}, {Gallant}, {Garrigoux},
  {Giavitto}, {Giebels}, {Glicenstein}, {Gottschall}, {Goyal}, {Grondin},
  {Hadasch}, {Hahn}, {Haupt}, {Hawkes}, {Heinzelmann}, {Henri}, {Hermann},
  {Hervet}, {Hinton}, {Hofmann}, {Hoischen}, {Holler}, {Horns}, {Ivascenko},
  {Jacholkowska}, {Jamrozy}, {Janiak}, {Jankowsky}, {Jankowsky}, {Jingo},
  {Jogler}, {Jouvin}, {Jung- Richardt}, {Kastendieck}, {Katarzy{\'n}ski},
  {Katz}, {Kerszberg}, {Kh{\'e}lifi}, {Kieffer}, {King}, {Klepser}, {Klochkov},
  {Klu{\'z}niak}, {Kolitzus}, {Komin}, {Kosack}, {Krakau}, {Kraus}, {Krayzel},
  {Kr{\"u}ger}, {Laffon}, {Lamanna}, {Lau}, {Lees}, {Lefaucheur}, {Lefranc},
  {Lemi{\`e}re}, {Lemoine-Goumard}, {Lenain}, {Leser}, {Lohse}, {Lorentz},
  {Liu}, {L{\'o}pez-Coto}, {Lypova}, {Marandon}, {Marcowith}, {Mariaud},
  {Marx}, {Maurin}, {Maxted}, {Mayer}, {Meintjes}, {Meyer}, {Mitchell},
  {Moderski}, {Mohamed}, {Mohrmann}, {Mor{\r{a}}}, {Moulin}, {Murach},
  {Naurois}, {Niederwanger}, {Niemiec}, {Oakes}, {O'Brien}, {Odaka},
  {{\"O}ttl}, {Ohm}, {Ostrowski}, {Oya}, {Padovani}, {Panter}, {Parsons},
  {Pekeur}, {Pelletier}, {Perennes}, {Petrucci}, {Peyaud}, {Piel}, {Pita},
  {Poon}, {Prokhorov}, {Prokoph}, {P{\"u}hlhofer}, {Punch}, {Quirrenbach},
  {Raab}, {Reimer}, {Reimer}, {Renaud}, {los Reyes}, {Rieger}, {Romoli},
  {Rosier-Lees}, {Rowell}, {Rudak}, {Rulten}, {Sahakian}, {Salek}, {Sanchez},
  {Santangelo}, {Sasaki}, {Schlickeiser}, {Sch{\"u}ssler}, {Schulz},
  {Schwanke}, {Schwemmer}, {Settimo}, {Seyffert}, {Shafi}, {Shilon}, {Simoni},
  {Sol}, {Spanier}, {Spengler}, {Spies}, {Stawarz}, {Steenkamp}, {Stegmann},
  {Stinzing}, {Stycz}, {Sushch}, {Takahashi}, {Tavernet}, {Tavernier},
  {Taylor}, {Terrier}, {Tibaldo}, {Tiziani}, {Tluczykont}, {Trichard}, {Tuffs},
  {Uchiyama}, {van der Walt}, {Eldik}, {Rensburg}, {Soelen}, {Vasileiadis},
  {Veh}, {Venter}, {Viana}, {Vincent}, {Vink}, {Voisin}, {V{\"o}lk}, {Volpe},
  {Vuillaume}, {Wadiasingh}, {Wagner}, {Wagner}, {Wagner}, {White},
  {Wierzcholska}, {Willmann}, {W{\"o}rnlein}, {Wouters}, {Yang}, {Zabalza},
  {Zaborov}, {Zacharias}, {Zdziarski}, {Zech}, {Zefi}, {Ziegler}, \&
  {{\.Z}ywucka}}]{hess2018c}
{H.E.S.S. Collaboration}, {Abdalla}, H., {Abramowski}, A., {et~al.} 2018, \aap,
  612, A6

\bibitem[{{H.E.S.S.~Collaboration} {et~al.}(2018){H.E.S.S.~Collaboration},
  {Abdalla}, {Abramowski}, {Aharonian}, {Benkhali}, {Ang{\"u}ner}, {Arakawa},
  {Arrieta}, {Aubert}, {Backes}, \& et~al.}]{hess2018a}
{H.E.S.S.~Collaboration}, {Abdalla}, H., {Abramowski}, A., {et~al.} 2018, \aap,
  612, A1

\bibitem[{{Hobbs} {et~al.}(2006){Hobbs}, {Edwards}, \&
  {Manchester}}]{tempo22006}
{Hobbs}, G.~B., {Edwards}, R.~T., \& {Manchester}, R.~N. 2006, \mnras, 369, 655

\bibitem[{{Holder}(2012)}]{holder2012}
{Holder}, J. 2012, Astroparticle Physics, 39, 61

\bibitem[{{Holler} {et~al.}(2017){Holler}, {Berge}, {Hahn}, {Khangulyan},
  {Parsons}, \& {for the H.~E.~S.~S.~collaboration}}]{holler2017}
{Holler}, M., {Berge}, D., {Hahn}, J., {et~al.} 2017, ArXiv e-prints
  [\eprint[arXiv]{1707.04196}]

\bibitem[{Hunter(2007)}]{hunter2007}
Hunter, J.~D. 2007, Computing In Science \& Engineering, 9, 90

\bibitem[{{H{\"u}tten} {et~al.}(2016){H{\"u}tten}, {Combet}, {Maier}, \&
  {Maurin}}]{huetten2016}
{H{\"u}tten}, M., {Combet}, C., {Maier}, G., \& {Maurin}, D. 2016, \jcap, 9,
  047

\bibitem[{{H{\"u}tten} \& {Maier}(2018)}]{huetten2018}
{H{\"u}tten}, M. \& {Maier}, G. 2018, \jcap, 8, 032

\bibitem[{{Kn{\"o}dlseder} {et~al.}(2016{\natexlab{a}}){Kn{\"o}dlseder},
  {Mayer}, {Deil}, {Buehler}, {Bregeon}, \& {Martin}}]{ctools2016}
{Kn{\"o}dlseder}, J., {Mayer}, M., {Deil}, C., {et~al.} 2016{\natexlab{a}},
  Astrophysics Source Code Library [\eprint{ascl:1601.005}]

\bibitem[{{Kn{\"o}dlseder} {et~al.}(2011){Kn{\"o}dlseder}, {Mayer}, {Deil},
  {Cayrou}, {Owen}, {Kelley-Hoskins}, {Lu}, {Buehler}, {Forest}, {Louge},
  {Siejkowski}, {Kosack}, {Gerard}, {Schulz}, {Martin}, {Hassan}, \&
  {Brau-Nogu'e}}]{gammalib2011}
{Kn{\"o}dlseder}, J., {Mayer}, M., {Deil}, C., {et~al.} 2011, Astrophysics
  Source Code Library [\eprint{ascl:1110.007}]

\bibitem[{{Kn{\"o}dlseder} {et~al.}(2016{\natexlab{b}}){Kn{\"o}dlseder},
  {Mayer}, {Deil}, {Cayrou}, {Owen}, {Kelley-Hoskins}, {Lu}, {Buehler},
  {Forest}, {Louge}, {Siejkowski}, {Kosack}, {Gerard}, {Schulz}, {Martin},
  {Sanchez}, {Ohm}, {Hassan}, \& {Brau-Nogu{\'e}}}]{knoedlseder2016}
{Kn{\"o}dlseder}, J., {Mayer}, M., {Deil}, C., {et~al.} 2016{\natexlab{b}},
  \aap, 593, A1

\bibitem[{{Landoni} {et~al.}(2019){Landoni}, {Romano}, {Vercellone},
  {Kn{\"o}dlseder}, {Bianco}, {Tavecchio}, \& {Corina}}]{landoni2019}
{Landoni}, M., {Romano}, P., {Vercellone}, S., {et~al.} 2019, \apjs, 240, 32

\bibitem[{{Li} \& {Ma}(1983)}]{li1983}
{Li}, T.-P. \& {Ma}, Y.-Q. 1983, \apj, 272, 317

\bibitem[{{Ma} {et~al.}(1998){Ma}, {Arias}, {Eubanks}, {Fey}, {Gontier},
  {Jacobs}, {Sovers}, {Archinal}, \& {Charlot}}]{ma1998}
{Ma}, C., {Arias}, E.~F., {Eubanks}, T.~M., {et~al.} 1998, \aj, 116, 516

\bibitem[{{Mattox} {et~al.}(1996){Mattox}, {Bertsch}, {Chiang}, {Dingus},
  {Digel}, {Esposito}, {Fierro}, {Hartman}, {Hunter}, {Kanbach}, {Kniffen},
  {Lin}, {Macomb}, {Mayer-Hasselwander}, {Michelson}, {von Montigny},
  {Mukherjee}, {Nolan}, {Ramanamurthy}, {Schneid}, {Sreekumar}, {Thompson}, \&
  {Willis}}]{mattox1996}
{Mattox}, J.~R., {Bertsch}, D.~L., {Chiang}, J., {et~al.} 1996, \apj, 461, 396

\bibitem[{{Nigro} {et~al.}(2019){Nigro}, {Deil}, {Zanin}, {Hassan}, {King},
  {Ruiz}, {Saha}, {Terrier}, {Br{\"u}gge}, {N{\"o}the}, {Bird}, {Lin},
  {Aleksi{\'c}}, {Boisson}, {Contreras}, {Donath}, {Jouvin}, {Kelley-Hoskins},
  {Khelifi}, {Kosack}, {Rico}, \& {Sinha}}]{nigro2019}
{Nigro}, C., {Deil}, C., {Zanin}, R., {et~al.} 2019, accepted by \aap
  [\eprint[arXiv]{1903.06621}]

\bibitem[{{Patricelli} {et~al.}(2018){Patricelli}, {Stamerra}, {Razzano},
  {Pian}, \& {Cella}}]{patricelli2018}
{Patricelli}, B., {Stamerra}, A., {Razzano}, M., {Pian}, E., \& {Cella}, G.
  2018, \jcap, 5, 056

\bibitem[{{Petropoulou} {et~al.}(2017){Petropoulou}, {Vasilopoulos}, \&
  {Giannios}}]{petropoulou2017}
{Petropoulou}, M., {Vasilopoulos}, G., \& {Giannios}, D. 2017, \mnras, 464,
  2213

\bibitem[{{Romano} {et~al.}(2018){Romano}, {Vercellone}, {Foschini},
  {Tavecchio}, {Landoni}, \& {Kn{\"o}dlseder}}]{romano2018}
{Romano}, P., {Vercellone}, S., {Foschini}, L., {et~al.} 2018, \mnras, 481,
  5046

\bibitem[{{Tavecchio} {et~al.}(2019){Tavecchio}, {Romano}, {Landoni}, \&
  {Vercellone}}]{tavecchio2019}
{Tavecchio}, F., {Romano}, P., {Landoni}, M., \& {Vercellone}, S. 2019, \mnras,
  483, 1802

\bibitem[{{Weekes} {et~al.}(1989){Weekes}, {Cawley}, {Fegan}, {Gibbs},
  {Hillas}, {Kowk}, {Lamb}, {Lewis}, {Macomb}, {Porter}, {Reynolds}, \&
  {Vacanti}}]{weekes1989}
{Weekes}, T.~C., {Cawley}, M.~F., {Fegan}, D.~J., {et~al.} 1989, \apj, 342, 379

\bibitem[{Wilks(1938)}]{wilks1938}
Wilks, S.~S. 1938, Ann. Math. Statist., 9, 60

\bibitem[{{Yang} \& {Razzaque}(2019)}]{yang2019}
{Yang}, L. \& {Razzaque}, S. 2019, \prd, 99, 083007

\end{thebibliography}

\begin{appendix}

\section{Residuals computation}
\label{sec:residual_computation}

We express residuals throughout this paper as significance in units of Gaussian $\sigma$, computed
using a log-likelihood-ratio test for Poisson statistics.
The two hypotheses in the test are that the model is sufficient to describe the data (null hypothesis) with
log-likelihood given by
\begin{equation}
\ln L_0 = \exp{(-m)} \,  \frac{m^n}{n!} ,
\end{equation}
where $n$ is the number of observed counts, and $m$ the number of expected counts based on
the model,
or that the data comprise the model plus an unknown residual component (alternative hypothesis),
the sum of which is assumed to match the observed number of counts $n$, so that the log-likelihood is
\begin{equation}
\ln L_{\rm test} = \exp{(-n)} \, \frac{n^n}{n!} .
\end{equation}
Based on Wilks' theorem \citep{wilks1938} twice the logarithm of the likelihood ratio is distributed
as a $\chi^2$ with one degree of freedom (for one additional parameter, the unknown number of
residual counts set to match the data), and, therefore, its square root is distributed as a normal variable.
The final formula for the residual $r$ expressed as significance in Gaussian $\sigma$ is
\begin{equation}
r = \sgn(n-m) \sqrt{ 2 \left( n \ln \frac{n}{m} + m - n \right)} ,
\label{eq:sigma}
\end{equation}
where the sign term indicates whether the measured number of counts is larger or smaller than
the number of counts predicted by the model.
Some special cases need to be treated separately.
Namely, if $n = 0$ the residual significance is
\begin{equation}
r = \sgn(n-m) \sqrt{2 m} ,
\label{eq:sigma0}
\end{equation}
while if $m=0$ the significance cannot be computed and we set $r=0$.
This implies that the method becomes inaccurate in the low-counting regime.
The residual computation is implemented in the \texttt{csresmap} and \texttt{csresspec} scripts.

\section{Maximum likelihood estimation}
\label{sec:ctlike}

The central method behind the ctools data analysis is the maximum likelihood estimation of the parameters
of a given model.
The method obtains the parameter estimates by finding the parameter values that maximise the likelihood
function $L(M)$.
The likelihood function quantifies the probability that the data are drawn from a particular model $M$.
The formula used for the likelihood depends on whether the data are unbinned or binned and on the assumed
underlying statistical law.
For the 3D analyses, the reader should refer to \citet{knoedlseder2016} for the likelihood formulae;
for the On-Off analyses they are given in Appendix \ref{sec:wstat_cstat}.

We used an iterative Levenberg-Marquardt algorithm for the estimation of the maximum of the
likelihood function $L(M)$.
Since the Levenberg-Marquardt algorithm minimises a function, $-\ln L(M)$ is used as the function to
minimise.
The Levenberg-Marquardt algorithm starts with an initial guess of the model parameters $a_k$ and iteratively
replaces this estimate by a new estimate $a_k+\Delta a_k$.
The $\Delta a_k$ are determined by solving
\begin{equation}
\sum_l \alpha_{kl} (1 + \delta_{kl} \lambda) \Delta a_l = \beta_k ,
\end{equation}
where
\begin{equation}
\alpha_{kl} = \frac{\partial^2 (-\ln L(M))}{\partial a_k \partial a_l}
\end{equation}
is the curvature matrix,
\begin{equation}
\beta_k = \frac{\partial (-\ln L(M))}{\partial a_k}
\end{equation}
is the gradient, and $\delta_{kl}$ is the Kronecker delta that is $1$ for $k=l$ and $0$ otherwise.
$\lambda$ is a damping parameter that initially is set to $0.001$.
If a Levenberg-Marquardt iteration leads to an increase of the log-likelihood function $\ln L(M)$,
$\lambda$ is decreased by a factor of 10.
If the log-likelihood function $\ln L(M)$ does not improve, $\lambda$ is increased by a factor of 10 and
the iteration is repeated.
The iterations are stopped when the log-likelihood increase is less than a small value, typically $0.005$; the
optimiser status is then set to {\tt converged}.
The iterations are also stopped if the log-likelihood function does not increase for (typically) ten iterations;
the optimiser status is then set to {\tt stalled}.

The matrix equation is solved using a sparse matrix Cholesky decomposition.
Parameters are constrained within their parameter limits in case they have been specified by the user.
Gradients for background model parameters are computed analytically.
If energy dispersion is neglected (option {\tt edisp=no}), gradients for spectral source model parameters
are also computed analytically.
Otherwise, analytical computation is not possible, and gradients are computed numerically using a
two-point formula.
Numerical gradients are also used for spatial source model parameters.

Statistical errors on the model parameters $\delta a_k$ are determined by computing the square root of the
diagonal elements of the covariance matrix $C$ which is the inverse of the curvature matrix:
\begin{equation}
\delta a_k = \sqrt{C_{kk}}
\end{equation}
with
\begin{equation}
C = [\alpha]^{-1} .
\end{equation}
Inversion of $[\alpha]$ is again performed using a sparse matrix Cholesky decomposition.
Maximum-likelihood estimation is implemented by {\tt ctlike}.

If gamma-ray emission from a source is not detected, an upper limit for the flux can be derived by determining
the flux $F_\mathrm{up}$ that leads to a log-likelihood decrease of $\Delta \ln L$ with respect to the
maximum log-likelihood estimate $F_\mathrm{0}$:
\begin{equation}
\ln L(F_\mathrm{up}) = \ln L(F_\mathrm{0}) - \Delta \ln L .
\end{equation}
The log-likelihood decrease $\Delta \ln L$ is computed from the chance probability (p-value) using
\begin{equation}
\Delta \ln L = (\mathrm{erf}^{-1}(p))^2 .
\end{equation}
Upper limit computation is implemented by {\tt ctulimit}.

\section{On-Off spectral analysis}\label{sec:onoff_methods}
\subsection{On-Off spectra and response}

The data for the On-Off spectral analysis were prepared using the \texttt{csphagen} script which
generates the necessary files from one or several event lists
in the OGIP format\footnote{
  \url{https://heasarc.gsfc.nasa.gov/docs/heasarc/ofwg/docs/spectra/ogip_92_007/node5.html}} 
normally used in X-ray astronomy.
This format is composed of Pulse Height Analyser spectral files (PHA), an Auxiliary Response
File (ARF) and a Redistribution Matrix File (RMF).

Data were either analysed in joint or stacked mode.
In joint mode, PHA, ARF, and RMF files are generated for each observation $i$.
PHA files are generated for the On and Off regions by binning the events in both
regions for each observation $i$ as a function of reconstructed energy $k$, resulting in vectors
$n^\mathrm{on}_{k,i}$ and $n^\mathrm{off}_{k,i}$, respectively.
The ARF is given by
\begin{equation}
ARF_i(E) = \int_\mathrm{on} \int_{\vec{p}} A_{\rm eff}(\vec{p},E) \times PSF(\vec{p'} | \vec{p},E) \times M_s(\vec{p},E) \, d\vec{p} \, d\vec{p'} ,
\label{eq:arf}
\end{equation}
where
$A_{\rm eff}(\vec{p},E)$ is the effective area,
$PSF(\vec{p'} | \vec{p},E)$ is the point-spread function,
$M_s(\vec{p},E)$ is the source model,
$\vec{p}$ and $\vec{p'}$ are the true and reconstructed photon arrival directions, respectively,
and $E$ is the true energy
(see \citet{knoedlseder2016} for the definition of the terms).
The integration in $\vec{p'}$ is done over the On region, and the integration in $\vec{p}$ is done over all
$\vec{p}$ that contribute events within the On region.
In practice, the ARF is stored as a vector for a specified number of true energies.
The RMF is given by
\begin{equation}
RMF_{k,i}(E) = \frac{\int_\mathrm{on} \int_{E'_k} A_{\rm eff}(\vec{p},E) \times E_{\rm disp}(E' | \vec{p}, E) \, dE' \, d\vec{p}}
                                {\int_\mathrm{on} A_{\rm eff}(\vec{p},E)\, d\vec{p}} ,
\end{equation}
where
$E_{\rm disp}(E' | \vec{p}, E)$ is the energy dispersion.
The integration in $\vec{p}$ is done over the On region to allow for possible variations of
the energy dispersion over that region, but given the typically small size of the On region we can
approximate
\begin{equation}
RMF_{k,i}(E) \approx \int_\mathrm{on} \int_{E'_k} E_{\rm disp}(E' | \vec{p}, E) \, dE' \, d\vec{p} .
\end{equation}
The integration over reconstructed energy $E'$ is done over the width of the energy bin $k$.

{\tt csphagen} also computes the background scaling factors $\alpha_{k,i}$ that are stored in
the {\tt BACKSCAL} column of each On PHA file, and background response vectors $b_{k,i}$ that
are stored in the {\tt BACKRESP} column of each Off PHA file.
The background scaling factors are computed using
\begin{equation}
\alpha_{k,i} = \frac{\int_\mathrm{on} M_b(\vec{p'}, E') \, d\vec{p'}}{\int_\mathrm{off} M_b(\vec{p'}, E') \, d\vec{p'}} ,
\end{equation}
where $M_b(\vec{p'}, E')$ is a background acceptance model, specified either using a model
definition XML file, or the template background found in the IRF.
If no background acceptance model is provided, $M_b(\vec{p'}, E')=1$, and $\alpha_{k,i}$ gives
the solid angle ratio between On and Off regions.
The background response vectors are computed using
\begin{equation}
b_{k,i} = \int_\mathrm{off} M_b(\vec{p'}, E') \, d\vec{p'} ,
\end{equation}
where $M_b(\vec{p'}, E')$ are evaluated at the reconstructed energy bins $k$.
If no background acceptance model is provided, $M_b(\vec{p'}, E')=1$, and $b_{k,i}$
gives the solid angle of the Off region.

In stacked mode, events from all observations $i$ are combined into a single On and Off PHA
spectrum and the effective response functions are computed using
\begin{equation}
n^\mathrm{on}_{k} = \sum_i n^\mathrm{on}_{k,i}
\end{equation}
\begin{equation}
n^\mathrm{off}_{k} = \sum_i n^\mathrm{off}_{k,i}
\end{equation}
\begin{equation}
ARF(E) = \frac{\sum_i ARF_i(E) \times \tau_i}{\sum_i \tau_i}
\end{equation}
\begin{equation}
RMF_{k}(E) = \frac{\sum_i RMF_{k,i}(E) \times ARF_i(E) \times \tau_i}{\sum_i ARF_i(E) \times \tau_i}
\end{equation}
\begin{equation}
\alpha_{k} = \frac{\sum_i \alpha_{k,i} \times b_{k,i} \times \tau_i}{\sum_i b_{k,i} \times \tau_i}
\end{equation}
\begin{equation}
b_{k} = \frac{\sum_i b_{k,i} \times \tau_i}{\sum_i \tau_i} ,
\end{equation}
where
$\tau_i$ is the live time (or exposure) of observation $i$.

\subsection{Likelihood for On-Off spectral analysis}
\label{sec:wstat_cstat}

On-Off data were analysed using the maximum likelihood method (see Appendix \ref{sec:ctlike}).
For energy-binned On-Off data following the Poisson distribution the log-likelihood function $L_i(M)$
can be expressed as 
\begin{equation}
\label{eq:likeonoff}
-\ln L_i (M) = \sum_k m^\mathrm{on}_{k,i} (M) - n^\mathrm{on}_{k,i} \ln m^\mathrm{on}_{k,i} (M)
  + m^\mathrm{off}_{k,i}(M) - n^\mathrm{off}_{k,i} \ln m^\mathrm{off}_{k,i}(M) ,
\end{equation}
where the sum is performed over energy bins $k$,
$n^\mathrm{on}_{k,i}$ and $n^\mathrm{off}_{k,i}$ are the number of events observed in the On and
Off regions, for energy bin $k$ and observation $i$, respectively,
and $m^\mathrm{on}_{k,i}$ and $m^\mathrm{off}_{k,i}$ are the numbers of events expected in these
bins based on the model $M$ in the On and Off regions, respectively.

It is convenient to split the model in signal (gamma rays) and background, $M = M_s + M_b$. 
Let $\alpha_{k,i} (M_b)$ be the scale factor that transforms the number of expected background counts in 
the Off region into the number of expected background counts in the On region.
The log-likelihood in Equation~\ref{eq:likeonoff} can then be reformulated as
\begin{equation}
\label{eq:cstat}
\begin{split}
-\ln L_i (M) = \sum_k & s_{k,i} (M_s)+ \alpha_{k,i}(M_b)\, b_{k,i} (M_b) -\\
& n^\mathrm{on}_{k,i} \ln  [s_{k,i} (M_s)+ \alpha_{k,i}(M_b)\, b_{k,i} (M_b)] +\\
& b_{k,i} (M_b)  - n^\mathrm{off}_{k,i} \ln b_{k,i} (M_b) ,
\end{split}
\end{equation}
where $s_{k,i} (M_s)$ is the number of expected signal counts in the
On region, and $b_{k,i} (M_b)$ is the number of expected background
counts in the Off region of energy bin $k$ and observation $i$.
For consistency with XSPEC this statistic is called cstat.\footnote{
  See Poisson data with Poisson background at
  \url{https://heasarc.nasa.gov/xanadu/xspec/manual/XSappendixStatistics.html}.}

In practice a background model is not always available.
The likelihood can be reformulated in this case by adopting as single assumption that the background
rates per unit solid angle and background spectrum are the same in the On and Off regions (this matches
the general expectations for reflected Off regions, which is the main On-Off method implemented in ctools).
Thus, the factors  $\alpha_{k,i}$ can be calculated as the ratio of solid angle subtended by the On
region over that subtended by the Off region.\footnote{
  For a single observation the scale factor under this assumption is independent of energy.
  However, for an On-Off observation derived by stacking multiple observations with varying
  energy thresholds the scale factor will still depend on the energy bin; thus we keep the notation
  $\alpha_{k,i}$.}
Furthermore, the terms $b_{k,i}$ can be treated as nuisance parameters that is derived by solving
the equations
\begin{equation}
\frac{\partial \ln L}{\partial b_{k,i}} = 0.
\end{equation}
This yields quadratic equations for $b_{k,i}$
\begin{equation}
1 + \alpha_{k,i} - \frac{ \alpha_{k,i} \, n^\mathrm{on}_{k,i}}{s_{k,i} (M_s)+ \alpha_{k,i} b_{k,i} } - \frac{n^\mathrm{off}_{k,i}}{b_{k,i}} = 0
\end{equation}
which have the general solution
\begin{equation}\label{eq:bkgwstat}
b_{k,i} (M_s) = \frac{C_{k,i} (M_s) +D_{k,i} (M_s)}{2 \alpha_{k,i} (\alpha_{k,i} + 1)} ,
\end{equation}
where
\begin{equation}
C_{k,i} (M_s) =  \alpha_{k,i} (n^\mathrm{on}_{k,i} + n^\mathrm{off}_{k,i}) - ( \alpha_{k,i} +1) s_{k,i} (M_s) ,
\end{equation}
and 
\begin{equation}
D_{k,i} (M_s) = \sqrt{C^2_{k,i} (M_s) + 4 \alpha_{k,i} ( \alpha_{k,i} +1)\,  n^\mathrm{off}_{k,i}\, s_{k,i} (M_s)} .
\end{equation}
By replacing the values of $b_{k,i}$ obtained from Eq.~\ref{eq:bkgwstat} in Eq.~\ref{eq:cstat} one obtains a
formulation of the log-likelihood that depends only on $M_s$:
\begin{equation}
\label{eq:wstat}
\begin{split}
-\ln L_i(M_s) = \sum_k & s_{k,i} (M_s)+ \alpha_{k,i} b_{k,i} (M_s) -\\
& n^\mathrm{on}_{k,i} \ln  [s_{k,i} (M_s)+ \alpha_{k,i}\, b_{k,i} (M_s)] +\\
& b_{k,i} (M_s)  - n^\mathrm{off}_{k,i} \ln b_{k,i} (M_s) - \\
& n^\mathrm{on}_{k,i} (1-\ln n^\mathrm{on}_{k,i}) - n^\mathrm{off}_{k,i} (1-\ln n^\mathrm{off}_{k,i}) .
\end{split}
\end{equation}
The terms in the last row do not depend on the model $M_s$, and are added for consistency with current
practice so that, in the limit of large number of counts, $2 \ln L_i (M_s)$ is asymptotically distributed as a
$\chi_p^2$ distribution, where $p$ is the number of degrees freedom, equivalent to the difference between
number of energy bins and number of free model parameters.
We call this statistic wstat (in XSPEC this statistic is automatically used in lieu of cstat when no background
model is specified).
We stress that, although not explicitly noted, this formulation of the likelihood relies on the assumption that 
the background rates per solid angle unit and the background spectrum are the same in the On and Off regions.

Some special cases need to be handled separately in wstat. 
If $n^\mathrm{on}_{k,i} = 0$ but $n^\mathrm{off}_{k,i} > 0$, then one finds
\begin{equation}
b_{k,i} = \frac{ n^\mathrm{off}_{k,i}}{\alpha_{k,i} +1}
\end{equation}
and the contribution to the log-likelihood from the energy bin $k$ is
\begin{equation}
-\ln L_{k,i}(M_s) = s_{k,i} (M_s) + n^\mathrm{off}_{k,i} \ln(\alpha_{k,i}+1) .
\end{equation}
If $n^\mathrm{off}_{k,i} = 0$, then one finds
\begin{equation}
b_{k,i}(M_s) = \frac{ \alpha_{k,i} +1}{\alpha_{k,i}} n^\mathrm{on}_{k,i} - s_{k,i} (M_s) .
\end{equation}
For
\begin{equation}
n^\mathrm{on}_{k,i} > s_{k,i} (M_s) \frac{\alpha_{k,i} + 1}{\alpha_{k,i}}
\end{equation}
$b_{k,i}$ is positive and the contribution to the log-likelihood from the energy bin $k$ is
\begin{equation}
-\ln L_{k,i}(M_s) = -\frac{s_{k,i} (M_s)}{\alpha_{k,i}} - n^\mathrm{on}_{k,i} \ln\left(\frac{\alpha_{k,i}}{\alpha_{k,i}+1}\right) .
\end{equation}
However, for smaller $n^\mathrm{on}_{k,i}$, the value of $b_{k,i} (M_s)$ is null or negative.
Since a negative number of background counts is unphysical, the number of background counts is forced 
to be zero.
This yields the following expression for the log-likelihood in the energy bin $k$:
\begin{equation}
-\ln L_{k,i}(M_s) = s_{k,i} (M_s) + n^\mathrm{on}_{k,i} \left( \ln n^\mathrm{on}_{k,i} - \ln s_{k,i} (M_s) - 1 \right) ,
\end{equation}
or, if also $n^\mathrm{on}_{k,i} = 0$,
\begin{equation}
-\ln L_{k,i}(M_s) = s_{k,i} (M_s) .
\end{equation}
Forcing the number of expected background counts to zero biases the likelihood estimator. Therefore, wstat is 
known to be inaccurate if there are energy bins with zero Off counts.

In ctools there is also the possibility to use wstat in {\tt ctlike} with the $\alpha_{k,i}(M_b)$ coefficients based on 
a background model ({\tt use\_model\_bkg = yes} in {\tt csphagen}).
This can be useful if the Off region is chosen such that the background rates per solid angle unit are 
not expected to be the same as in the On region and there is a background model that is deemed to be 
sufficiently accurate in the spatial component, but there is not an acceptable background spectral model.

\section{Generation of ring-background sky map}
\label{sec:ring-skymap}

For the ring-background method, the number of excess counts $r_j$ in an On region centred on
sky map pixel $j$ is computed using
\begin{equation}
r_j = n^{\rm on}_j - \alpha_j n^{\rm off}_j ,
\end{equation}
where
\begin{equation}
n^{\rm on}_j = \sum_{k \in \{{\rm on}\}} n_k
\end{equation}
is the number of events in the On region, defined as all sky map pixels whose centres are within
the radius {\tt roiradius} around pixel $j$,
\begin{equation}
n^{\rm off}_j = \sum_{k \in \{{\rm off}\} \, \land \, k \notin \{X_l\}} n_k
\label{eq:ring_noff}
\end{equation}
is the number of events in the Off region, defined as all sky map pixels whose centres are within
a ring with inner radius {\tt inradius} and outer radius {\tt outradius} around pixel $j$ and that do not
fall within an exclusion region $\{X_l\}$,
and
\begin{equation}
\alpha_j = \frac{\sum_{k \in \{{\rm on}\}} B_k}{\sum_{k \in \{{\rm off}\} \, \land \, k \notin \{X_l\}} B_k}
\label{eq:ring_alpha}
\end{equation}
is the ratio between the background acceptance in the On and Off regions, where $B_k$ is the
background acceptance for sky map pixel $k$.
If a template background model is available in the IRF files, this template background is used
to compute $B_k$ as the integral over the energy range of the sky map,
specified by the {\tt emin} and {\tt emax} parameters.
Otherwise, a constant background rate $B_k=1$ is assumed for each sky map pixel,
which is the case for the sky map shown in the right panel of Fig.~\ref{fig:rx_skymap}.

The significance $\sigma_j$ of the excess emission in an On region centred on sky map pixel $j$ is
computed using Eq.~17 of \citet{li1983}:
\begin{equation}
\begin{split}
\sigma_j = \sgn(r_j) \Biggl\{
& 2 n^{\rm on}_j \ln \Biggl( \frac{1+\alpha_j}{\alpha_j} \frac{n^{\rm on}_j}{n^{\rm on}_j + n^{\rm off}_j} \Biggr) + \\
& 2 n^{\rm off}_j \ln \Biggl( (1+\alpha_j) \frac{n^{\rm off}_j}{n^{\rm on}_j + n^{\rm off}_j} \Biggr) \Biggr\}^{\!1/2}
\end{split}
\end{equation}
which is simplified to
\begin{equation}
\sigma_j = - \sqrt{2 n^{\rm off}_j \ln (1+\alpha_j)}
\end{equation}
for the case $n^{\rm on}_j=0$ and
\begin{equation}
\sigma_j = \sqrt{2 n^{\rm on}_j \ln \frac{1+\alpha_j}{\alpha_j}}
\end{equation}
for the case $n^{\rm off}_j=0$ (as mentioned already earlier, however, the significances become
inaccurate in the low-count regime, hence in particular for $n^{\rm on}_j=0$ and $n^{\rm off}_j=0$).

The exclusion region $\{X_l\}$ that appears in Eqs.~(\ref{eq:ring_noff}) and (\ref{eq:ring_alpha}) can either be
specified using a ds9 region file or an exclusion map via the {\tt inexclusion} parameter, or it may be computed
internally by iteratively adding significant sky map pixels to the exclusion region and repeating the computations.
This is achieved by setting the {\tt iterations} parameter to a positive value (typically, 3 iterations are sufficient),
and by specifying a significance {\tt threshold} parameter above which pixels are added to the exclusion
region.

\section{Residual plots}
\label{sec:residuals}

\begin{figure*}
\centering
\includegraphics[width=\textwidth]{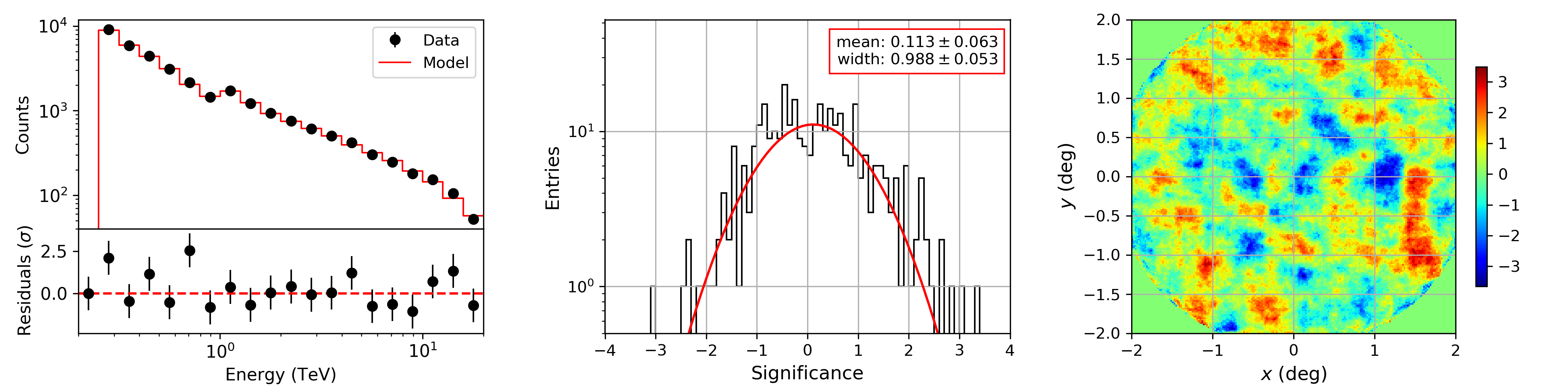}
\caption{
Counts spectra and residuals, significance histogram and residual map for set 1 of the stacked empty-field
observations.
\label{fig:off_stacked_residuals_set1}
}
\end{figure*}

\begin{figure*}
\centering
\includegraphics[width=\textwidth]{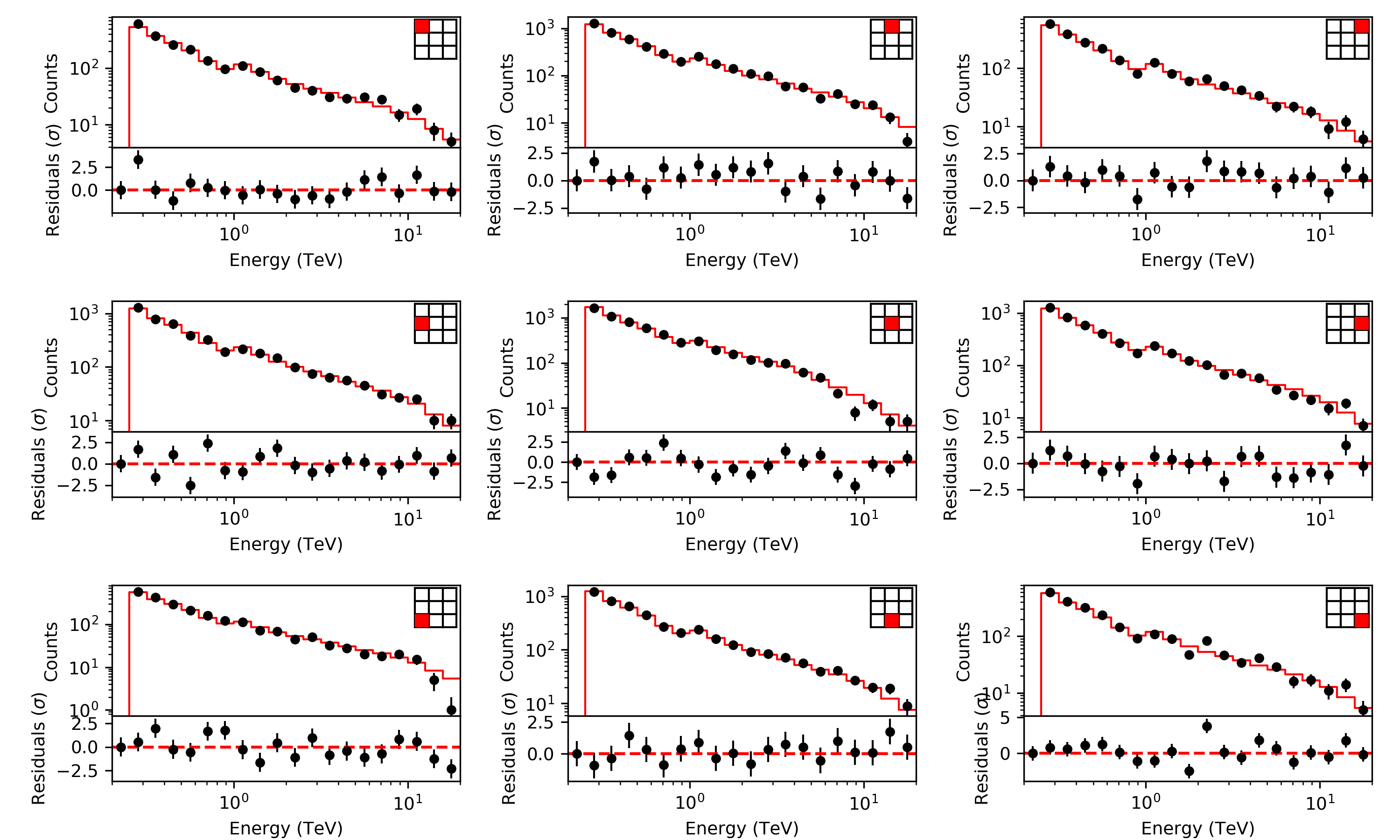}
\caption{
Sub-region counts spectra and residuals for set 1 of the stacked empty-field observations.
\label{fig:off_stacked_residuals_sectors_set1}
}
\end{figure*}

\begin{figure*}
\centering
\includegraphics[width=\textwidth]{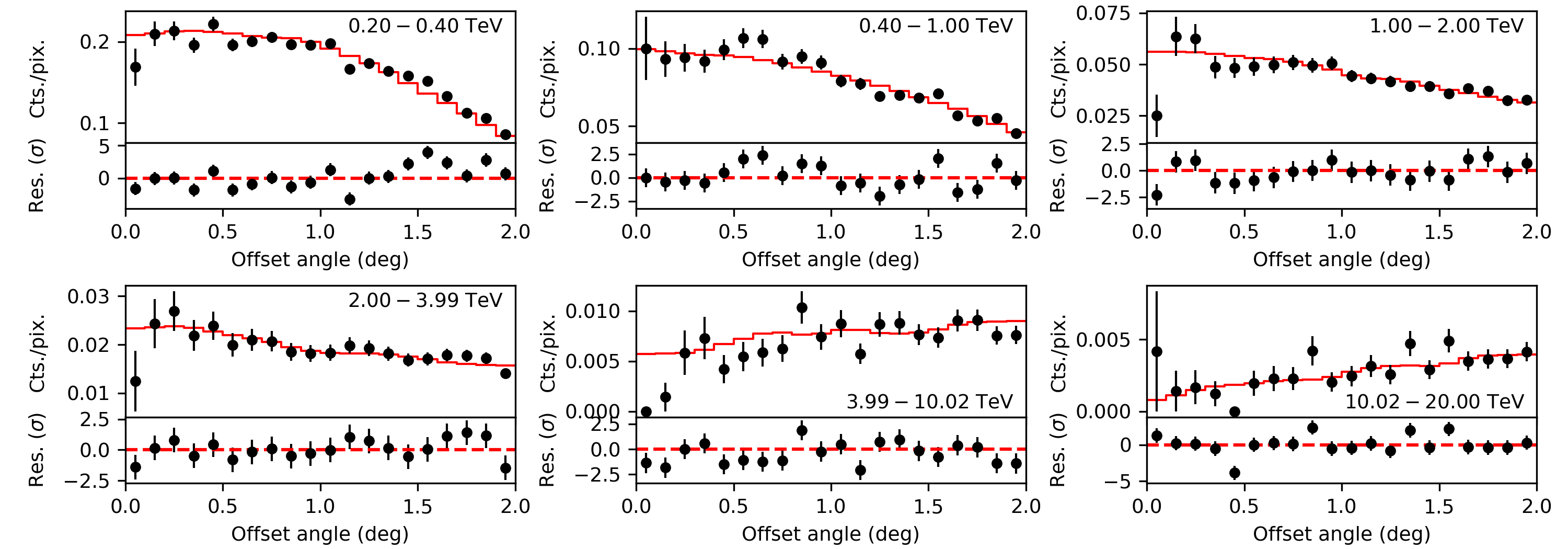}
\caption{
Radial counts profiles and residuals as a function of offset angle $\theta$ for set 1 of the stacked empty-field
observations.
\label{fig:off_stacked_residuals_profiles_set1}
}
\end{figure*}

\begin{figure*}
\centering
\includegraphics[width=\textwidth]{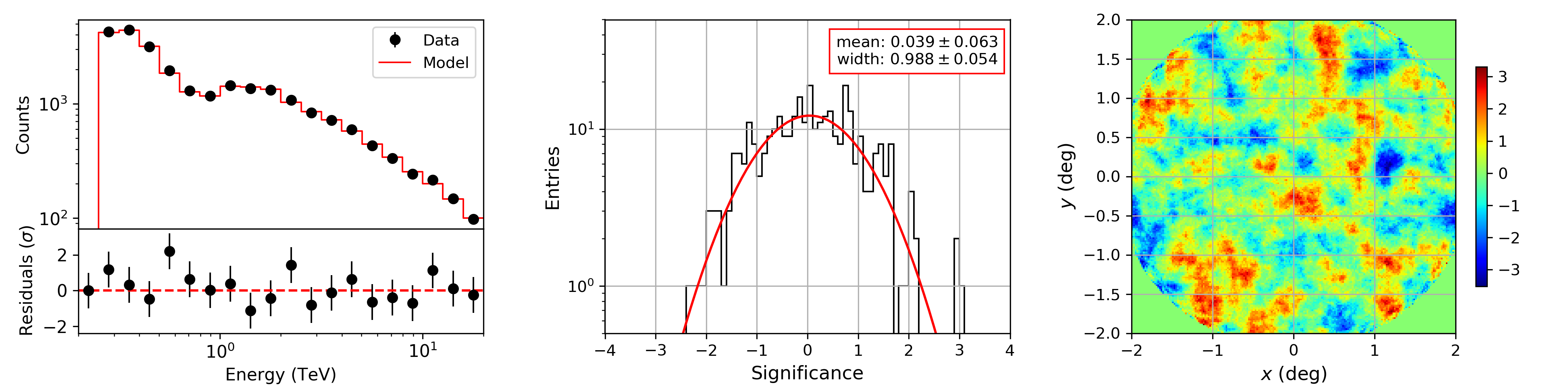}
\caption{
Counts spectra and residuals, significance histogram and residual map for set 2 of the stacked empty-field
observations.
\label{fig:off_stacked_residuals_set2}
}
\end{figure*}

\begin{figure*}
\centering
\includegraphics[width=\textwidth]{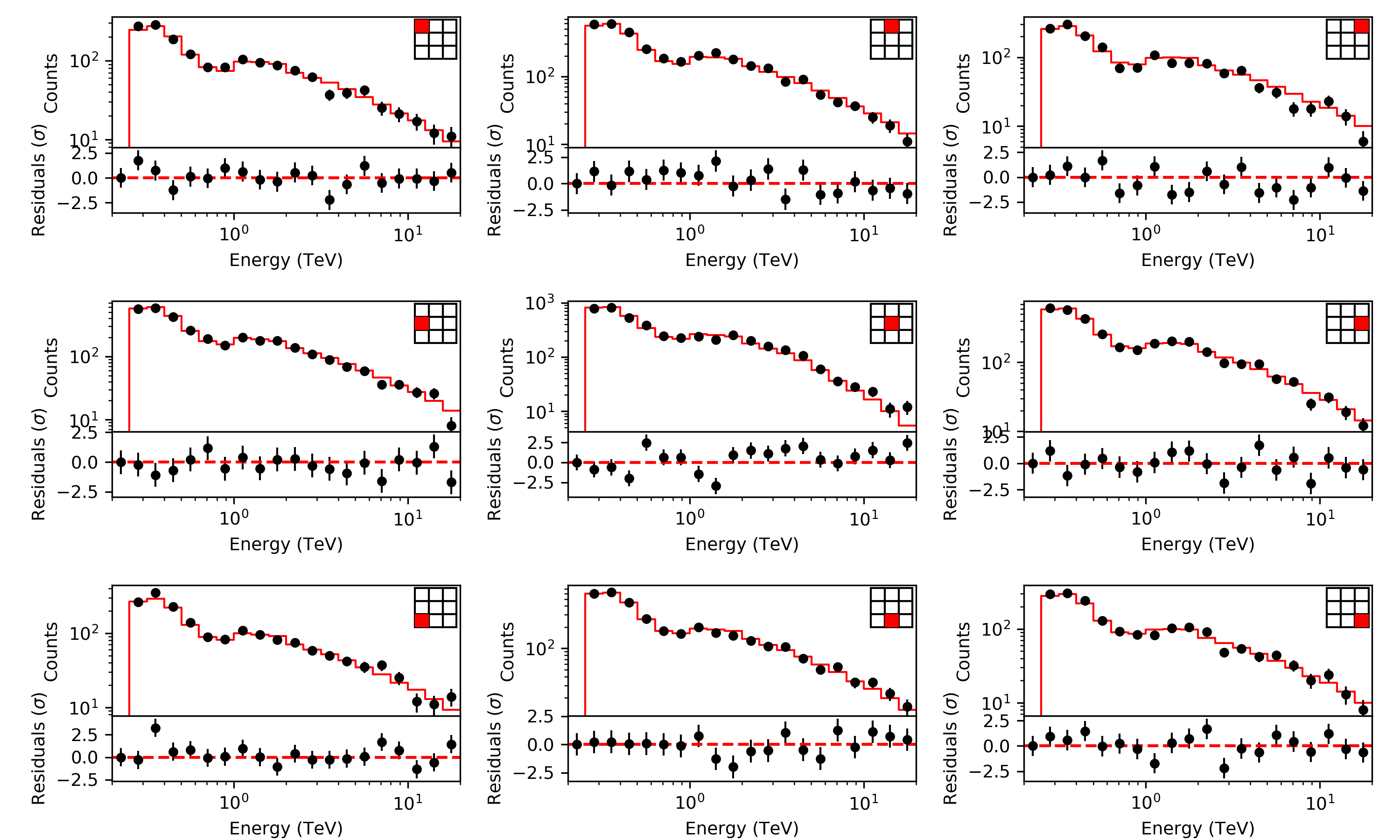}
\caption{
Sub-region counts spectra and residuals for set 2 of the stacked empty-field observations.
\label{fig:off_stacked_residuals_sectors_set2}
}
\end{figure*}

\begin{figure*}
\centering
\includegraphics[width=\textwidth]{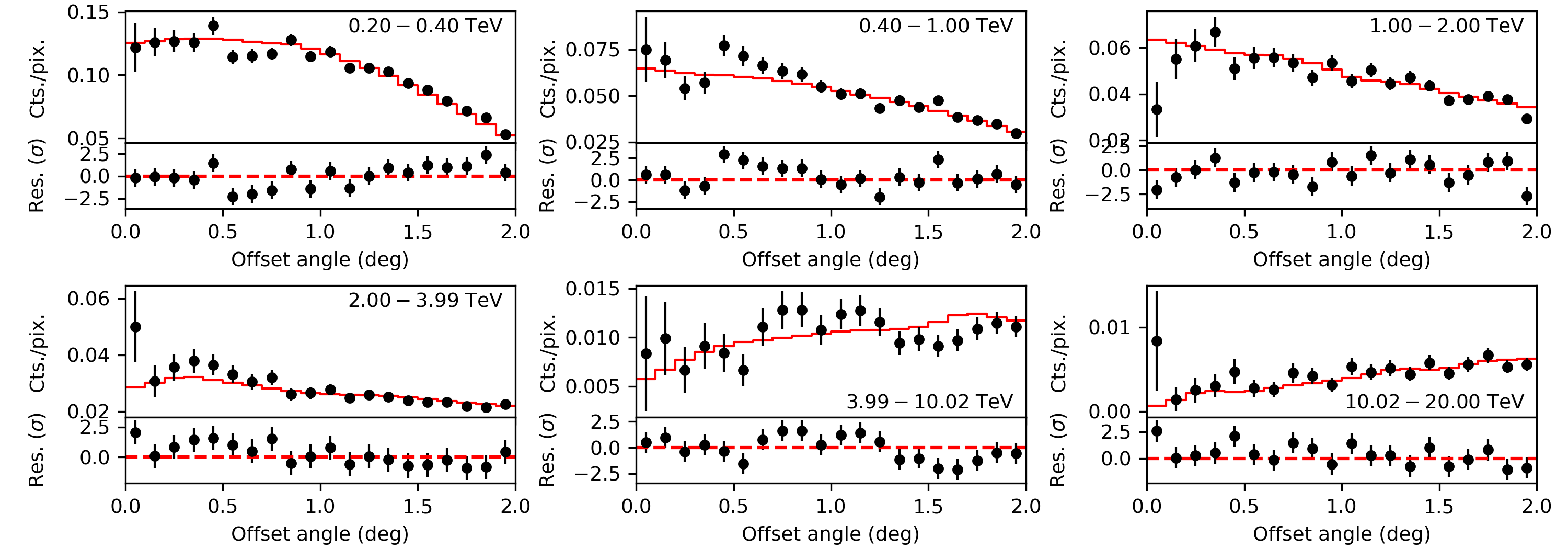}
\caption{
Radial counts profiles and residuals as a function of offset angle $\theta$ for set 2 of the stacked empty-field
observations.
\label{fig:off_stacked_residuals_profiles_set2}
}
\end{figure*}

\begin{figure*}
\centering
\includegraphics[width=\textwidth]{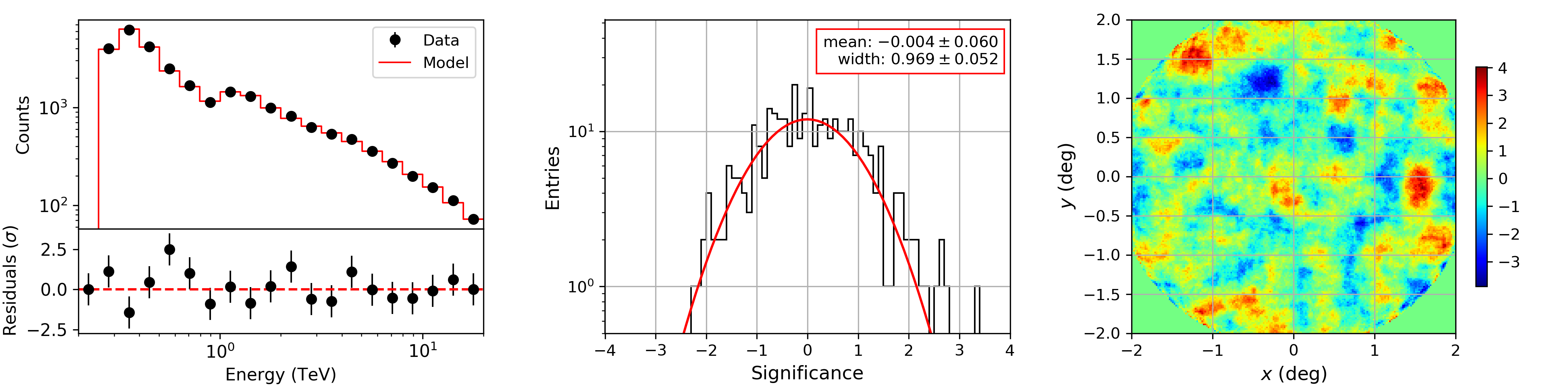}
\caption{
Counts spectra and residuals, significance histogram and residual map for set 3 of the stacked empty-field
observations.
\label{fig:off_stacked_residuals_set3}
}
\end{figure*}

\begin{figure*}
\centering
\includegraphics[width=\textwidth]{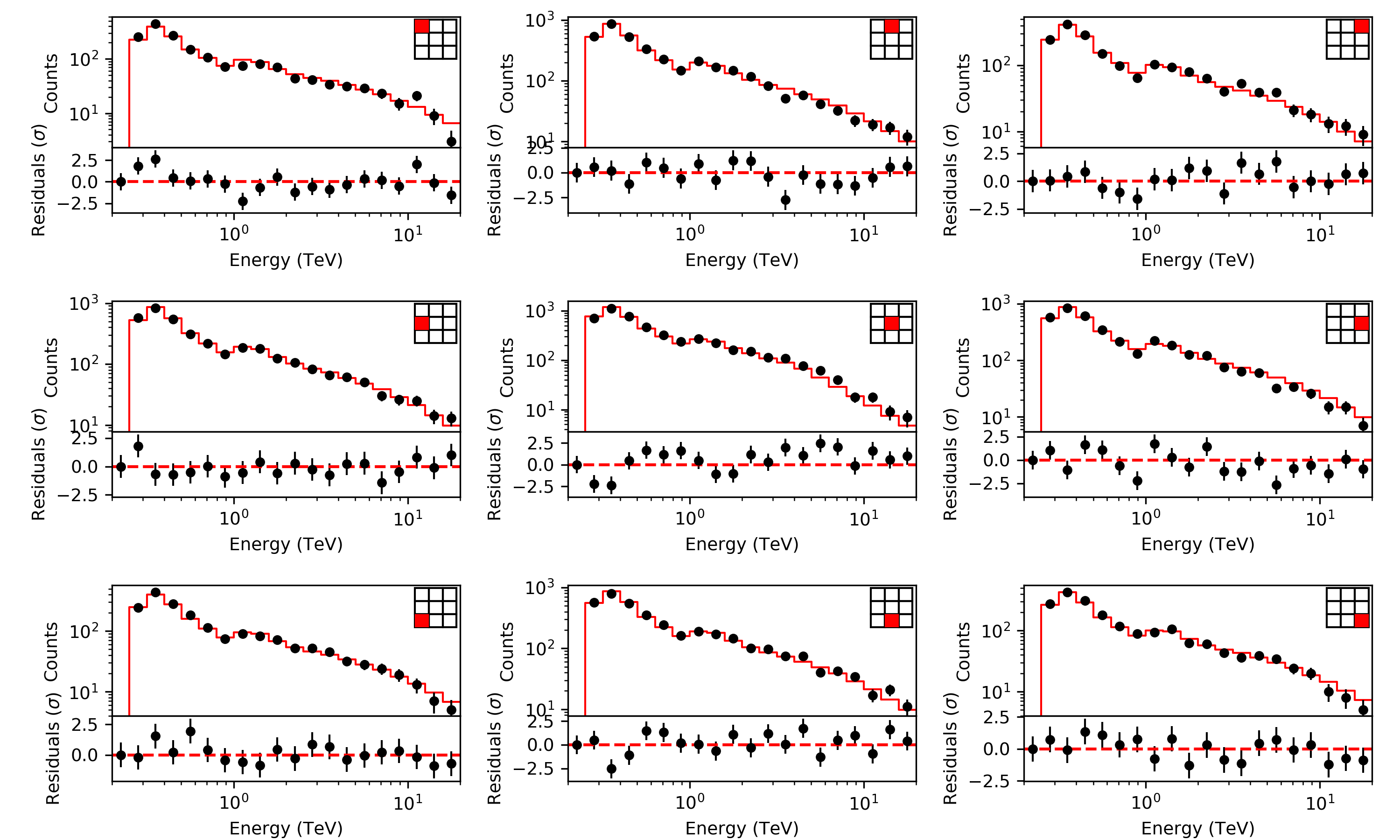}
\caption{
Sub-region counts spectra and residuals for set 3 of the stacked empty-field observations.
\label{fig:off_stacked_residuals_sectors_set3}
}
\end{figure*}

\begin{figure*}
\centering
\includegraphics[width=\textwidth]{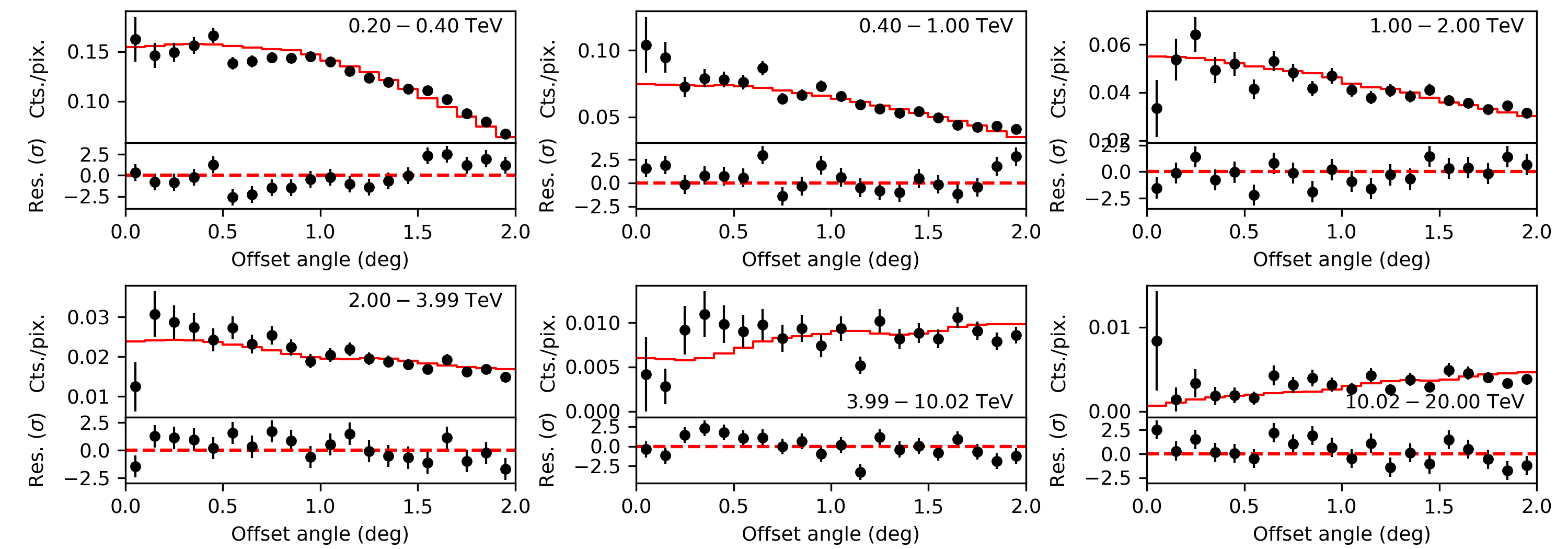}
\caption{
Radial counts profiles and residuals as a function of offset angle $\theta$ for set 3 of the stacked empty-field
observations.
\label{fig:off_stacked_residuals_profiles_set3}
}
\end{figure*}

\begin{figure*}
\centering
\includegraphics[width=\textwidth]{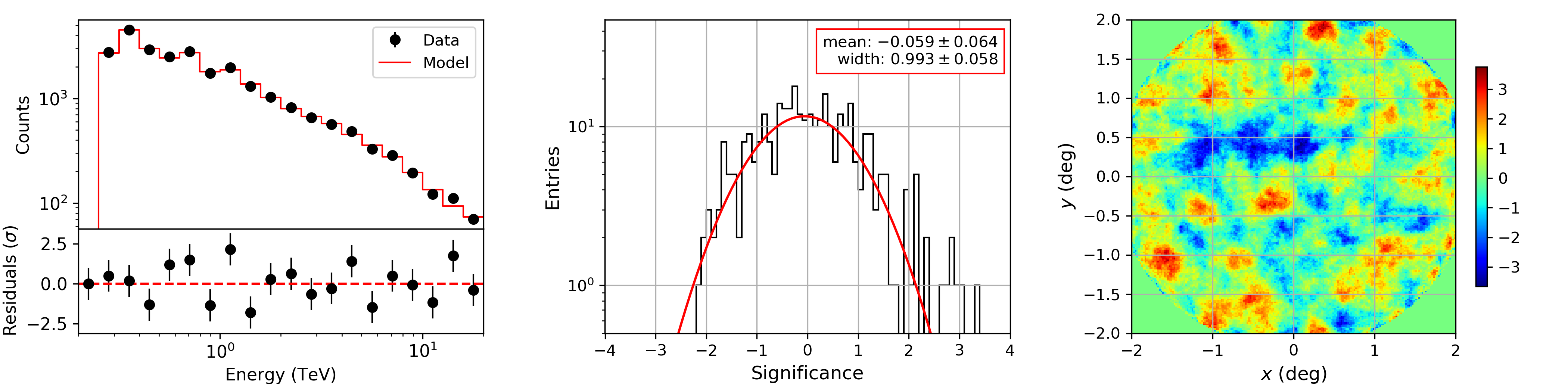}
\caption{
Counts spectra and residuals, significance histogram and residual map for set 4 of the stacked empty-field
observations.
\label{fig:off_stacked_residuals_set4}
}
\end{figure*}

\begin{figure*}
\centering
\includegraphics[width=\textwidth]{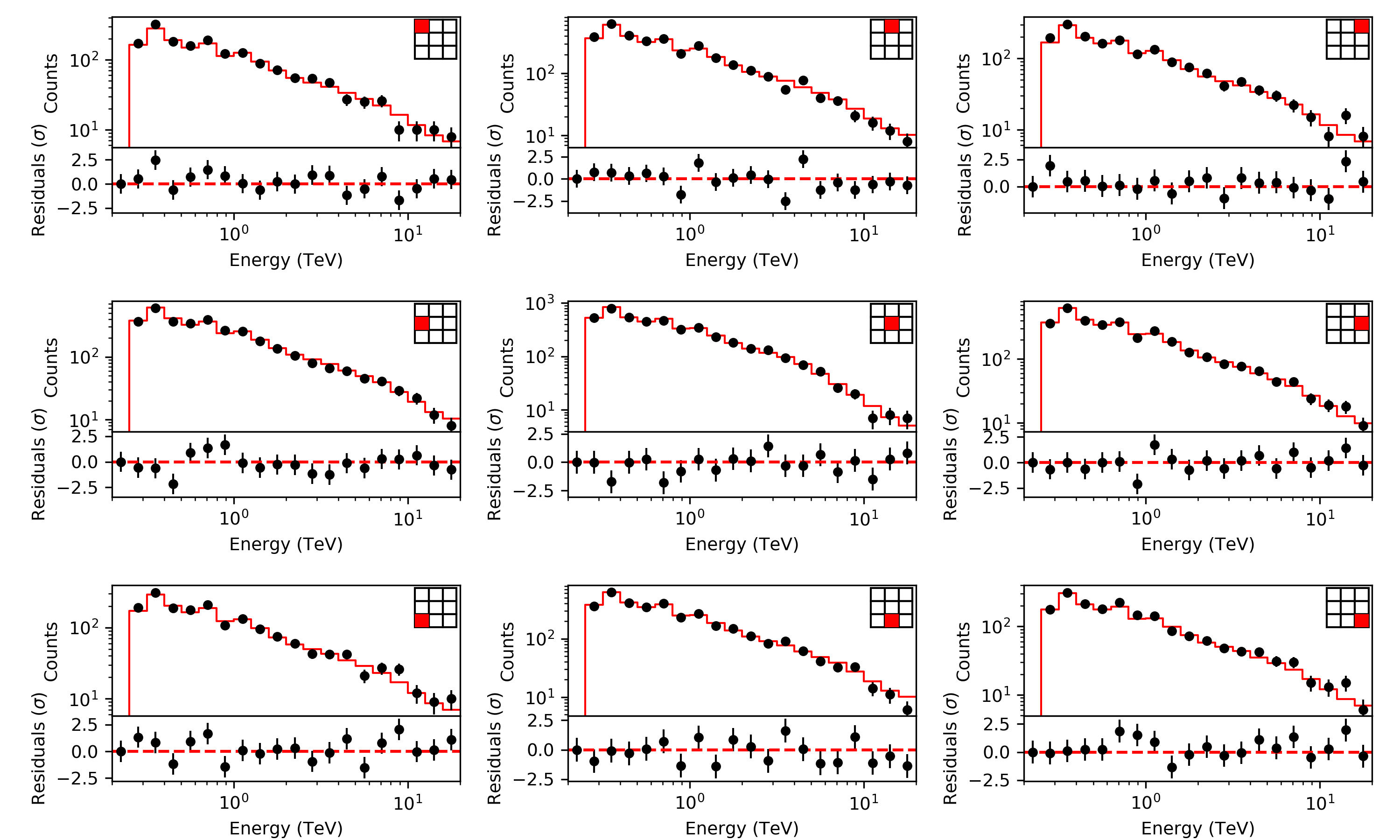}
\caption{
Sub-region counts spectra and residuals for set 4 of the stacked empty-field observations.
\label{fig:off_stacked_residuals_sectors_set4}
}
\end{figure*}

\begin{figure*}
\centering
\includegraphics[width=\textwidth]{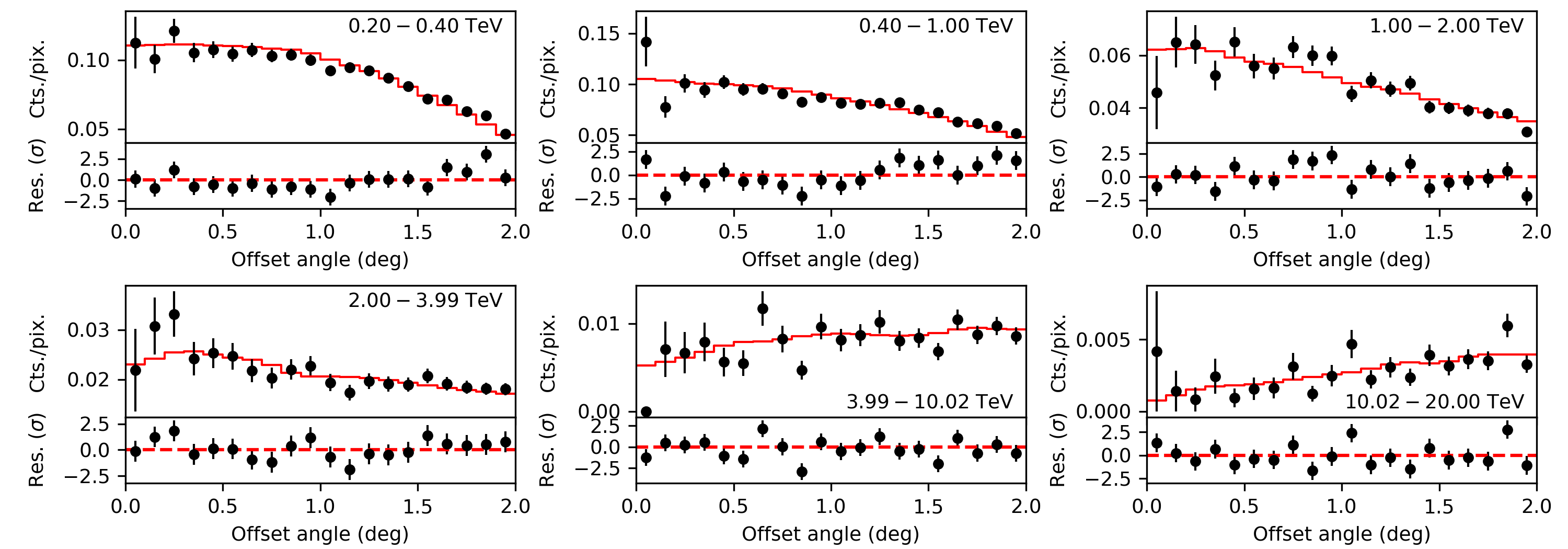}
\caption{
Radial counts profiles and residuals as a function of offset angle $\theta$ for set 4 of the stacked empty-field
observations.
\label{fig:off_stacked_residuals_profiles_set4}
}
\end{figure*}

\begin{figure*}
\centering
\includegraphics[width=\textwidth]{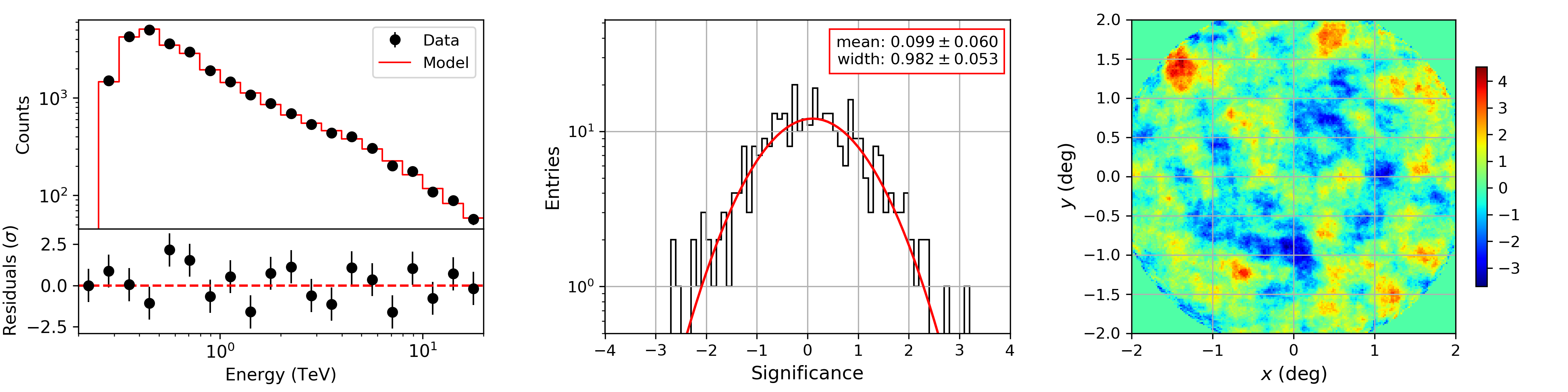}
\caption{
Counts spectra and residuals, significance histogram and residual map for set 5 of the stacked empty-field
observations.
\label{fig:off_stacked_residuals_set5}
}
\end{figure*}

\begin{figure*}
\centering
\includegraphics[width=\textwidth]{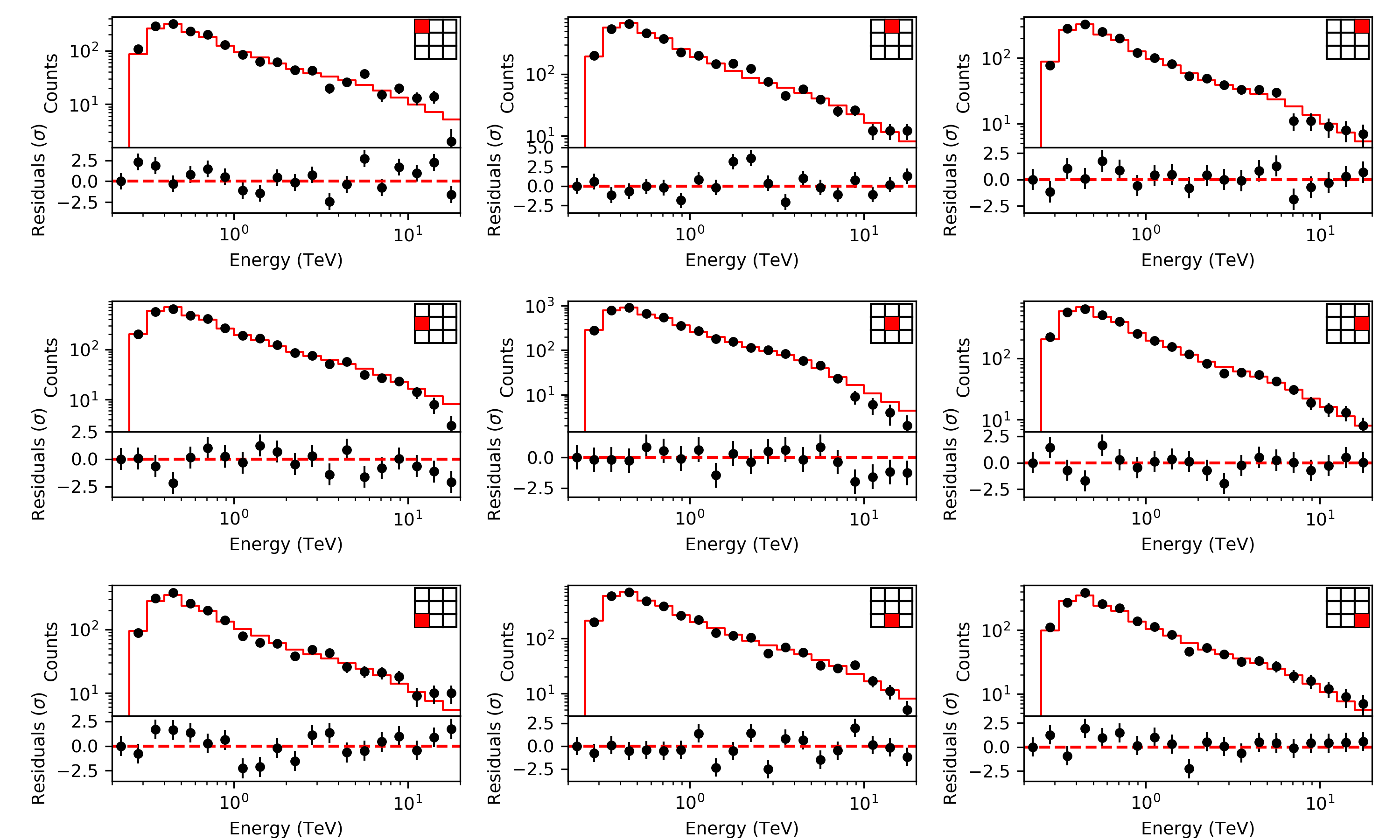}
\caption{
Sub-region counts spectra and residuals for set 5 of the stacked empty-field observations.
\label{fig:off_stacked_residuals_sectors_set5}
}
\end{figure*}

\begin{figure*}
\centering
\includegraphics[width=\textwidth]{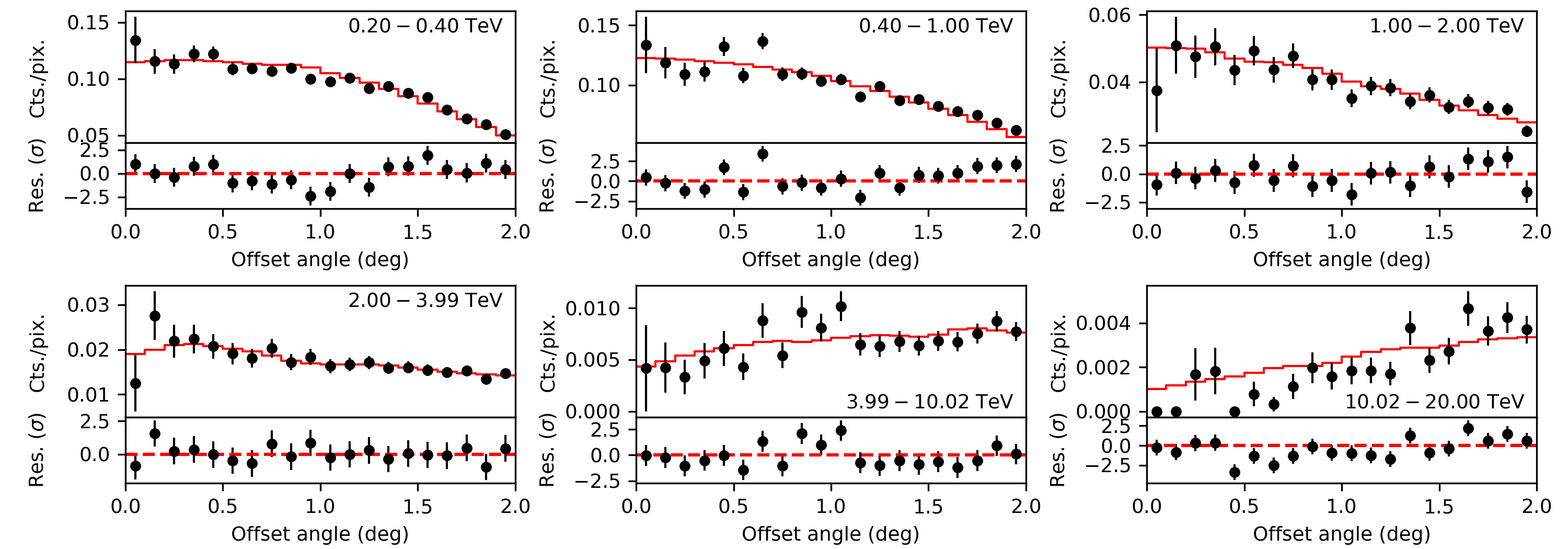}
\caption{
Radial counts profiles and residuals as a function of offset angle $\theta$ for set 5 of the stacked empty-field
observations.
\label{fig:off_stacked_residuals_profiles_set5}
}
\end{figure*}

This section provides residuals for five sets of ten randomly selected empty-field observations
stacked together in the field-of-view coordinate system.
The observation identifiers for the five sets are given in Table \ref{tab:off_sets}.
We generated a background model for each of the ten observations using a lookup table that
was based on the remaining 35 empty-field observations, and hence the selected empty-field
observations are statistically independent of the background lookup table.
For each set, the first figure shows the field-of-view integrated spectral residuals and the
energy-integrated spatial residuals, in the form of a significance histogram and a significance
map.
The second figure shows the spectral residuals for a grid of $3\times3$ sub-regions, and the
third figure shows the radial residual profiles for six energy bands.
Figures \ref{fig:off_stacked_residuals_set1} -- \ref{fig:off_stacked_residuals_profiles_set1}
are for set 1,
Figs.~\ref{fig:off_stacked_residuals_set2} -- \ref{fig:off_stacked_residuals_profiles_set2} for
set 2,
Figs.~\ref{fig:off_stacked_residuals_set3} -- \ref{fig:off_stacked_residuals_profiles_set3} for
set 3,
Figs.~\ref{fig:off_stacked_residuals_set4} -- \ref{fig:off_stacked_residuals_profiles_set4} for
set 4,
and Figs.~\ref{fig:off_stacked_residuals_set5} -- \ref{fig:off_stacked_residuals_profiles_set5} for
set 5.

\begin{table}[!h]
\caption{
Observation identifiers for the five sets of ten randomly selected empty-field observations.
\label{tab:off_sets}}
\centering
\begin{tabular}{c c c c c}
\hline\hline
Set 1 & Set 2 & Set 3 & Set 4 & Set 5 \\
\hline
{\tt 020339} & {\tt 021824} & {\tt 020561} & {\tt 021753} & {\tt 020275} \\
{\tt 021824} & {\tt 022022} & {\tt 021753} & {\tt 021807} & {\tt 021824} \\
{\tt 023246} & {\tt 023736} & {\tt 021824} & {\tt 021851} & {\tt 022997} \\
{\tt 023635} & {\tt 025443} & {\tt 026791} & {\tt 023246} & {\tt 023040} \\
{\tt 026850} & {\tt 026850} & {\tt 026850} & {\tt 023736} & {\tt 023651} \\
{\tt 027939} & {\tt 027044} & {\tt 026964} & {\tt 025345} & {\tt 025443} \\
{\tt 027987} & {\tt 027987} & {\tt 027121} & {\tt 026077} & {\tt 026077} \\
{\tt 029177} & {\tt 029024} & {\tt 029118} & {\tt 026850} & {\tt 027044} \\
{\tt 029487} & {\tt 029433} & {\tt 029177} & {\tt 028341} & {\tt 029072} \\
{\tt 029683} & {\tt 029526} & {\tt 029683} & {\tt 028967} & {\tt 029118} \\
\hline
\end{tabular}
\end{table}

\begin{figure*}
\centering
\includegraphics[width=\textwidth]{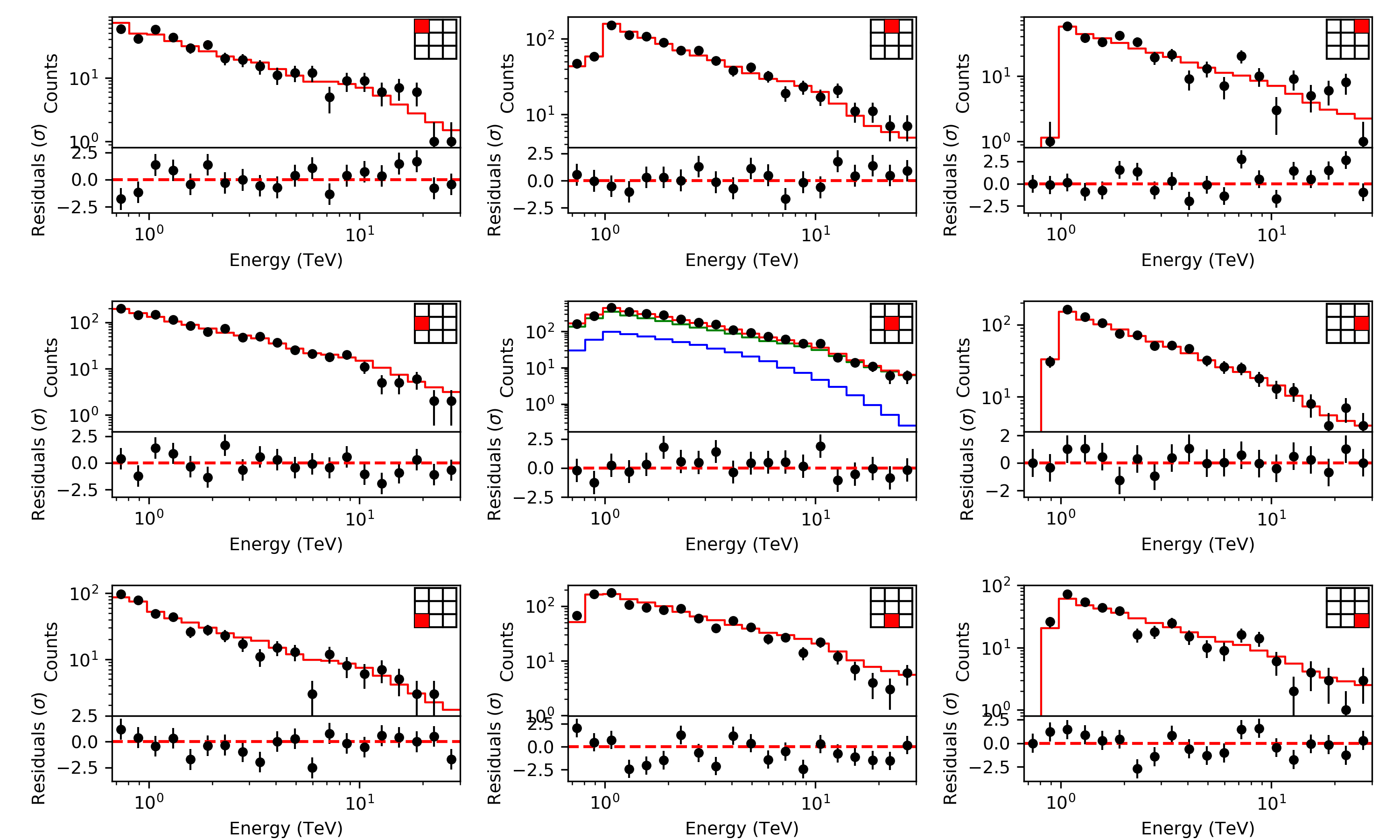}
\caption{
Sub-region counts spectra and residuals for the Crab observations.
Red lines represent the total predicted model counts, blue lines the predicted source counts and
green lines the predicted background counts.
\label{fig:crab_residuals_sectors}
}
\end{figure*}

\begin{figure*}
\centering
\includegraphics[width=\textwidth]{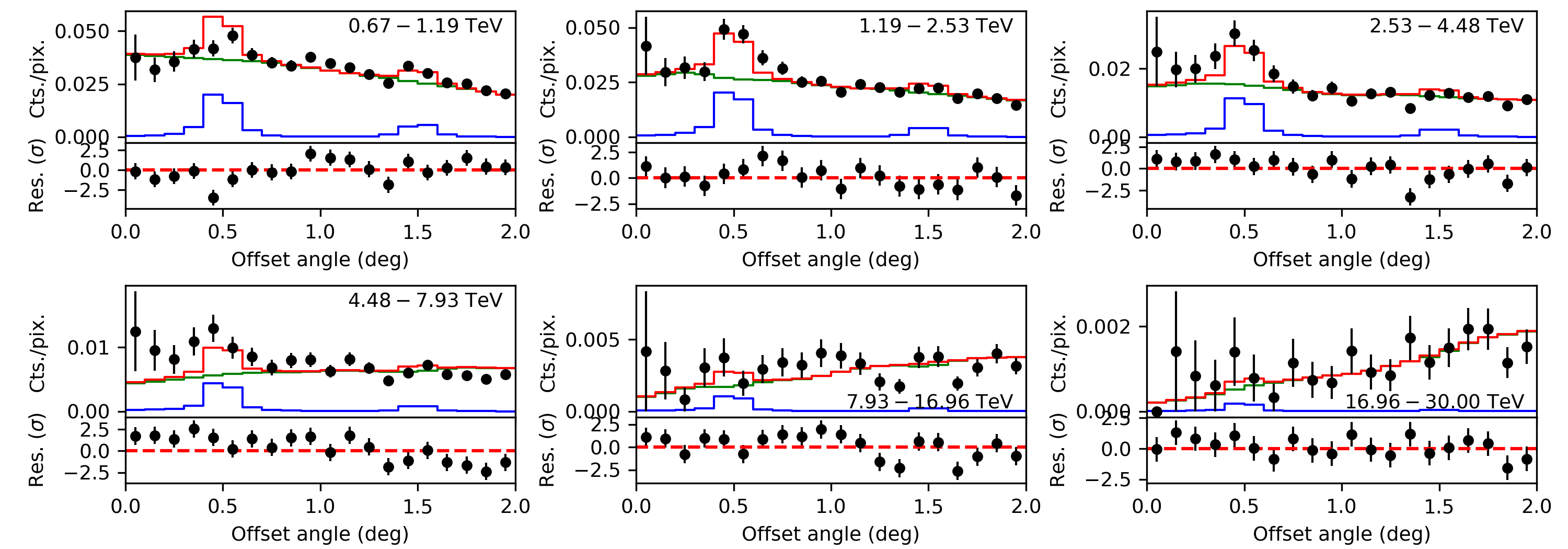}
\caption{
Radial counts profiles and residuals as a function of offset angle $\theta$ for the Crab observations.
Red lines represent the total predicted model counts, blue lines the predicted source counts and green
lines the predicted background counts.
To determine the profiles, the Crab observations were stacked in the field-of-view coordinate system;
the source counts are located around the offset angles $\theta=0.5\degree$ and
$1.5\degree$ under which the source was observed.
\label{fig:crab_residuals_profiles}
}
\end{figure*}

\begin{figure*}
\centering
\includegraphics[width=\textwidth]{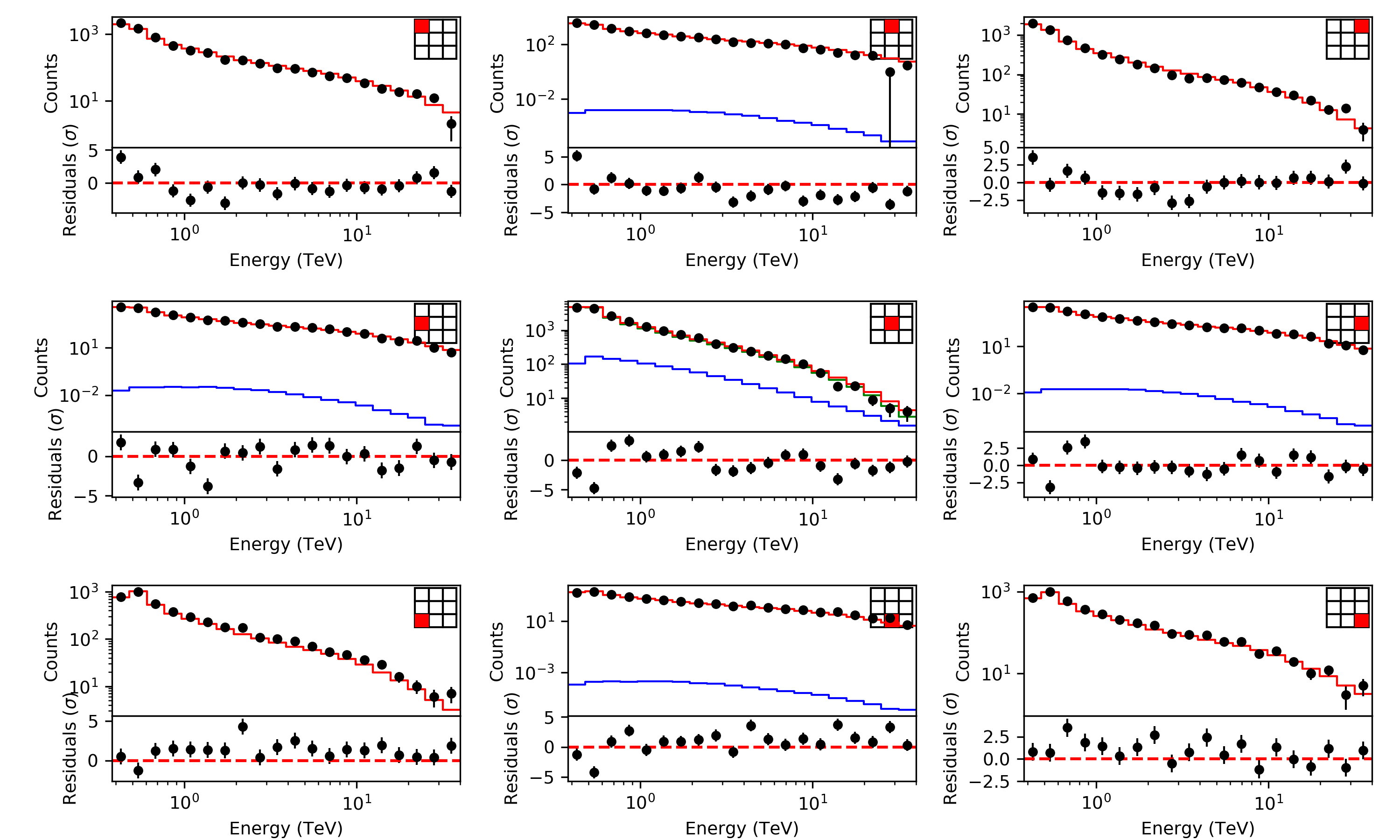} 
\caption{
Sub-region counts spectra and residuals spectra for the MSH 15--52 observations fitted using
a source model with a power-law spectrum.
Red lines represent the total predicted model counts, blue lines the predicted source counts and green
lines the predicted background counts.
\label{fig:msh_residuals_sectors}
}
\end{figure*}

\begin{figure*}
\centering
\includegraphics[width=\textwidth]{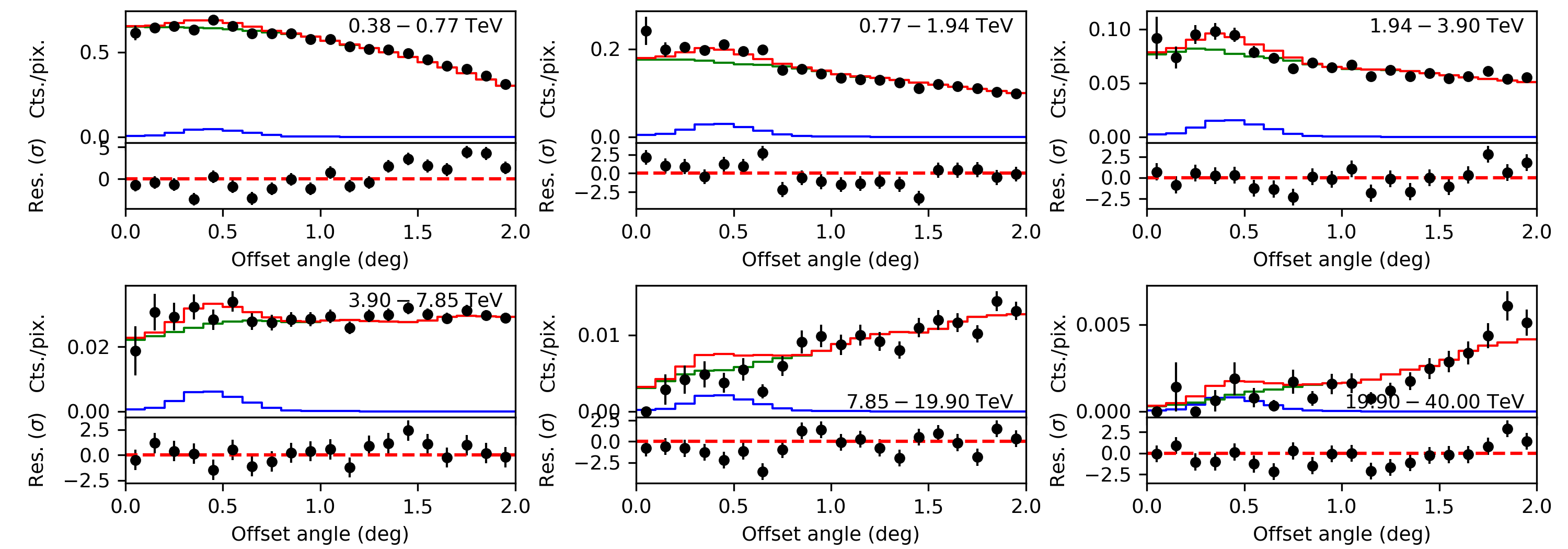} 
\caption{
Radial counts profiles and residuals for the MSH 15--52 observations fitted using
a source model with a power-law spectrum.
Red lines represent the total predicted model counts, blue lines the predicted source counts and green
lines the predicted background counts.
To determine the profiles, the MSH 15--52 observations were stacked in the field-of-view coordinate
system; the source counts are located around the offset angle $\theta=0.5\degree$ under
which the source was observed.
\label{fig:msh_residuals_profiles}
}
\end{figure*}

\begin{figure*}
\centering
\includegraphics[width=\textwidth]{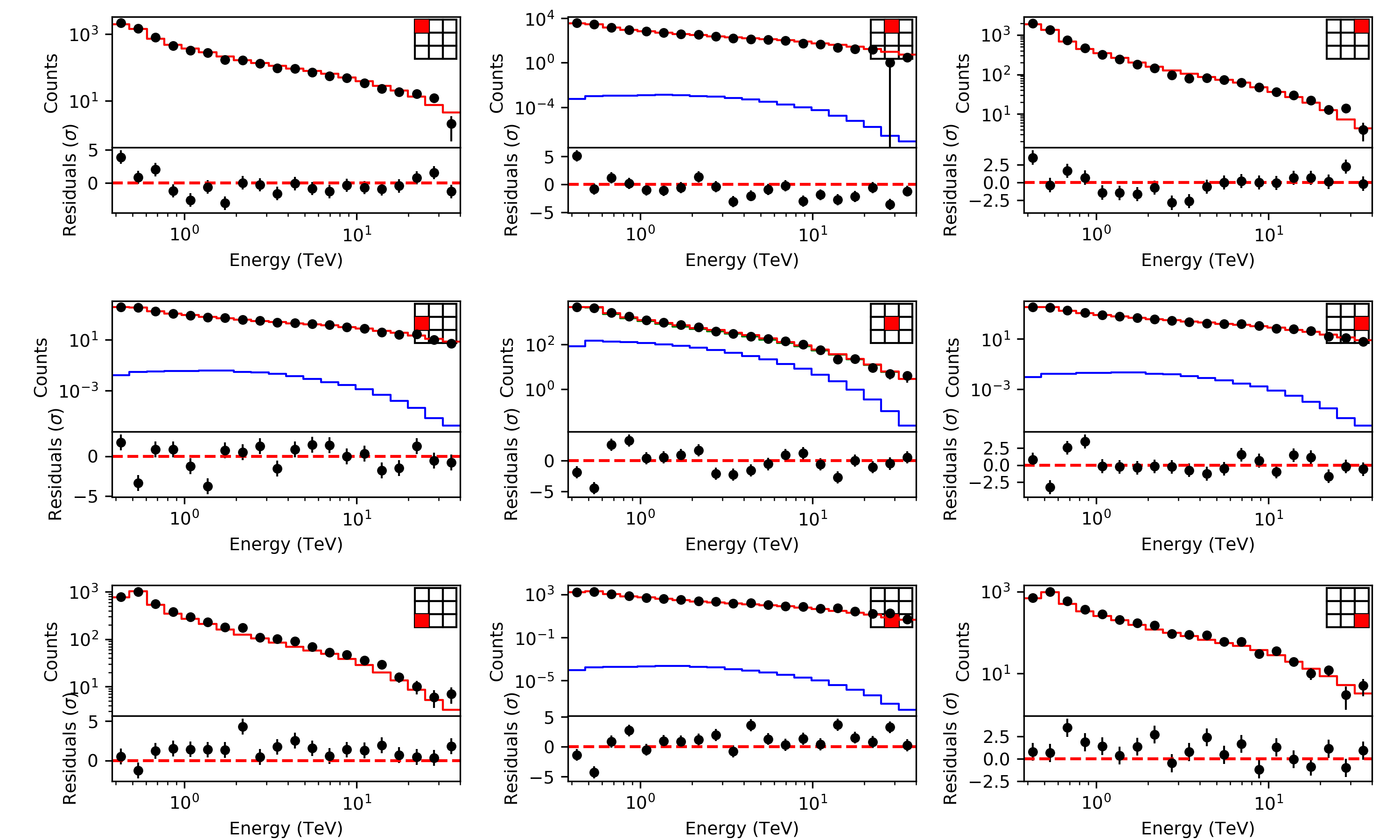} 
\caption{
Sub-region counts spectra and residuals spectra for the MSH 15--52 observations fitted using
a source model with an exponentially cut-off power-law spectrum.
Red lines represent the total predicted model counts, blue lines the predicted source counts and
green lines the predicted background counts.
\label{fig:msh_residuals_sectors_eplaw}
}
\end{figure*}

\begin{figure*}
\centering
\includegraphics[width=\textwidth]{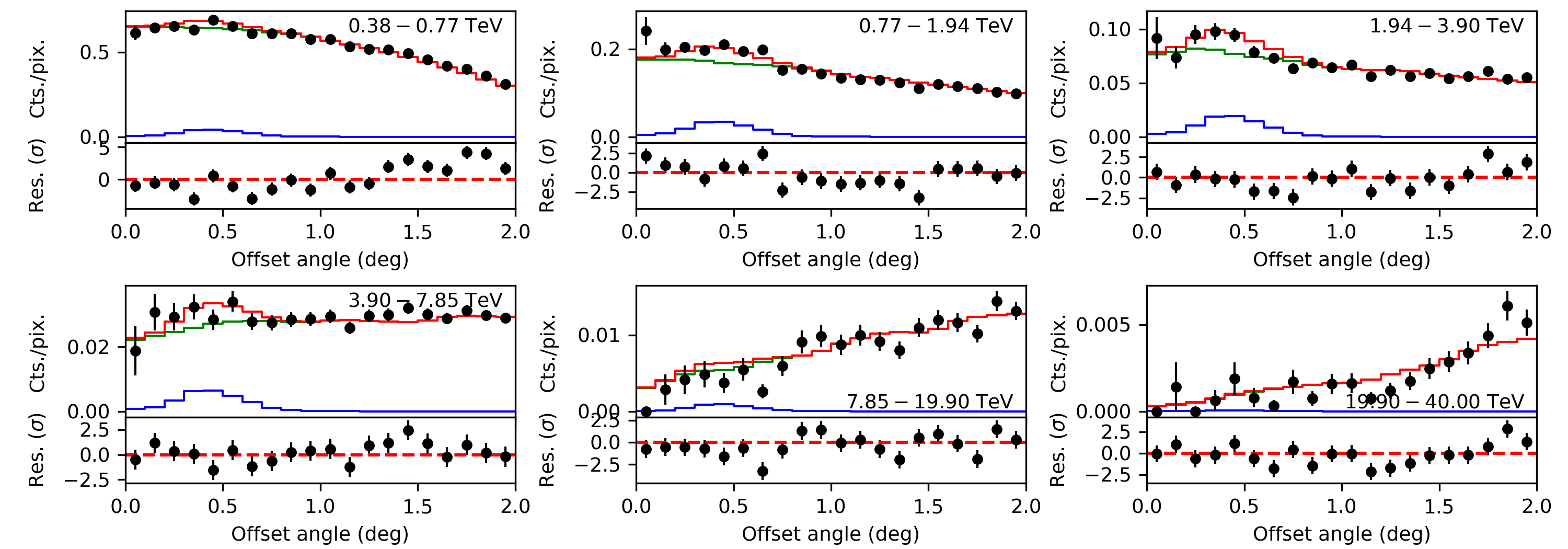} 
\caption{
Radial counts profiles and residuals for the MSH 15--52 observations fitted using
a source model with an exponentially cut-off power-law spectrum.
Red lines represent the total predicted model counts, blue lines the predicted source counts and green 
lines the predicted background counts.
To determine the profiles, the MSH 15--52 observations were stacked in the field-of-view coordinate
system; the source counts are located around the offset angle $\theta=0.5\degree$ under
which the source was observed.
\label{fig:msh_residuals_profiles_eplaw}
}
\end{figure*}

\begin{figure*}
\centering
\includegraphics[width=\textwidth]{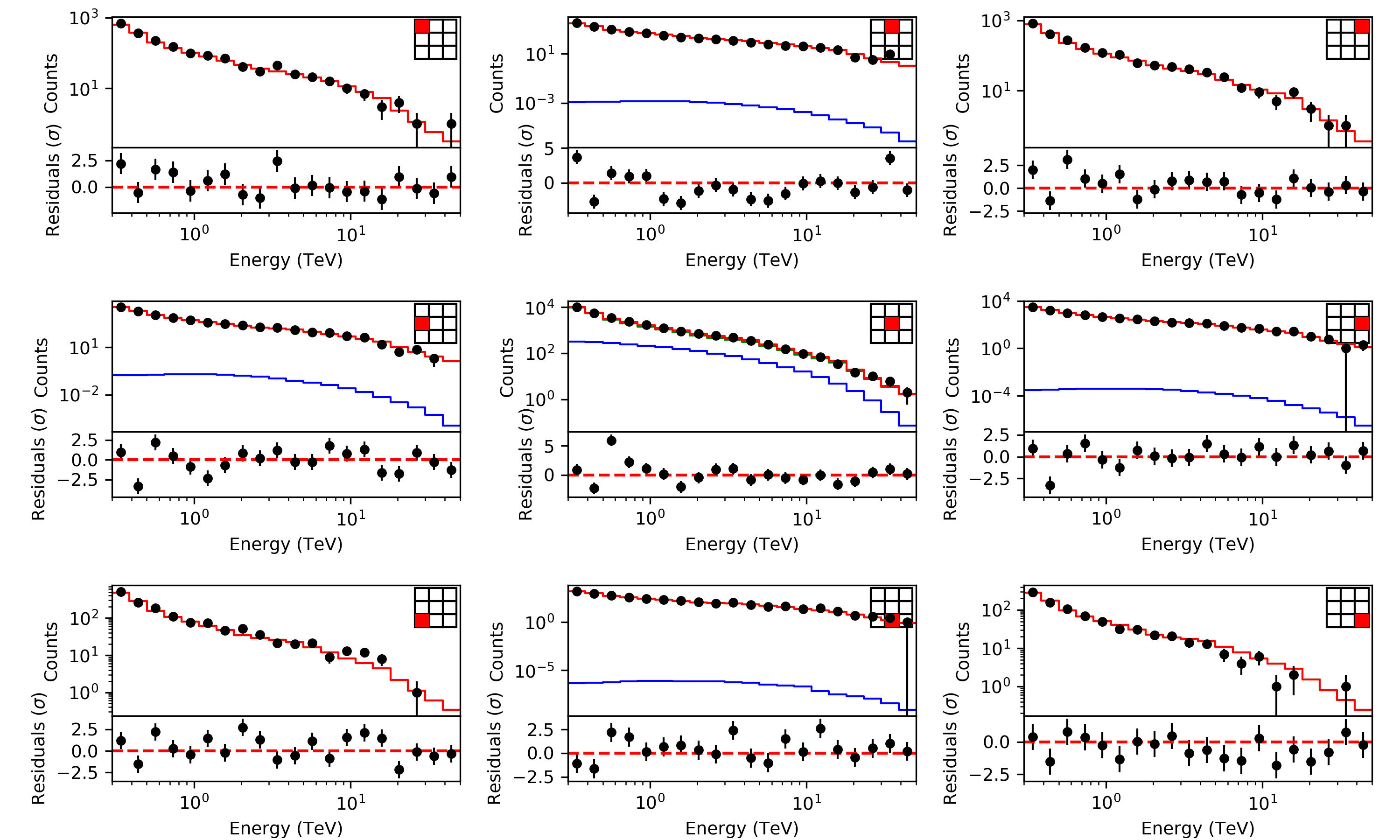} 
\caption{
Sub-region counts spectra and residuals spectra for the RX J1713.7--3946 observations.
Red lines represent the total predicted model counts, blue lines the predicted source counts and
green lines the predicted background counts.
\label{fig:rx_residuals_sectors}
}
\end{figure*}

\begin{figure*}
\centering
\includegraphics[width=\textwidth]{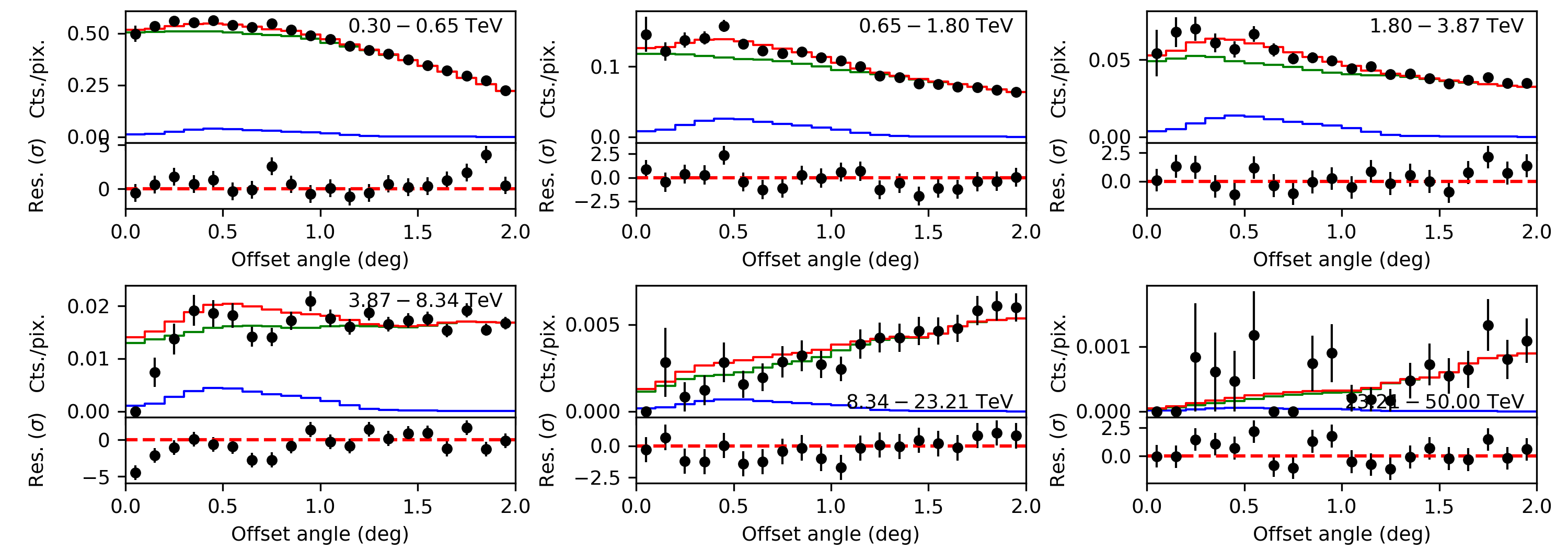} 
\caption{
Radial counts profiles and residuals for the RX J1713.7--3946 observations.
Red lines represent the total predicted model counts, blue lines the predicted source counts and green 
lines the predicted background counts.
To determine the profiles, the RX J1713.7--3946 observations were stacked in the field-of-view coordinate
system; the source counts are spread over the entire offset angle range with a maximum
of source events near the offset angle $\theta=0.5\degree$.
\label{fig:rx_residuals_profiles}
}
\end{figure*}

\begin{figure*}
\centering
\includegraphics[width=\textwidth]{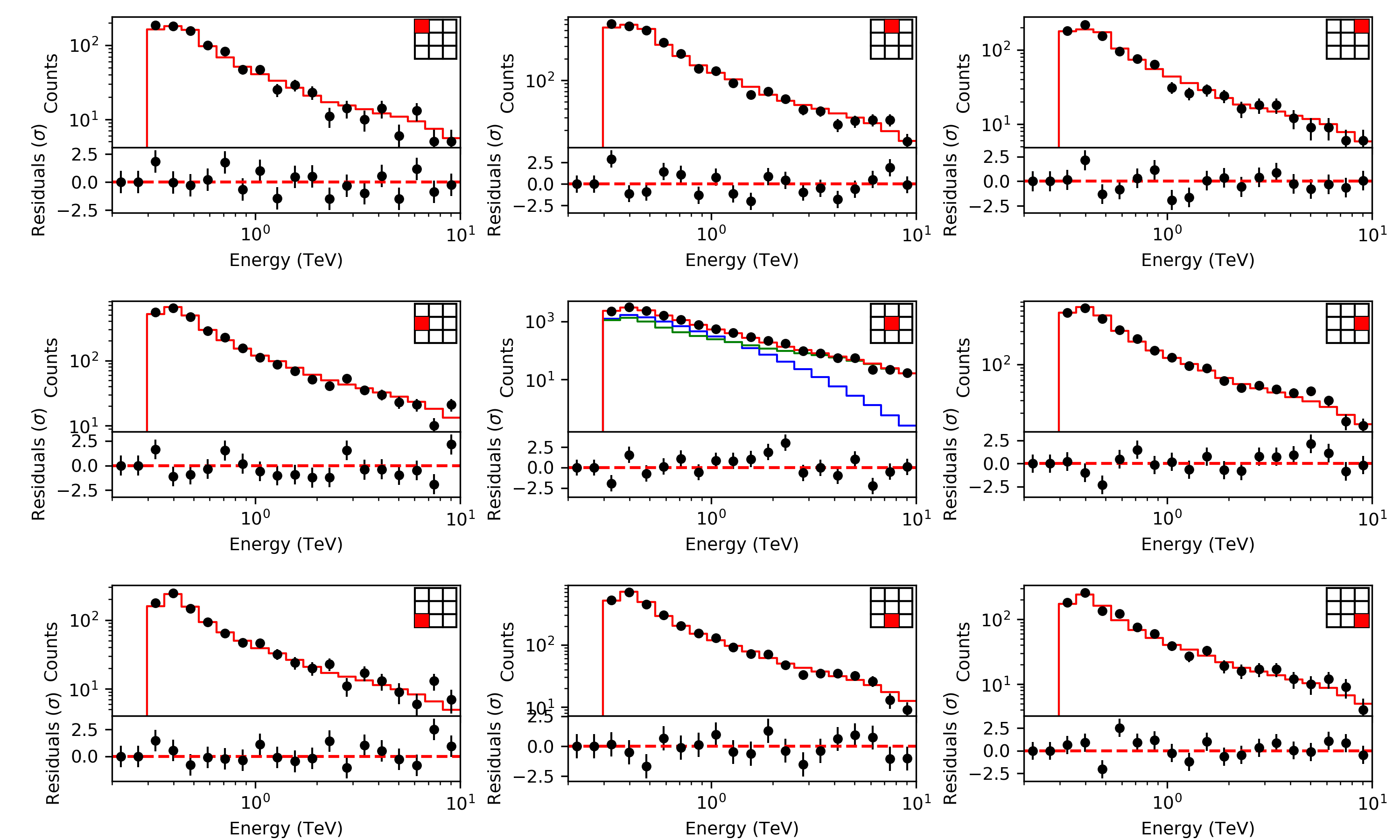}
\caption{
Sub-region counts spectra and residuals spectra for the PKS 2155--304 observations.
Red lines represent the total predicted model counts, blue lines the predicted source counts and green 
lines the predicted background counts.
\label{fig:pks_residuals_sectors}
}
\end{figure*}

\begin{figure*}
\centering
\includegraphics[width=\textwidth]{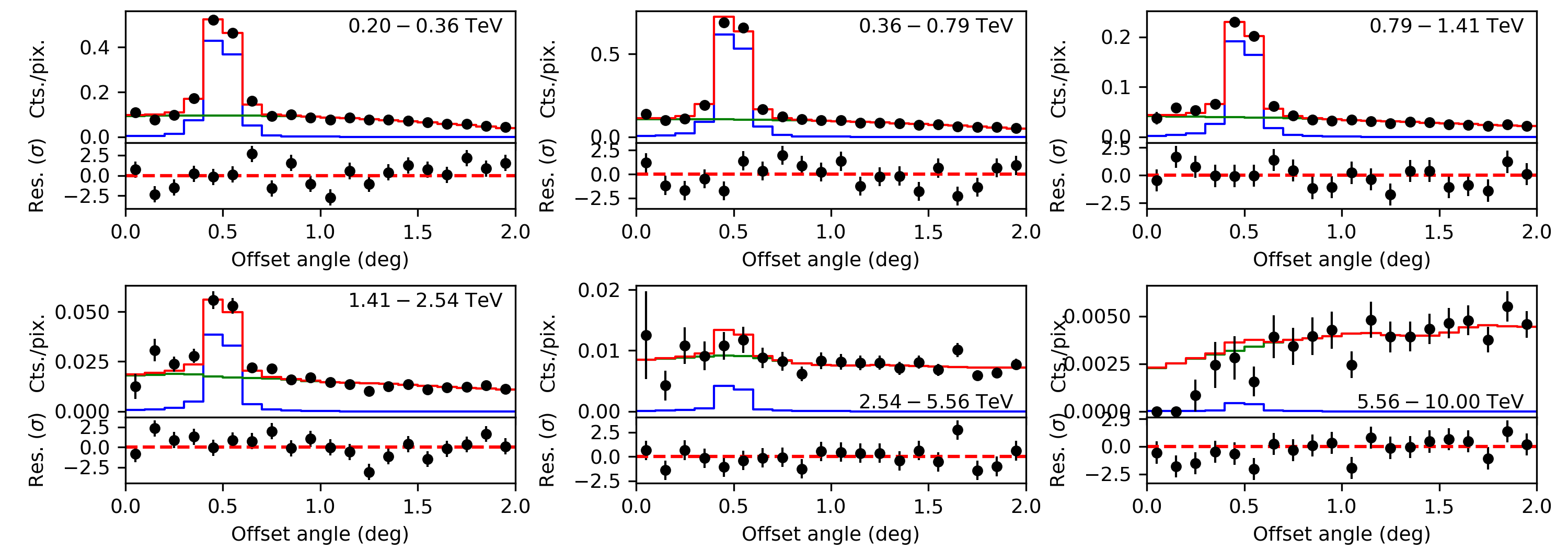}
\caption{
Radial counts profiles and residuals for the PKS 2155--304 observations.
Red lines represent the total predicted model counts, blue lines the predicted source counts and green 
lines the predicted background counts.
To determine the profiles, the PKS 2155--304 observations were stacked in the field-of-view coordinate
system; the source counts are located around the offset angle $\theta=0.5\degree$ under
which the source was observed.
\label{fig:pks_residuals_profiles}
}
\end{figure*}

In addition, residual spectra for a grid of $3\times3$ sub-regions and radial residual profiles
for each of the four sources are shown in Figs.~\ref{fig:crab_residuals_sectors} to \ref{fig:pks_residuals_profiles}.

\clearpage

\clearpage

\clearpage

\clearpage

\clearpage






\end{appendix}

\end{document}